\documentclass[twocolumn, amsmath,amssymb, superscriptaddress]{revtex4}

\usepackage{dcolumn}
\usepackage{bm}
\usepackage{amsmath}
\usepackage{amsfonts}
\usepackage{graphics}
\usepackage[pdftex]{graphicx}
\usepackage{times}
\usepackage{setspace}
\usepackage{color}   

\begin{document}

\title{Statistical regularities in the rank-citation profile of scientists}

\author{Alexander M. Petersen }
\affiliation{IMT Lucca Institute for Advanced Studies, Lucca, Italy}
\affiliation{Center for Polymer Studies and Department of Physics, Boston University, Boston, Massachusetts 02215, USA}
\author{H. Eugene Stanley}
\affiliation{Center for Polymer Studies and Department of Physics, Boston University, Boston, Massachusetts 02215, USA}
\author {Sauro Succi}
\affiliation{Istituto Applicazioni Calcolo C.N.R.,  Rome, IT} 
\affiliation{Freiburg Institute for Advanced Studies, Albertstrasse, 19, D-79104, Freiburg, Germany} 
\date{\today}

\begin{abstract}
Recent {\it ``science of science'' } research  shows that  scientific impact measures for  journals and individual articles  have quantifiable regularities across both  time and discipline. 
However, little is known about the scientific impact distribution at the scale of an individual scientist.
We analyze the aggregate scientific production and impact of individual careers
using the rank-citation profile $c_{i}(r)$ of 200 distinguished professors and 100 assistant professors. 
For the entire range of paper rank $r$, we fit each $c_{i}(r)$ 
to a common distribution function that is parameterized by two scaling exponents. Since two scientists with equivalent {\it Hirsch} $h$-index  can have significantly different $c_{i}(r)$ profiles, our results demonstrate the
utility of the $\beta_{i}$ scaling parameter in conjunction with $h_{i}$ for quantifying individual publication impact. We show that the total number of citations $C_{i}$ tallied from a  scientist's  $N_{i}$ papers scales as $C_{i} \sim h_{i}^{1+\beta_{i}}$.
Such statistical regularities in the input-output patterns of scientists can be used as  benchmarks for theoretical models of career progress.
\end{abstract}

\maketitle

A scientist's career path is subject to a myriad of  decisions and unforeseen events, e.i. Nobel Prize worthy discoveries \cite{citationboosts}, that can significantly alter an individual's career trajectory. 
As a result,  the career path can be
difficult to analyze since there are potentially many factors (e.g. individual, mentor-apprentice, institutional, coauthorship, field) \cite{Matthew1, Matthew2, socialstratification, TeamAssembly, mentoreffect, UnivCite, Scientists} to account for in the statistical analysis of scientific panel data. 

The rank-citation profile, $c_{i}(r)$, represents the number of citations of individual $i$ to his/her paper $r$, ranked in decreasing order 
$c_{i}(1) \ge c_{i}(2) \ge \dots c_{i}(N)$, and provides a quantitative synopsis of  a given scientist's publication career. 
Here, we analyze the rank-ordered citation  distribution $c_{i}(r)$ for $300$ scientists in order to better understand  patterns of success and to characterize scientific production at the individual scale using a common framework. 
The review of scientific achievement for post-doctoral selection, tenure review, award and academy selection,  at all stages of the career is becoming largely based on quantitative publication impact measures. Hence, understanding quantitative  patterns in production are important for developing a transparent and unbiased review system.
Interestingly, we observe  statistical regularities in $c_{i}(r)$ that  are remarkably robust despite the idiosyncratic details of scientific achievement and career evolution. 
Furthermore, empirical regularities in scientific achievement suggest that there are fundamental social forces governing career progress \cite{CareerTrajectory,BB2,GrowthDynamicsH,GrowthCareers}.

We group the 300 scientists that we analyze into three sets of $100$, referred to as datasets A, B and C, so that we can  analyze and compare the complete publication careers of each individual, as well as across the three groups:
\renewcommand{\labelenumi}{[\Alph{enumi}]}
\begin{enumerate}
\item 100 highly-profile  scientists with  average $h$-index $\langle h \rangle = 61 \pm 21$. These scientists were selected using the
citation shares metric \cite{Scientists} to quantify cumulative career impact in the journal  {\it Physical Review Letters} (PRL).
\item 100  additional ``control" scientists with average $h$-index $\langle h \rangle = 44 \pm 15$. 
\item 100  current Assistant professors with average $h$-index $\langle h \rangle = 14 \pm 7$. We   selected two scientists from each of the top-50 US physics departments (departments ranked according to the magazine  {\it U.S. News}). 
\end{enumerate} 
In the  methods section we  describe the selection procedure for datasets A, B and C in more detail.  

\begin{figure*}
\centering{\includegraphics[width=0.6\textwidth]{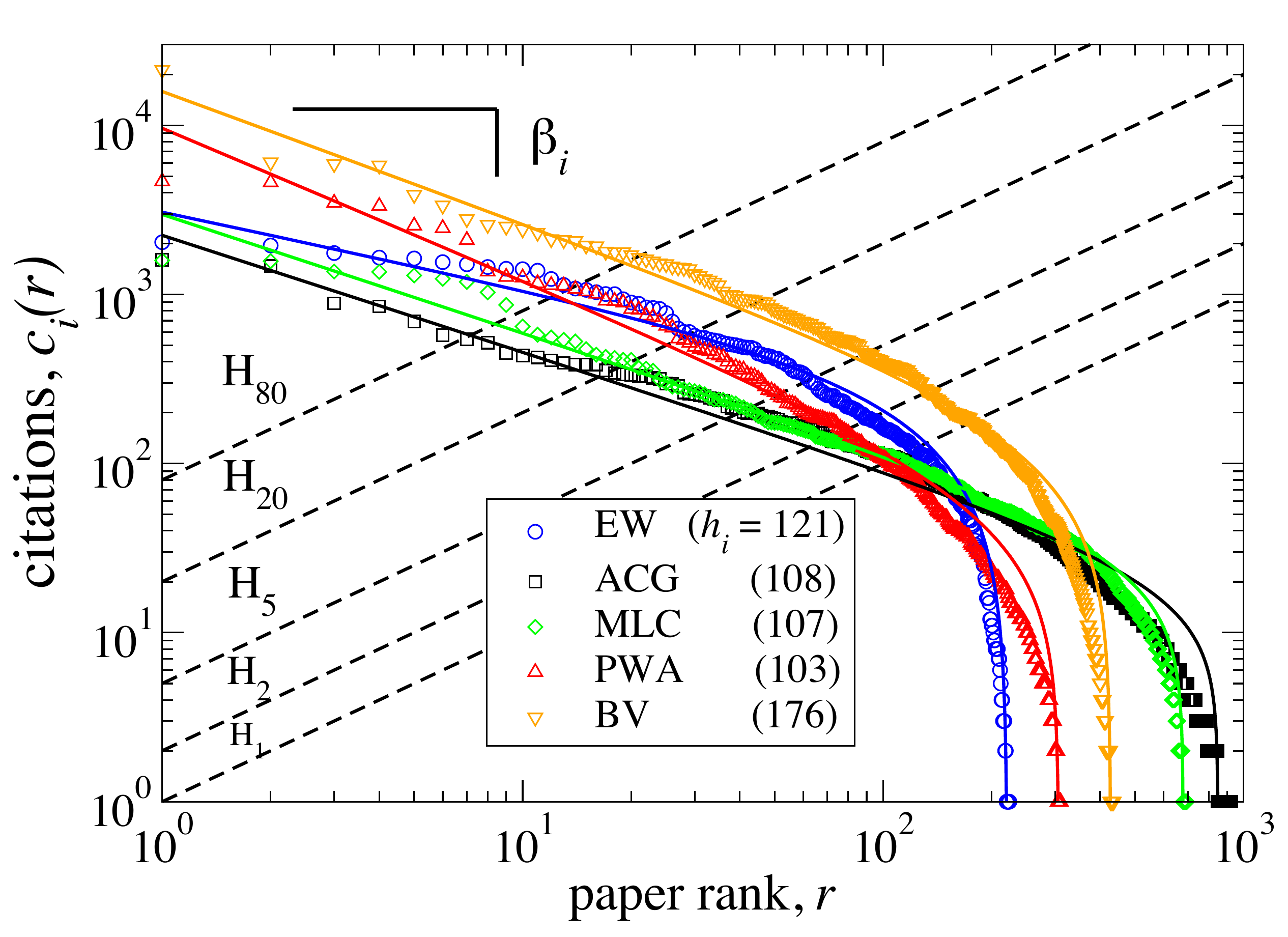}}
\caption{\label{Ztop5}  {\bf The citation distribution of individual scientists is heavy-tailed.}
 We show 5 empirical rank-citation $c_{i}(r)$ profiles  corresponding to extremely high-impact scientists whose initials and $h$-index (as of Jan. 2010) are listed in the legend. 
The hierarchical scaling pattern in $c_{i}(r)$ for small $r$ values indicate that  the pillar contributions of top scientists are ``off-the-charts'' since they have no characteristic scale. Furthermore, this statistical regularity demonstrates the utility of the $\beta_{i}$ scaling exponent  in characterizing the highly cited papers of a given scientist $i$.
 Interestingly, each  scientist has coauthored a significant number of papers that are 
 significantly lower impact than their $c_{i}(1)$ pillar paper. 
 The $c_{i}(r)$ distributions show significant  variability in both the high-rank ($\beta$) and low-rank ($\gamma$) regimes. Moreover, for $c_{i}(r)$ with  similar $h$ values, the $h$-index (a single point on each curve) is insufficient to adequately distinguish  career profiles. 
The solid curves are the best-fit DGBD functions (see Eq. \ref{Cr}) for each corresponding $c_{i}(r)$ over the entire rank range in each case. 
The intersection of $c_{i}(r)$ with the line  $H_{p}(r)$ corresponds to the generalized $h$-index $h_{p}$, which together  uniquely quantify the $c_{i}(r)$ profile.
Five $H_{p}(r)$ lines are provided for reference, with $p=\{1,2,5,20,80\}$. } 
\end{figure*}

There are many conceivable ways to quantify the impact of a scientist's $N_{i}$ publications. The
$h$-index \cite{H}  is a widely acknowledged  single-number measure that 
serves as a proxy for production and  impact simultaneously.
The $h$-index $h_{i}$ of scientist $i$ is defined by a single point on the rank-citation 
profile $c_{i}(r)$  satisfying the condition
\begin{equation}
\label{H} c_{i}(h_{i}) = h_{i} \ .
\end{equation} 
To address the shortcomings of  the $h$-index, numerous remedies  have been 
proposed in the bibliometric sciences \cite{ComparisonIndex}. 
For example, Egghe proposed the $g$-index, where the most cited $g$ papers 
cumulate $g^2$ citations overall \cite{G}, and Zhang proposed the $e$-index which 
complements the $h$ and $g$ indices quantitatively \cite{E}.  

 To justify the importance of analyzing the entire profile $c_{i}(r)$, consider a scientist $i = 1$  with rank-citation profile $c_{1}(r)
\equiv$ [100, 50, 33, 25, 20, 16, 14, 12, 11, 10, 9...]  and a scientist $i = 2$ with 
$c_{2}(r) \equiv$ [10, 10, 10, 10, 10, 10, 10, 10, 10, 10, 9 ...]. Both scientists have the 
same $h$-index value $h=10$, although $c_{1}(r)$ tallies 2.9 times as many citations as $c_{2}(r)$ from his/her most-cited 10 papers. 
Hence, an additional parameter $\beta_{i}$  is necessary in order to distinguish these two example careers. 
Specifically, the $\beta_{i}$ parameter quantifies the scaling slope in $c_{i}(r)$ for the high-rank papers corresponding to small $r$ values. In this simple illustration,  $\beta_{1}\approx 1$ while  $\beta_{2} \approx 0$.

In Fig. \ref{Ztop5} we plot  $c_{i}(r)$ for 5 extremely high-impact scientists.
The individuals EW,  ACG, MLC, and PWA are physicists with the largest $h_{i}$ values in our data set; BV is a prolific molecular biologists who we include in this graphical illustration in order to demonstrate the generality of the statistical regularity we find,  which likely exists across discipline. To demonstrate how the singe point $c_{i}(h_{i})$ is
an arbitrary point along the $c_{i}(r)$ curve,  we also plot the lines $H_{p}(r) \equiv p \ r$ for 5 values of $p =\
\{1,2,5,20, 80\}$. The value $p\equiv 1$ recovers the $h$-index $h_{1} = h$ proposed  by Hirsch.  The intersection  of any given line $H_{p}(r)$ with $c_{i}(r)$ corresponds to the ``generalized $h$-index" $h_{p}$, 
\begin{equation}
c(h_{p})= p h_{p} \ ,
\end{equation}
proposed 
in \cite{genH} and further analyzed in \cite{genH2}, with the relation $h_{p} \le h_{q}$ for $p > q$. 
Since the value $p\equiv 1$ is chosen somewhat arbitrarily, we take an alternative approach  which is to quantify the entire 
$c_{i}(r)$ profile at once (which is also equivalent to knowing the entire $h_{p}$ spectrum).  Surprisingly, because 
we find regularity in the functional form $c_{i}(r)$ for all 300 scientists analyzed, we can relate the relative impact of 
a scientist's publication career using the small set of parameters that specify the $c_{i}(r)$ profile for the entire set of papers ranging from rank $r = 1...N_{i}$. Using a much smaller
parameter space than the $h_{p}$ spectrum, we can begin to analyze the statistical regularities in the career accomplishments of scientists. 

The aim of this analysis is not to add another level of scrutiny to the review of scientific careers, but rather, to highlight the 
regularities across careers and to seed further 
exploration into the mechanisms  that underlie career success. The aim of this brand of quantitative social science is to utilize the vast amount of information available to 
develop an academic framework that is sustainable, efficient and fruitful. Young scientific careers are like ``startup'' companies that need
appropriate venture funding to support the career trajectory through lows as well as highs  \cite{GrowthCareers}.

\section{Results}
\subsection{A Quantitative Model for $c_{i}(r)$}

\begin{figure}
\centering{\includegraphics[width=0.45\textwidth]{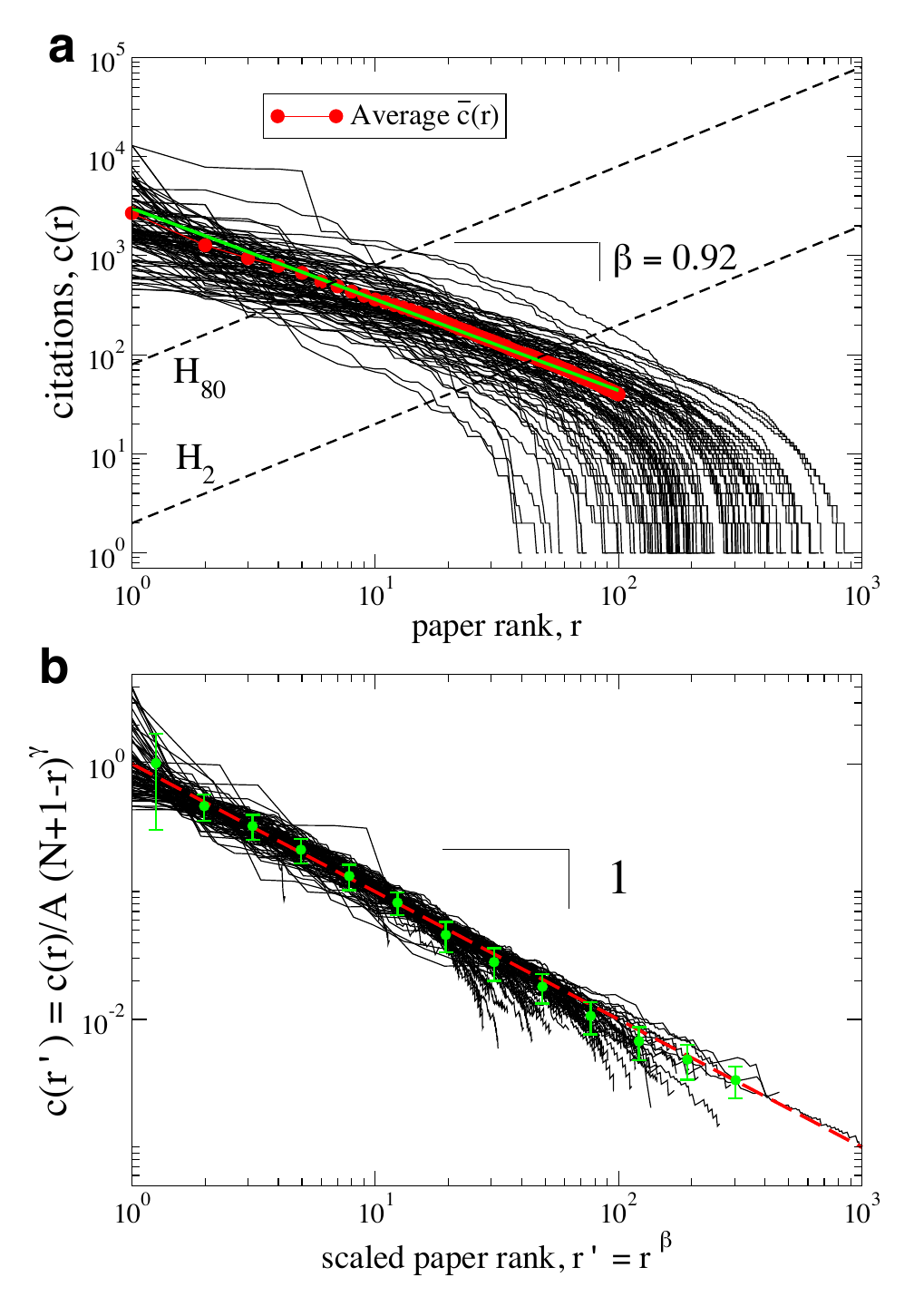}}
\caption{\label{ZAveTop100} {\bf Data collapse of each $c_{i}(r)$ along a universal curve}. A comparison of 100 rank-citation profiles $c_{i}(r)$  demonstrates the statistical regularity in  career publication output. Each scientist produces a cascade of papers of varying impact between the $c_{i}(1)$ pillar paper down to the least-known paper $c_{i}(N_{i})$. {\bf (a)} Zipf rank-citation profiles $c_{i}(r)$ for 100 scientists listed in dataset [A]. 
For reference, we plot the average $\overline{c}(r)$ of these 100 curves and find $\overline{c}(r) \sim r^{-\beta}$ with $\beta =
0.92 \pm 0.01$. The solid green line is a least-squares fit to $\overline{c}(r)$ over the range $1 \leq r
\leq 100$. 
 We also plot the $H_{2}(r)$ and $H_{80}(r)$ lines for reference. 
   {\bf (b)} We re-scale the curves in panel (a), plotting $c_{i}(r') \equiv c_{i}(r)/ A(r_{1}+1-r)^{\gamma}$,  where we use
the best-fit $\gamma_{i}$ and $A_{i}$ parameter values for each individual $c_{i}(r)$ profile. Using the  rescaled rank value $r' \equiv
r^{\beta_{i}}$, we
show excellent data collapse onto the expected curve $c(r') = 1/r'$. (see Figs. \ref{ZAveR100} and \ref{ZAveAsst100} 
for analogous plots for dataset [B] and [C] scientists). Green data points correspond to the average $c(r')$ value with 1$\sigma$ error bars  calculated using all 100 $c_{i}(r')$ curves separated into logarithmically spaced bins. 
 }
\end{figure}

For each scientist $i$, we find that $c_{i}(r)$  can be approximated by 
a scaling regime for small $r$ values, followed by a truncated scaling regime for large $r$ values. 
Recently a novel distribution, the discrete generalized beta distribution (DGBD)
\begin{equation}
c_{i}(r) \equiv A_{i} r^{-\beta_{i}} (N_{i}+1-r)^{\gamma_{i}} 
\label{Cr}
\end{equation}
 has been proposed as a model for rank profiles  in the social and natural sciences that 
exhibit such truncated scaling behavior  \cite{DGBfunc, RankOrder}. 
 The parameters $A_{i}$, $\beta_{i}$, $\gamma_{i}$ and $N_{i}$ are each defined for a  given $c_{i}(r)$ corresponding to an
individual scientists $i$, however we suppress the index $i$ in some equations to keep the notation concise. We estimate the
two scaling parameters $\beta_{i}$ and $\gamma_{i}$ using {\it Mathematica} software to perform a multiple linear regression of $\ln c_{i}(r) = \ln A_{i} - \beta_{i} \ln r + \gamma_{i} \ln (N_{i}+1-r)$ in the base functions $\ln r$ and $\ln (N_{i}+1-r)$. In our fitting procedure we replace $N$ with
$r_{1}$,  the largest value of $r$ for which $c(r) \geq 1$ (we find that $r_{1} / N_{i} \approx 0.84 \pm 0.01$ for careers in datasets A and B).   Figs. \ref{Ztop5} and \ref{ZAveTop100} demonstrate the utility of the DGBD to represent 
$c_{i}(r)$, for both large and small $r$. The regression correlation coefficient $R_{i}>0.97$ for all $\ln
c_{i}(r)$ profiles analyzed.

The DGBD proposed in \cite{DGBfunc} is an improvement over the Zipf law (also called the generalized power-law or Lotka-law \cite{EggheLotka}) model and the stretched exponential
model \cite{H} since it reproduces the varying curvature in $c_{i}(r)$ for both small and large $r$. 
Typically,  an exponential cutoff is imposed in the power-law model, and justified
as a finite-size effect. The DGBD does not require this assumption, but rather, introduces a second scaling exponent
$\gamma_{i}$ which controls the curvature  in $c_{i}(r)$ for large $r$ values. The DGBD  has been
successfully  used to model numerous rank-ordering profiles analyzed in \cite{DGBfunc, RankOrder} which arise in the
natural and socio-economic sciences.  The relative values of the $\beta_{i}$ and
$\gamma_{i}$ exponents are thought to capture two distinct mechanisms that contribute to the evolution of $c_{i}(r)$
\cite{DGBfunc, RankOrder}. Due to the data limitations in this study, we are not able to study the dynamics in $c_{i}(r)$ through time. Each $c_{i}(r)$ is a  ``snapshot'' in time, and so we can only conjecture on the evolution of $c_{i}(r)$ throughout the career. Nevertheless, we believe that there is likely a positive feedback  effect between the ``heavy-weight'' papers and ``newborn'' papers, whereby the reputation of the ``heavy-weight'' papers  can increase the exposure and impact the perceived significance of ``newborn'' papers during their infant phase. 
Moreover, the 2-regime power-law behavior of $c_{i}(r)$ suggests that the reinforcement dynamics can be quantified by the scale-free parameters $\beta$ and $\gamma$.

The $\beta_{i}$ value determines the relative change in the $c_{i}(r)$ 
values for the high-rank papers, and thus it can be used to further distinguish 
the careers of two scientists with the same $h$-index.
In particular,  smaller $\beta$ values characterize flat profiles with relatively low contrast between the high and low-rank regions of any given profile, while larger $\beta$ values indicate a sharper separation between the two regions. 

In Fig. \ref{ZAveTop100}(a) we plot $c_{i}(r)$ for each scientist from dataset [A] as well as
 the average of the 100 individual curves $\overline{c}(r) \equiv\frac{1}{100}\sum_{i=1}^{100} c_{i}(r)$ (see Figs. \ref{ZAveR100} and \ref{ZAveAsst100}
for analogous plots for datasets [B] and [C]). 
We find robust power-law scaling 
\begin{equation}
\overline{c}(r) \sim r^{-\beta} \ \ \  [\beta \approx 0.92 \pm 0.01]
\end{equation}
for $10^{0} \leq r  \leq 10^{2}$. 
Interestingly, this  $\beta$ value is similar to the scaling exponents calculated for other  rank-size (Zipf) distributions
in the social and economic sciences, e.g.  word frequency \cite{ZipfLawWord} 
and city size \cite{DGBfunc, RankOrder,ZipfLawCity}. 

We  calculate each $\beta_{i}$ value using a multilinear least-squares regression of $\ln c_{i}(r)$ for $ 1 \leq r \leq
r_{1}$ using the DGBD model defined in Eq.~[\ref{Cr}]. 
To properly weight the data points for better regression fit over the entire 
range, we  use only $20$ values of $c_{i}(r)$ data points that are equally spaced on 
the logarithmic scale in the range $r \in [1,r_{1}]$. We elaborate the details  of this fitting technique in the methods section.
We plot five empirical $c_{i}(r)$ along with their corresponding best-fit DGBD functions in Fig. \ref{Ztop5} to demonstrate the goodness of fit for the
entire range of $r$. 

In order to demonstrate the common functional form of  the DGBD model, we 
 collapse  each $c_{i}(r)$ along a universal 
scaling function $c(r') = 1/r'$, by using the rescaled rank values $r' \equiv r^{\beta_{i}}$ defined for each curve. 
In Figs. \ref{ZAveTop100}(b),  \ref{ZAveR100}(b) and \ref{ZAveAsst100}(b), we plot the quantity $c_{i}(r') \equiv c_{i}(r)/ A(r_{1}+1-r)^{\gamma}$,  
using the best-fit $\gamma_{i}$ and $A_{i}$  parameter values for each individual
$c_{i}(r)$ profile. While the curves in Fig. \ref{ZAveTop100}(a) are jumbled and distributed over a large range of $c(r)$ values, 
the rescaled $c_{i}(r)$ curves in Fig. \ref{ZAveTop100}(b) all lie approximately along the predicted curve $c(r') = 1/r'$. 

\subsection{Using $c_{i}(r)$ to quantify career production and impact} 

A  main advantage of the $h$-index is the simplicity in which it is 
calculated, e.g. {\it ISI Web of Knowledge} \cite{WofK} readily 
provides this quantity online for distinct authors.  
Another  strength of the $h$-index is  its stable growth with respect to changes in $c_{i}(r)$ due to time and information-dependent factors \cite{Hstability}. Indeed, the $h$-index is 
a ``fixed-point" of the citation profile. This time stability is evident in the observed growth rates of $h$ for  scientists. 
Average growth rates, calculated here as $h/L$, where $L$ is the duration in years between a given author's first and
most recent paper, 
typically lie in the range of one to three units per year (this annual growth rate corresponds to 
the quantity $m$ introduced by Hirsch \cite{H}). Annual growth rates $h/L \approx 3$ correspond to 
exceptional scientists (for the histogram of $P(h/L)$ see the Fig. \ref{H1byCLPDF} and  for $h/L$ values see the SI text (Tables S1-S4)). 
As a result,  $h/L$ is a good predictor for future achievement along with $h$ \cite{H2}.

 It is truly remarkable how a single number, $h_{i}$, correlates with other measures of impact. Understandably, being just a single number, the $h$-index cannot
fully account for other factors, such as variations in citation standards and coauthorship patterns across discipline \cite{hindexResearchers,hindexFields, ProConH}, nor can $h_{i}$
incorporate the full information contained in the entire $c_{i}(r)$ profile.
As a result, it is widely appreciated that the $h$-index can  underrate the value of the best-cited papers, since once a paper transitions into the  region $ r \leq h_{i}$, 
its citation record is discounted, until other less-cited papers with 
$r>h_{i}$ eventually overcome the rank ``barrier"  $r = h_{i}$. Moreover, as noted in \cite{H}, the papers for which $r > h_{i}$ do not contribute any additional credit.

\begin{figure}
\centering{\includegraphics[width=0.49\textwidth]{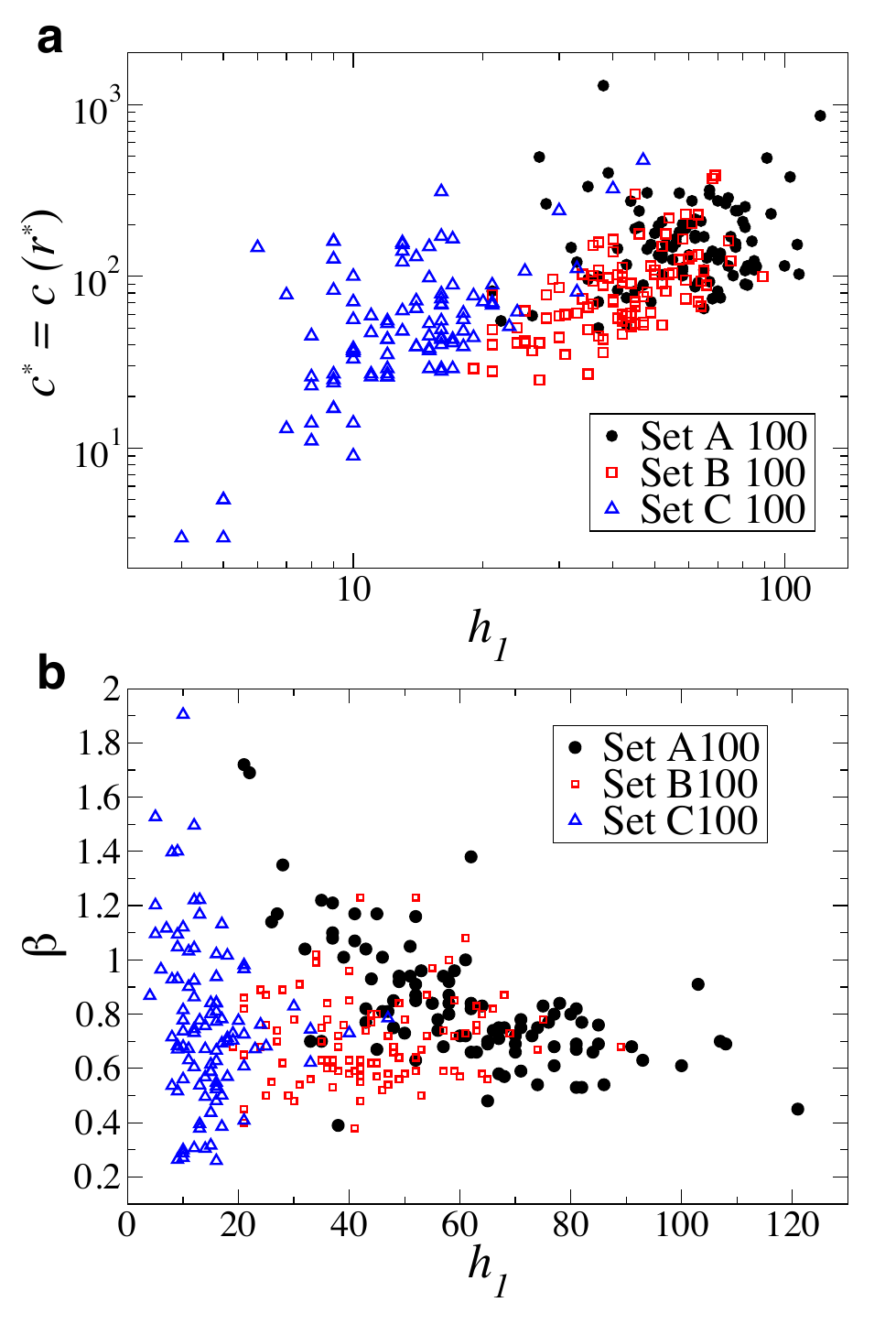}}
\caption{\label{CrstarH} 
{\bf Limitations to the use of the $h$-index alone.}
The $h$-index can be insufficient in comprehensively representing $c_{i}(r)$.  
{\bf (a)} The $h$-index does not contain any information about $c_{i}(r)$ for $r < h_{i}$, and can shield a scientist's most successful accomplishments which are the basis for much of a scientist's reputation.
This is evident in the cases where $c(r_{i}^{*}) \gg h_{i}$, in which case the $h$-index cannot account for the stellar impact of the papers. {\bf (b)}
For a given $h_{i}$ value, prolific careers are characterized by a large $\beta_{i}$ value, as it is harder to maintain large $\beta_{i}$ values for large $h_{i}$. As a result, the 
$\beta_{i}$ vs $h_{i}$ parameter space can be used to identify anomalous careers and to better compare two scientists with similar $h_{i}$ indices. We find that a third career metric $C_{i}$, the total number of citations to the papers of author $i$, can be calculated with high accuracy by the scaling relation $C_{i} \sim h_{i}^{1+\beta_{i}}$, which we illustrate in Fig. \ref{CbetaH1}(b). } 
\end{figure}

Instead of choosing an arbitrary $h_{p}$ as an productivity-impact indicator,
we use the analytic properties 
of the DGBD to calculate a  crossover value $r_{i}^{*}$.
In the methods section, we derive an exact expression for $r_{i}^{*}$ 
which highlights the  distinguished papers of a given author. 
To calculate $r_{i}^{*}$, we use the logarithmic derivative $\chi(r) \equiv d\ln c(r) / dr$ 
to quantify the relative change in $c_{i}(r)$  with increasing $r$. 
We defined papers as ``distinguished'' if they satisfy the  inequality $c_{i}(r)/c_{i}(r+1) > \exp (\overline{\chi})$, where $\overline{\chi}$ is the average value of  $\chi(r)$ over the entire range of $r$ values. 
This inequality selects the peak papers  which are significantly more cited than their  neighbors.
The peak region  $r \in [1,r_{i}^{*}]$ corresponds to a ``knee" in $c_{i}(r)$ when plotted on log-linear axes.  The dependence of  $\overline{\chi}$ and  $r_{i}^{*}$  on the three DGBD parameters $\beta_{i}$, $\gamma_{i}$ and $N_{i}$ are provided in the methods section.  

The advantage of $r_{i}^{*}$ is that
this characteristic 
rank value is a comprehensive representation of the stellar papers in the  high-rank scaling regime since it depends on the DGBD parameter values   $\beta_{i}$, $\gamma_{i}$ and $N_{i}$, and thus 
probes the entire citation profile. 
  Fig. \ref{CrstarH} shows a scatter plot of the ``$c$-star'' $c_{i}^{*} \equiv c_{i}(r_{i}^{*})$ and $h_{i}$
values calculated for each scientist and demonstrates that there is a 
non-trivial relation  between these two
single-value indices. It also shows  that for scientists within a small range of $c^{*}$ there is a large variation in the corresponding $h$ values, in some cases straddling across 
all three sets of scientists. Also, there are several $c_{i}^{*}$ values which significantly 
deviate from the  trend in Fig. \ref{CrstarH}, which is plotted on log-log axes. 
These results reflect the fact that 
the $h$-index cannot completely incorporate the entire $c_{i}(r)$ profile. 
We plot the histogram of $c_{i}^{*}$ and $r_{i}^{*}$ values in Figs.  \ref{CstarPDF} and \ref{rstarPDF}, respectively. 

To further contrast the values of $c_{i}^{*}$ and the $h$-index, we propose the 
``peak indicator" ratio $\Lambda_{i} \equiv c_{i}^{*} / h_{i}$, which corrects specifically
for the $h$-index penalty on the stellar papers in the peak region of $c_{i}(r)$. 
Thus, all papers in the peak region of $c_{i}(r)$ satisfy the condition $c_{i}(r)\geq h_{i} \Lambda_{i}$.
In an extreme example, R. P. Feynman has a peak value $\Lambda \approx 36$, indicating that his best papers are
monumental pillars with respect to his other papers which contribute to his $h$-index. Fig. \ref{PeakPDF} shows the histogram of $\Lambda_{i}$ values, with typical values for
dataset [A] scientists $\langle \Lambda \rangle \approx 3.4 \pm 3.9$, and for dataset [B] scientists $\langle \Lambda \rangle \approx 2.2 \pm 1.1$. 
This indicator can only be used to compare scientists with similar $h$ values, since a small $h_{i}$ can result in a large $\Lambda_{i}$.

An alternative  ``single number'' indicator is $C_{i}$, an author's  total number of citations
\begin{equation}
C_{i}=\sum_{r=1}^{N} c_{i}(r) \ ,
\label{Csum}
\end{equation}
 which incorporates the entire $c_{i}(r)$ profile. However, it has been shown that $\sqrt{C_{i}}$ correlates well with $h_{i}$ \cite{Hredner}, a result which we will demonstrate in Eq. [\ref{Cbh}] to follow directly  from a $c_{i}(r)$ with   $\beta_{i} \approx 1$.

\begin{figure}
\centering{\includegraphics[width=0.49\textwidth]{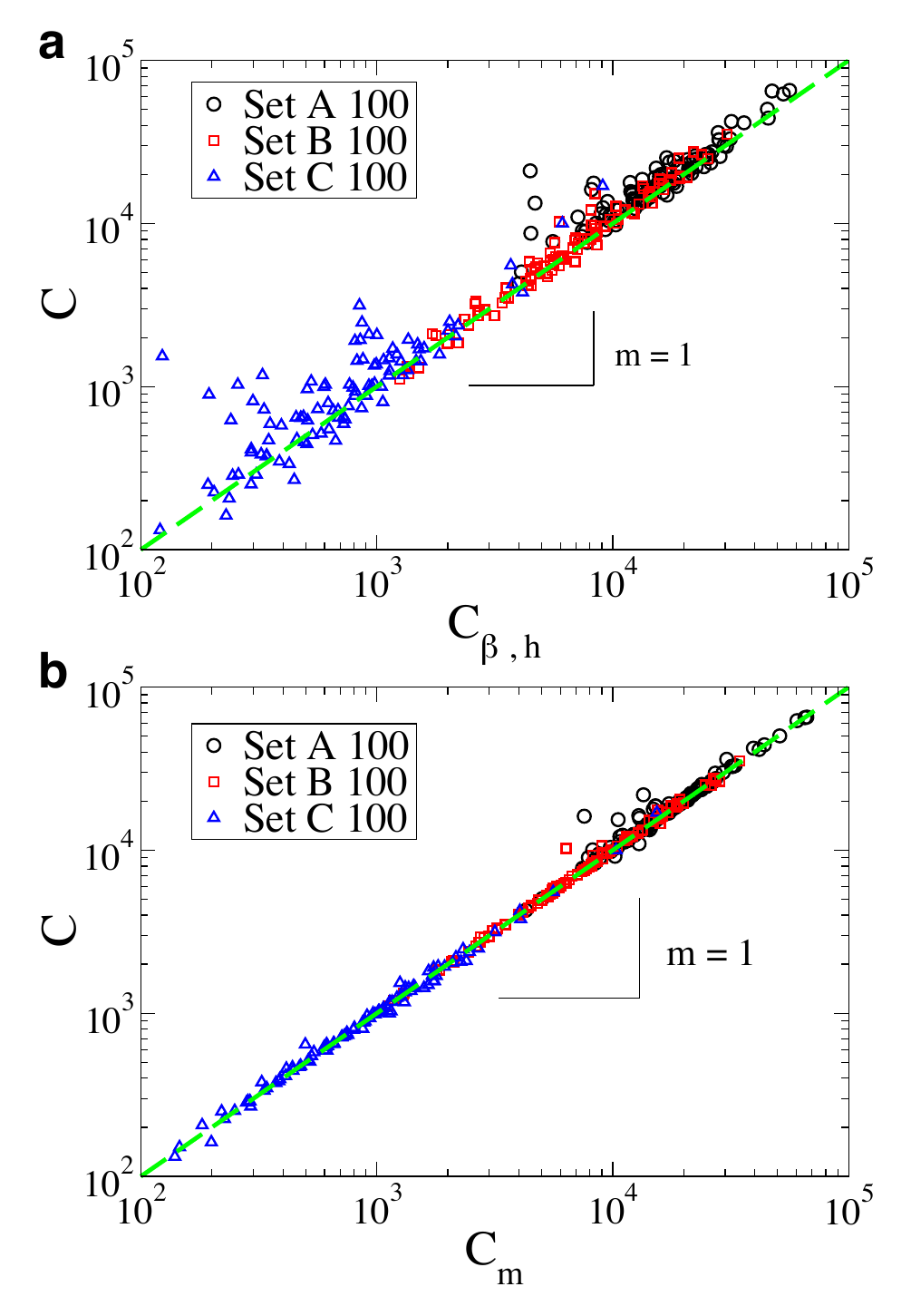}}
\caption{\label{CbetaH1} 
{\bf Aggregate publication impact $C$.}
The total number of citations $C_{i}\sim h_{i}^{1+\beta_{i}}$ is also comprehensive productivity-impact measure.
For most best-fit DGBD model curves, the $C_{i}$ value is preserved with high precision.
This shows that the difference between a given $c_{i}(r)$ and the corresponding best-fit DGBD model function are negligible on the macroscopic scale.
{\bf (a)} The exact aggregate number of citations $C_{i}$, calculated from $c_{i}(r)$ using Eq. [\ref{Csum}], can be analytically approximated  by $C_{\beta, h}$ using Eq. [\ref{Cbh}] which depends only on the scientist's $\beta_{i}$ and $h_{i}$ values. 
 {\bf (b)} 
We justify the use of the DGBD model defined in Eq. [\ref{Cr}] for the approximation of $c_{i}(r)$ by comparing the aggregate citations $C_{i}$ with the expected aggregate citations $C_{m} = \sum_{r=1}^{r_{1}} c_{m}(r)$ calculated from the best-fit DGBD model $c_{m}(r)$. Including the extra scaling-parameter, as in the DGBD model, improves the agreement between the theoretical and empirical $C_{i}$ values in (a) and (b). We plot the line $y=x$ (dashed-green line) for visual reference. } 
\end{figure}

We test the aggregate properties of $c_{i}(r)$ by
calculating 
the aggregate number of citations $C_{\beta, h}$ for a given profile,
\begin{equation}
C_{\beta, h} \equiv \sum_{r=1}^{N} A r^{-\beta} \approx  h^{1+\beta} \sum_{r=1}^{N'} r^{-\beta} = h^{1+\beta}H_{N',
\beta} \sim h^{1+\beta} 
\label{Cbh}
\end{equation}
where $H_{N', \beta}$ is the {\it generalized harmonic number} and is of order  $O(1)$ for $\beta \approx 1$. 
We neglect the $\gamma_{i}$ scaling regime since the low-rank papers do not significantly contribute  to an author's $C_{i}$ tally.
We approximate the coefficient $A$ in Eq.~[\ref{Cbh}] using the definition $c(h)\equiv h$, which implies that
$A/h^{\beta} \approx h$. 
We use the value $N' \equiv 3h$, so that $C_{\beta, h}$ can be approximated by only the two parameters $h_{i}$ and $\beta_{i}$ for
any given  author. 
We justify this choice of $N'$ by examining the  rescaled $c_{i}(r/h)$, which we consider to be 
negligible beyond  rank $r = 3 h_{i}$ for most scientists. 
In Fig.~\ref{CbetaH1}(a), we plot for each scientist the predicted $C_{\beta, h}$ value versus the empirical $C_{i}$ value,  
and we find excellent agreement with our theoretical prediction $C_{i}\sim h_{i}^{1+\beta_{i}}$ given by Eq. [\ref{Cbh}]. In Fig.~\ref{CbetaH1}(b), we plot
for each scientist the 
total number of citations $C_{m} = \sum_{r=1}^{r_{1}} c_{m}(r)$ using the  best-fit DGBD model $c_{m}(r) \equiv c_{i}(r; \beta_{i}, \gamma_{i}, A_{i}, r_{1})$  to approximate $c_{i}(r)$. The excellent agreement demonstrates that the fluctuations in the residual difference  $c_{m}(r) - c_{i}(r)$ cancel out  on the aggregate level. Furthermore, a comparison of the quality of agreement between the theoretical $C_{i}$ values and the empirical  $C_{i}$ values in Fig.~\ref{CbetaH1}(a) and (b) shows the importance of the additional $\gamma_{i}$ scaling regime in  the DGBD model.

\section{Discussion}

  
We use the DGBD model  to provide an
 analytic description of $c_{i}(r)$ over the entire range of $r$, and provide  a deeper quantitative understanding of 
scientific impact arising from an author's career publication works.  The DGBD model exhibits scaling
behavior for both large and small $r$, where the 
 scaling for small $r$ is quantified by the exponent $\beta_{i}$, which for many scientists analyzed, can be approximated
using only 
 two values of the generalized $h$-index $h_{p}$ (see SI text).  
In particular, we show that for a given $h$-value, a larger $\beta_{i}$ value corresponds to a more prolific publication career, since $C_{i}\sim h_{i}^{1+\beta_{i}}$.

 Many studies analyze only the high rank values of 
 generic Zipf ranking profiles $c(r)$, e.g. computing the scaling regime for  $r<r_{c}$ below some some rank cutoff $r_{c}$.
 However, these
 studies cannot quantitatively relate the large observations to the small observations within the system of interest. 
To account for this shortcoming, our method for calculating the crossover
values $r_{i}^{*}\equiv \overline{r}_{-}$, $r_{\mathsf x }$, and $\overline{r}_{+}$,  which we elaborate in the methods section, can be used in general to quantitatively distinguish relatively large
observations and relatively small observations within the entire set of observations. 
Moreover, the DGBD model has been shown to have wide application in 
quantifying the Zipf rank profiles in various phenomena \cite{RankOrder}.

To measure the upward mobility of a scientist's career, in the SI text we address the question: given that a scientist  has index $h$, what
is her/his most likely $h$-index value $\Delta t$ years in the future? 
In consideration of the bulk of $c_{i}(r)$, and following from the regularity of $c_{i}(r)$ for $r\approx h$,
we propose  a model-free gap-index
 $G(\Delta h)$ as both an estimate and a target for future achievement which can be used in the review of career
advancement.  The gap index $G(\Delta h)$, defined as a proxy for the total number of
citations a scientist needs to reach a target value $h+\Delta h$, can  detect the potential for fast $h$-index growth by quantifying $c_{i}(r)$ 
around $h$. This estimator differs from other estimators for the time-dependent $h$-index \cite{DynamicH, StochasticModel, SimulatingH} in that $G(\Delta h)$ is model independent. 
 
 Even though the productivity of scientists can vary substantially \cite{Scientists, ShockleyProductivity, ProductDiff,
huber98, PetersonPNAS}, and despite the complexity of  success in academia, we find remarkable statistical regularity in the
functional form of $c_{i}(r)$ for the scientists analyzed here from the physics community. Recent work in
\cite{Scientists, UnivCite, Rad2} calculates the citation distributions of papers from various disciplines and shows that proper normalization of impact measures can 
 allow for comparison  across time and discipline. 
 Hence, it is likely that the publication careers of productive scientists in many disciplines obey  the statistical
regularities observed here for the set of 300 physicists. Towards developing a  model for career evolution, it is still unclear how 
the relative strengths of two contributing factors  (i) the extrinsic cumulative advantage effect \cite{Matthew1, Matthew2, Scientists} versus (ii) the intrinsic role
of the "sacred spark" in combination with intellectual genius \cite{ProductDiff} manifest in the parameters of the DGBD model.

 With little calculation,  the $\beta_{i}$ metric developed here, used in conjunction with the $h_{i}$, can better answer
the question, ``How popular are your papers?'' \cite{howpopular}. 
 Since the cumulative impact and productivity of individual scientists are also found to obey
statistical laws \cite{Scientists, BB2}, it is possible that the  competitive nature of scientific advancement can be
quantified and utilized in order to monitor career progress.
Interestingly, there is strong evidence for a governing mechanism of career progress based on cumulative advantage  \cite{Scientists,cumadvprocess, BB2}
coupled with the 
the inherent talent of an individual, which results in statistical regularities in the career achievements of scientists as well as professional athletes \cite{BB0,BB2,BB3}. 
Hence, whenever data are available \cite{CompSocScience, SocialDynamicsRev}, finding statistical regularities 
emerging from human endeavors is a first step towards better understand the dynamics of human productivity. 

\section{Methods}
\subsection{Selection of scientists and data collection}
We use  disambiguated  ``distinct author'' data from {\it ISI Web of Knowledge}. 
This online database is  host to comprehensive data that is well-suited for developing testable models for
 scientific impact \cite{DiffusionRanking,Scientists,Rad2} and career progress \cite{BB2}. 
In order to approximately control for discipline-specific publication and citation factors, we analyze 300 scientists from the field of physics. 

We aggregate all authors who published in {\it Physical Review Letters} (PRL) over the 50-year period 1958-2008 into a common dataset. From this dataset, we rank the scientists using the citations shares metric defined in 
\cite{Scientists}. This citation shares metric divides equally the total number of citations a paper receives among the
$n$ coauthors, and also normalizes the total number of citations by a  time-dependent factor to account for citation
variations across time and discipline. 

Hence, for each scientist in the PRL database, we calculate a cumulative number of citation shares received
from only their PRL publications. This tally serves as a proxy for his/her scientific impact in all journals.
The top 100 scientists according to this citation shares metric comprise  dataset [A].
As a control, we also choose  100
other dataset [B] scientists, approximately randomly, from our ranked PRL list. The selection criteria for the control 
dataset [B]  group are that an author must have published between 10 and 50 papers in PRL. This likely ensures that the
total publication history, in all journals, be on the order of 100 articles for each author selected. 
We compare the tenured scientists in datasets A and B with 100 relatively young assistant professors in dataset [C]. 
To select dataset [C] scientists, we chose two assistant professors from the top 50 U.S. physics and astronomy departments (ranked 
according to the magazine {\it U.S. News}).

For privacy reasons, we provide in the SI tables only the abbreviated initials for each scientist's name (last name initial, first and middle name initial, e.g. L, FM). Upon request we can provide full names.

We downloaded datasets A and B from ISI Web of Science in Jan.  2010 and  dataset C from ISI  
ISI Web of Science in Oct.  2010.
We used the ``Distinct Author Sets" function provided by ISI in order 
to increase the likelihood that only papers published by each given author are analyzed. On a case by case basis, we performed further
author disambiguation  for each author.

\subsection{Statistical significance tests for the $c(r)$ DGBD  model}

We  test the statistical significance of the DGBD model fit using the $\chi^{2}$ test 
between the 3-parameter best-fit DGBD $c_{m}(r)$ and the empirical $c_{i}(r)$. 
We calculate the $p$-value for the $\chi^{2}$ distribution with $r_{1}-3$ degrees of freedom and find, for each data set, the number $N_{> p_{c}}$ of $c_{i}(r)$  with $p$-value  $> p_{c}$: $N_{> p_{c}}$ =
4 [A], 19 [B], 22 [C] for $p_{c}=0.05$,  and 8  [A], 22 [B], 37 [C] for $p_{c} = 0.01$. 

The significant number of  $c_{i}(r)$ which
do not pass the $\chi^{2}$ test for $p_{c} = 0.05$, results from the fact that the DGBD is a scaling function over several orders of magnitude in both  $r$ and $c_{i}(r)$ values, and so the residual differences  $[c_{i}(r) - c_{m}(r)]$ are not expected to be normally distributed since there is no characteristic scale for scaling functions such as the DGBD. Nevertheless, the fact that so many  $c_{i}(r)$ do pass the $\chi^{2}$ test at such a high significance level, provides evidence for the quality-of-fit of the DGBD model. For comparison, none of the $c_{i}(r)$ pass the $\chi^{2}$ test using the power-law model at the $p_{c} = 0.05$ significance level. In the next section, we will also compare the macroscopic agreement in the total number of citations for each scientist and the total number of citations predicted by the DGBD model for each scientist, and find excellent  agreement. 
s

\begin{figure*}
\centering{\includegraphics[width=0.6\textwidth]{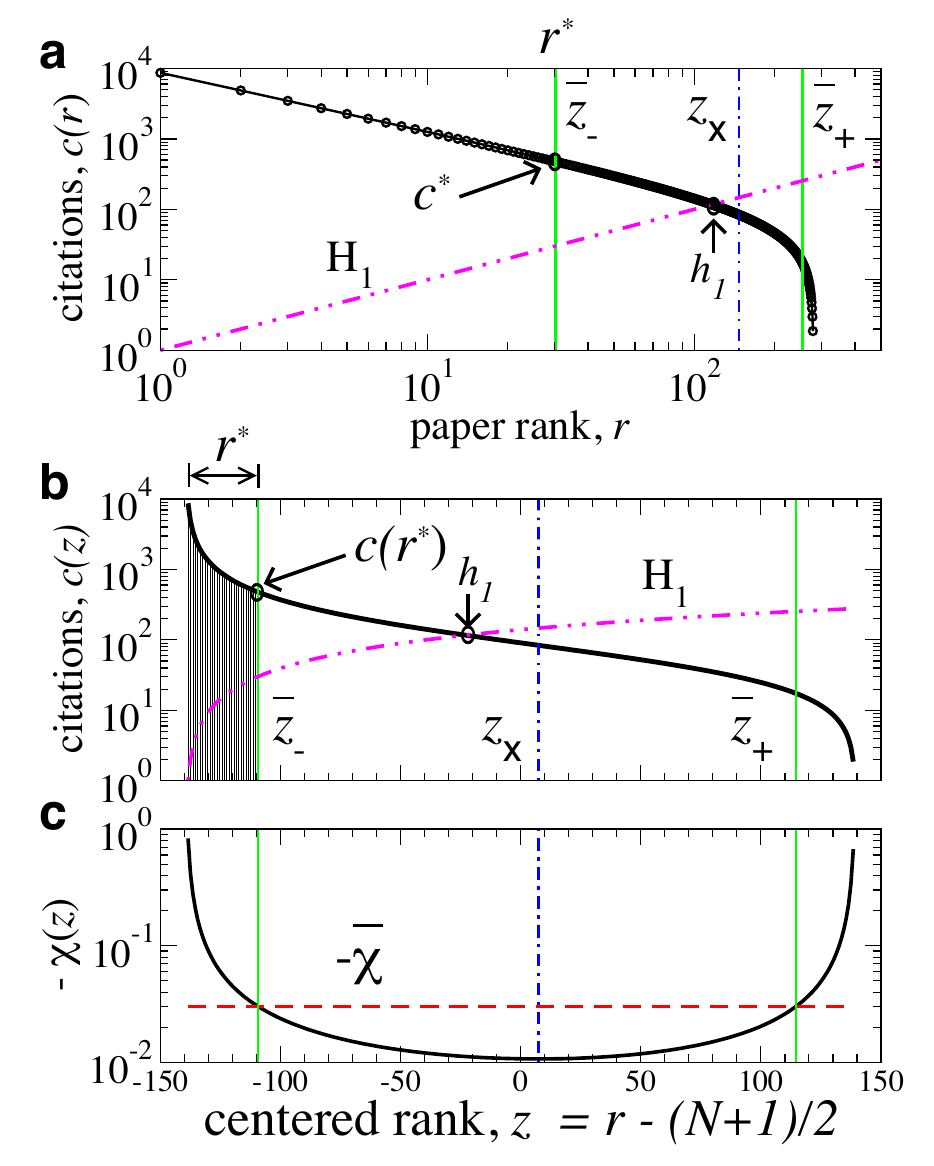}}
\caption{\label{demo} {\bf Characteristic properties of the DGBD.} 
We graphically illustrate the derivation of the characteristic $c_{i}(r)$ crossover values  that locate the two tail regimes of $c_{i}(r)$,  in particular, the
distinguished ``peak" paper regime corresponding to paper ranks $r \leq r^{*}$ (shaded region).
The crossover between two scaling regimes suggests a
complex reinforcement relation between the impact of a scientist's most famous papers and the impact of his/her other papers. 
{\bf(a)} The $c_{i}(r)$ plotted on log-log axes with $N=278$, $\beta=0.83$ and $\gamma=0.67$, corresponding to the average values of the Dataset [A] scientists.The hatched  magenta curve  is the $H_{1}(z)$ line on the log-linear
scale with corresponding $h$-index value $h = 104$.  The $r^{*}$ value for   $c_{i}(r)$ is not visibly obvious. 
{\bf(b)} 
We plot on log-linear axes the centered citation profile $c_{i}(z)$ (solid black curve) given by the symmetric rank transformation $z = r -z_{0}$ in Eq. [\ref{cz}].  This representation better highlights the peak paper regime, but fails to highlight the power-law $\beta$ scaling.
{\bf(c)} We plot the corresponding logarithmic derivative $\chi(z)$ of $c(z)$ (solid black curve),
which represents the relative change in $c(z)$. The dashed red line corresponds to $-\overline{\chi}$, where $\overline{\chi}
$ is the average value of $\chi(z)$ given by Eq. [\ref{avechi}]. The values of $\overline{z}_{\pm}$, indicated by the
solid vertical green lines, are defined as the intersection of $\overline{\chi}$ with $\chi(z)$ given by Eq. [\ref{zpm}].
The regime $z<\overline{z}_{-}$ corresponds to the best papers of a given author. The hatched blue line corresponds to
$z_{\mathsf x }^{-}$ which marks the crossover between the $\beta$ and $\gamma$ scaling regimes. } 
\end{figure*}

\subsection{Derivation of the characteristic DGBD $r$ values }

Here we use the analytic properties of the DGBD defined in Eq. [\ref{Cr}] to calculate 
the special $r$ values from the parameters $\beta$, $\gamma$ and $N$ which locate the two tail regimes of $c(z)$, and in particular, the distinguished paper regime.
 The scaling features of the DGBD do not 
readily convey any characteristic scales which distinguish the two scaling regimes. 
Instead, we use the properties of $\ln c_{i}(r)$ to characterize the crossover 
between the high-rank and the low-rank regimes of $c_{i}(r)$.

We begin by considering $c_{i}(r)$ under the centered rank transformation $z = r- z_{0}$, where $z_{0}=(N+1)/2$, then
\begin{equation}
c(z) = A \frac{(z_{0}-z)^{\gamma}}{(z_{0}+z)^{\beta}} \ ,
\label{cz}
\end{equation}
in the domain $z \in [ -(z_{0}-1), (z_{0}-1)]$. 
The logarithmic derivative of $c(z)$ expresses the relative change in $c(z)$, 
\begin{eqnarray}
\chi(z) &\equiv& \frac{d \ln c(z)}{dz} =  \frac{d c(z)/dz}{c(z)} \nonumber \\
 &=& - \Big( \frac{\gamma}{z_{0}-z} + \frac{\beta}{z_{0}+z}\Big)= - m \Big( \frac{1 + \theta x}{1-x^{2}}\Big)  \ ,
 \end{eqnarray}
  where $x=z/z_{0}$, $\theta = \frac{\gamma- \beta}{\gamma + \beta}$, and $m= \Big(\frac{\gamma + \beta}{z_{0}}\Big)$.
The extreme  values of $\frac{d \ln c(z)}{dz}$ for $z_{0} \gg 1$ are given by
\begin{eqnarray}
 \frac{d \ln c(z)}{dz}\Big|_{z = -(z_{0}-1)} \approx - \beta  \\
 \frac{d \ln c(z)}{dz}\Big|_{z = z_{0}-1} \approx - \gamma
 \end{eqnarray}
and the average value $\overline{\chi}$ is calculated by,
\begin{eqnarray}
\overline{\chi} &\equiv& \langle \frac{d \ln c(z)}{dz}  \rangle \nonumber \\
&=& \frac{-m}{(1-1/z_{0})-(1/z_{0}-1)}\int_{-(1-1/z_{0})}^{(1-1/z_{0})} dx \frac{(1+\theta x)}{1-x^{2}} \nonumber\\
&=& \frac{- m}{2} \ln N
\end{eqnarray}
The function $\chi(z)$ takes on the value of $\overline{\chi}$ twice at the values $\overline{z}_{\pm} = z_{0} \overline{x}_{\pm}$ corresponding to the solutions to the quadratic equation,\\
\begin{equation}
\overline{\chi} = - m \Big( \frac{1 + \theta x}{1-x^{2}}\Big) \ ,
\label{avechi}
\end{equation}
which has the solution 
\begin{eqnarray}
\overline{x}_{\pm} &=& -\frac{\theta}{\ln N}\pm\frac{\sqrt{(\ln N)^{2}-2\ln N + \theta^{2}}}{\ln N} \nonumber \\
& \approx & -\frac{\theta}{\ln N}\pm  \sqrt{1-2/\ln N} 
 \label{zpm}
\end{eqnarray}
for $\theta^{2}/\ln^{2}N  \ll 1$.
Converting back to rank, then
 \begin{eqnarray}
 \overline{r}_{\pm} 
 \approx \big(\frac{N}{2}\big)\big(1-\frac{\theta}{\ln N}\pm  \sqrt{1-2/\ln N}\big) \ ,
 \label{rminus}
 \end{eqnarray}
 and so the value $r^{*}\equiv \overline{r}_{-}$ is the special rank value which distinguishes the 
 set of excellent  papers of each given author. The $c$-star value $c_{i}(r^{*})$ is thus a characteristic value arising from the special analytic properties of $c_{i}(r)$.  This method for determining the 
 crossover value $r^{*}$  can be applied to any general Zipf profile which can be modeled by the DGBD.

Furthermore, the crossover $z_{\mathsf x }$ between the $\beta$ scaling regime and the $\gamma$ scaling regime is calculated from the inflection points of $\ln c(z)$, 
\begin{eqnarray}
0 = \frac{d^{2} \ln c(z)}{dz^{2}}\vert_{z=z_{\mathsf x }}  
=  \frac{-\gamma}{(z_{0}-z_{\mathsf x })^{2}}  + \frac{\beta}{(z_{0}+z_{\mathsf x })^{2}}
\end{eqnarray}
which has  2 solutions $z_{\mathsf x }^{\pm} = z_{0} \Big( \frac{1\pm \zeta}{1\mp  \zeta}\Big)$ , where $\zeta \equiv \sqrt{\gamma/\beta}$. Only $\vert z_{\mathsf x }^{-} \vert <z_{0}$ is a physical solution. Transforming back to rank values, we find $r_{\mathsf x } = z_{0}  + z_{\mathsf x }^{-} = z_{0} \frac{2}{ 1 + \zeta} =  \frac{N+1}{ 1 + \zeta}$. We illustrate these special $z$ values in Fig. \ref{demo}.

\section{Acknowledgments}
We thank J. E. Hirsch and J. Tenenbaum for helpful suggestions.

\clearpage
\newpage
\begin{widetext}

\pagebreak[4]

\clearpage
\newpage

\begin{center}

{\bf Supplementary Information} \\
\bigskip

{\bf Statistical regularities in the rank-citation profile of scientists} \\

Alexander M. Petersen,$^{1,2}$  H. Eugene Stanley$^{2}$, Sauro Succi $^{3,4}$\\
\bigskip
$^{1}$IMT Lucca Institute for Advanced Studies, Lucca 55100, Italy \\
$^{2}$Center for Polymer Studies and Department of Physics, Boston University, Boston, Massachusetts 02215, USA\\
$^{3}$Istituto Applicazioni Calcolo C.N.R.,  Rome, IT\\
$^{4}$Freiburg Institute for Advanced Studies, Albertstrasse, 19, D-79104, Freiburg, Germany\\
\bigskip
(2011)
\end{center}
\bigskip
\renewcommand{\theequation}{S\arabic{equation}
}
\renewcommand{\thefigure}{S\arabic{figure}}
\renewcommand{\thetable}{S\arabic{table}}
 
\setcounter{equation}{0}  
\setcounter{figure}{0}
\setcounter{table}{0}
\setcounter{section}{0}

\footnotetext[1]{ Corresponding author: Alexander M. Petersen \\
{\it E-mail}: \text{petersen.xander@gmail.com}
}

\section{Simple method for estimating the $\beta$ scaling of $c_{i}(r)$ using two $h_{p}$ values}

We analyze the citation profiles of 300 prolific scientists who published  {\it Physical Review Letters}, and find
statistical regularity in 
the functional form of $c_{i}(r)$ of each individual scientist $i$. 
Here  we further quantify $c_{i}(r)$ and discuss 
the  information contained in the
 ``generalized" $h$-index $h_p$, defined by the relation: 
\begin{equation}
c(h_p) = p h_p \ ,
\end{equation}
with $p>1$ a positive  integer. 
In analogy to the $h$-index, $h_p$ is the number of papers which are cited at least $p h_p$ times.  
By definition, $h/p < h_p < h$. Also, the index $h_p$ can be viewed as a 
functional transform in $p-$space of the citation profile $c_{i}(r)$. 
This transform exhibits a number of characteristic values, namely 
$h_{1} \equiv h$, the standard $h$ index, $h_0 = N$, the total number of 
papers, and $h_{\infty}=c(1)$, a scientist's top-cited paper. Therefore, by changing $p$ over the entire interval
$[0,\infty[$, one gains spectral information  of the entire citation profile $c_{i}(r)$ for a given scientist.

Also, for the high-rank  power-law regime $c(r) \sim \ r^{-\beta}$, there is a useful relation  between $h_{p}$ and
$h_{1/p}$. Since $h_p = h \ p^{-\mu}$ then the ratio for complementary $p$-values $h_{1/p}/h_p = p^{2\mu}$, where
$\mu=1/(1+\beta)$. Small values of $\mu \approx 1$ indicate 
slowly-decaying $c_{i}(r)$ corresponding to productive authors with potentially high mobility of the $h$-index. 
Hence, if $c_{i}(r)$ is power-law, then the relation
\begin{equation}
I_p \equiv  \frac{h_p h_{1/p}}{h^2} = 1 
\end{equation}
should hold independent of the value of $\beta$. For all scientists analyzed, we calculate  $I_{2}$ and find $I_{2} =
0.97 \pm 0.07$. This implies that $c_{i}(r)$  obeys a power law around $r = h$. 
This is visually confirmed by inspecting $c_{i}(r / h)$ in Figs. 2(b), \ref{ZAveR100}(b) and \ref{ZAveAsst100}(b) for $r / h
\approx 1$. Hence, we define two complementary mobility indices as
\begin{equation}
m_2 \equiv \frac{h_2}{h}
\end{equation}
and 
\begin{equation}
m_{1/2} \equiv \frac{h_{1/2}}{h}
\end{equation}
By definition, $0<m_2<1$ and $1<m_{1/2}< N$. The potential for high mobility of the $h$-index is associated with 
$m_2$ close to $1$ (low barrier on the high-cite side) and $m_{1/2}>>1$ (high propensity to change in the low-cite
side). We show the relation between $m_{1/2}$ and $m_{2}$ in Fig. \ref{m12m2} along with the expected relation $ m_{2}=
1/m_{1/2}$ for visual reference. 

 
 To test the small-$r$ scaling for each $c_{i}(r)$, we estimate $\beta$ using two methods: 

(i) We define an approximation to $\beta$ by assuming  $c(r) \sim \ r^{-\beta_{pq}}$ for $r<h$. Hence, two intersection values,  $h_{p}$ and $h_{q}$ are sufficient to calculate $\beta_{pq}$  using the relationship for power-law $c_{i}(r)$,
\begin{equation}
\beta_{pq} = \frac{\ln(q/p)}{\ln(h_{p}/h_{q})}-1 \ .
\label{mup1p2}
\end{equation}
We use the values $p \equiv 80$ and $q \equiv 2$ since these values generally enclose the scaling regime in the
$c_{i}(r)$ profiles (see Fig. \ref{ZAveTop100}). 

(ii) We  calculate $\beta$ using a multilinear least-squares regression of $\ln c_{i}(r)$ for $ 1 \leq r \leq
r_{1}$ using the DGBD model defined in Eq.~[\ref{Cr}]. 
To properly weight the data points for better regression fit over the entire 
range, we  use only $20$ values of $c_{i}(r)$ data points that are equally spaced on 
the logarithmic scale in the range $r \in [1,r_{1}]$. 
We plot four empirical $c_{i}(r)$ along with their corresponding best-fit DGBD functions in Fig. \ref{Ztop5} to demonstrate the goodness of fit for the
entire range of $r$. 

We compare  $\beta_{pq}$ and $\beta$ values  
calculated using methods (i) and (ii) in  Tables S1 -S6 and find good agreement. 
Furthermore, the average scaling exponent $\langle \beta \rangle $ is approximately 
equal to the value of $\overline{\beta}$
calculated for the average $\overline{c}(r)$ profile in Fig. \ref{ZAveTop100}. For the 
scientists analyzed in dataset [A] we find $ \langle \beta \rangle = 0.83 \pm 0.24$ 
as compared to $\overline{\beta} = 0.92 \pm 0.01$. 
For the scientists analyzed in dataset [B] we find $ \langle \beta \rangle = 0.70 \pm 0.17$ as compared
to $\overline{\beta} = 0.78 \pm 0.01$.  We plot the histograms of $\beta$ and $\gamma$ for datasets [A], [B], and [C] in Fig. \ref{BetaPDF}. 

\section{Decomposing the Hirsch $a$ factor to better understand efficiency}

Many alternative single-value indicators have been proposed to address the various criticisms of the $h$-index.
The $g$-index \cite{G} differs from the $h$-index in that it lends more weight to the more highly-cited papers.  
However, as with the $h$-index, the $g$-index does not immediately convey much more information than the total number of
citations 
 or the productivity coefficient $a = C / h^{2}$ introduced by Hirsch. In Fig. \ref{H1aPDF} we plot the histogram of both  $h$ and
$a$ values for dataset [A]  with $\langle a \rangle = 6.0 \pm 4.1$ and  for dataset [B] with $\langle a \rangle = 4.2
\pm 1.3$, indicating that most researchers do not fall into the `step-function' pathology $\beta \approx 0$ of scientist
$2$ above for which $a \approx 1$. Instead, most scientists have a significant number of citations arising from both
their high-cited ($r < h$) and low-cited ($r>h$) regions of $c_{i}(r)$. 

An interesting   decomposition is to write $a$ as the product of two factors,
\begin{equation}
\label{a} 
a = \Big(\frac{N}{h}\Big) \Big(\frac{\langle c \rangle}{h} \Big) \ ,
\end{equation}
where $\langle c \rangle \equiv C/N$.

(i) The first factor, by definition, is always greater than $1$, and represents the number of papers in units of $h$. 
Small values of $N/h$ correspond to  scientists who are very efficient  (or less productive), while large values
correspond to scientists who are very productive (or less efficient). Highly productive authors, who may  have a
substantial number of papers without a single citation, nevertheless can still have a large $h$-index. 

(ii) The second factor is the average number of citations per paper $\langle c \rangle$ in units of $h$. Relatively
large values of $\langle c \rangle / h > 1$ signal the presence of outstanding highly cited papers, i.e. papers with
$c(r)\gg h$. Many brilliant careers result from a combination of moderate $N/h \approx 3$ and  $\langle c \rangle / h >
1$. 
Fig. \ref{NAAveC},  which  plots $N/h$ versus $\langle c \rangle/h$ for the set of authors examined in this paper, 
shows a tendency for the two factors to occupy a narrow band of  hyperbolic curves 
$\langle c \rangle / h = a /(N/h)$ with $a
\in \{3,7\}$.

\section{Mobility of the h-index}
The $h$-index is taken  seriously by many research organizations, affecting important 
decisions such as tenure, promotions, honors. For instance, Hirsch noted that $h \sim 12$ seems to be appropriate for 
associate professor, $h \sim 18$ might be suitable for advancement to full professor, while $h \sim 45$ is the average
for NAS election \cite{H}. However, little work has been done to  measure the ``upward mobility" of the $h$-index with
time. Here we address the question: given that a scientist  has index $h$, what is her/his most likely $h$-index value
$\Delta t$ years in the future? This question is fundamentally related to the growth rate of $c_{i}(r)$ for $ r \gtrsim
h$. 
For productive scientists, Hirsch noted that $h$ grows at a rate of about one unit a 
year \cite{H}. However, a single-value indicator such as $h$ cannot quantify the probability of ``growth spurts", which
should also enter into evaluation 
criteria (based on the $h$-index, citation counts, etc.).

As a measure of upward mobility, we propose the gap index $G(\Delta h)$, defined as a proxy for the total
number of citations a scientist needs to reach a target value $h+\Delta h$, which is similar to the $w(q)$-index
proposed in \cite{genH2}. The merit of $G(\Delta h)$ is to detect the potential for fast growth by quantifying
$c_{i}(r)$ 
around $h$. 
For  $c_{i}(r)$ with index value $h$ at time $t$, we define $G(\Delta h)$ as the minimum number of citations,
distributed to papers $r = \{ 1,...h+\Delta h \}$, so that the $h$-index value at time $t+\Delta t$ becomes $h+\Delta
h$.   

Consider the citation gap  $g(r)=h^{+} -c(r)$ of each paper $r$ with $c(r)<h^{+}$.  Then $G(\Delta h)$ is given by the 
 exact relation which can be easily verified graphically,
\begin{equation}
\label{GAP} G(\Delta h) \equiv \sum_{r=h^{-}}^{h^{+}} g(r) =  h^{+}(h^{+}-h^{-}+1)-C(h^{-},h^{+}) \ ,
\end{equation}
where $h^{+}=h+\Delta h$, $h^{-}$ is  the smallest $r$ value for which $c(r)<h^{+}$, and
$C(m,n)=\sum_{r=m}^n c(r)$ is the number of  citations from paper $m$ up to paper $n$. Hence, $G(\Delta h)$ quantifies
the minimum number of citations, assuming perfect assignment, required 
to bring papers $r=h^{-}  \dots h \dots h^{+}$ up to citation level $h^{+}$. 

The gap index $G(\Delta h)$ establishes a characteristic time scale $\Delta t$ for the dynamics of $h(t)$. The estimated
amount of time for the transition  is $\Delta t=Max \lbrace t_r \equiv g(r)/\dot{c}(r) \rbrace$ for $h^{-}\leq r \leq 
h^{+}$, where $\dot c(r)$ is its average citation rate (citations/year).
This estimate does not take into account `rampant papers', papers with relatively large $r$ and $\dot{c}(r)$, which are 
either new or rejuvenated after a lengthy period with very few citations \cite{110PhsRev}. 
In practical terms, the short-term utility of $G(\Delta h)$ is for moderate 
values of $\Delta h$, say in multiples of $5$ or $10$. 
In other words, for a scientist with $h=12$, a plausible  target could be $h^+=17$ and 
a longer term target $h^+=22$.

 In Fig. \ref{GapPDF}  we plot the histograms for $G(5)$ and $G(10)$. 
The common distributions between authors in dataset [A] and dataset [B] indicate that the growth potential of $h$ does
not depend very strongly on prestige, but rather 
on the  publication patterns of individual authors. 
Indeed, the average annual growth rate $h / L$ are larger for dataset [A] physicists than dataset [B]  physicists, with
a significant number of exceptional ``outliers'' with $h / L>3$. 

For young careers corresponding to small $h$-values, there will be a correlation between $G(\Delta h)$ and $h$ because most new citations 
will contribute to the increase of $h$. However, for an advanced career, not all incoming citations will contribute to an increase in $h$. Hence, to 
test the dependence of $G(\Delta h)$ on $h$, we perform a linear regression $G_{i}(\Delta h) = g_{0}+g_{1} h_{i}$ for
both
datasets [A] and [B] and for $\Delta h = 5,10$. In each of the four regressions we calculate correlation values $R^{2} <0.05$ and
ANOVA (analysis of variance) F-statistics $F < 2.3$ for each case, indicating that we accept the null hypothesis that the
linear regression coefficient $g_{1} = 0$. Thus,
for significantly large $h$, the gap-index $G(\Delta h)$ is not dependent on $h$. 
A similar regression analysis between
$G(\Delta h)$ and $\beta$ results in the same conclusion, that
the gap index $G(\Delta h)$ is not dependent on $\beta$ for profiles with sufficiently large $h$. 
Hence, the gap index
can be used to estimate the mobility of $h$ and as a comparison index between $c_{i}(r)$.

\section{Characterizing the rank-citation $c_{i}(r)$ profile}

As many previous studies have shown, and further demonstrated here, there are many conceivable ways to quantify
$c_{i}(r)$.
 In Tables \ref{table:top100a}-\ref{table:random100b} we list $16$ values derived from $c_{i}(r)$ which can serve as
quantitative indicators of a scientific career:
\renewcommand{\labelenumi}{[\arabic{enumi}]}
\begin{enumerate}
\item the author's total number of papers $N$, 
\item the author's total number of citations $C \equiv \sum_{r=1}^{N}c(r) $,
\item the author's most-cited paper $c_{i}(1)$,
\item the author's $c$-star paper $c(r^{*})$, which distinguishes the minimum citation tally of his/her stellar papers
in the range $r\in [1,r^{*}]$.
\item the author's $r^{*}$ value calculated from his/her DGBD parameter values according to Eq. [14]
\item the author's original  $h$-index $ h_{1}\equiv h$ and the generalized $h$-index $h_{p}$ for $p=2,80$,
\item the author's scaling exponent $\beta_{pq}$ calculated using the values $p=2$ and $q=80$ corresponding to $h_{2}$ and
$h_{80}$,
\item the author's scaling exponents $\beta$ and $\gamma$ calculated using multilinear least-squares regression fit to the DGBD model
$c_{i}(r)$ in Eq. [3],
\item the author's peak-value $\Lambda$ given by the $c$-star  value $c^{*}$ in units of the $h$-index,  $\Lambda \equiv
c(r^{*}) / h$
\item the author's number of papers in units of the $h$-index $N' \equiv N/h$,
\item the author's average number of citations per paper in units of the $h$-index $\langle c \rangle' \equiv \langle c
\rangle/h$,
\item the author's  ``productivity" value proposed by Hirsch, $a \equiv C/h^{2}$,
\item the author's mobility estimator $G(\Delta h)$ quantifying the minimum number of citations needed to increase an
individual's h-value by $\Delta h$ units for $\Delta h = 5$ and $\Delta h = 10$,
\item the author's mobility indices $m_{1/2}$ and $m_2$ where $m_{1/2}m_{2} \approx 1$, 
\item the author's peak number $P = \frac{1}{h^{2}} \sum_{r = 1}^{h} c(r)$,
\item the author's average $h$-index growth rate  $h/L$ over the $L$-year time period between an author's first and most recent
paper.
\end{enumerate}

The $h$-index conveys a very informative 
one-number picture of productivity, however it does not tell the whole story, since it does not fully capture the
impact an author's most cited papers. 
Instead, we show the utility of the $c$-star value $c(r^{*})$, which is a better representative of an author's most
cited papers. Thus,  we introduce the peak-value indicator $\Lambda \equiv c(r^{*})/h$  in order to characterize the
most distinguished papers ($1 \le r \le r^*$) of each given author. The probability distribution of $\Lambda$ values is
given in Fig. \cite{PeakPDF}.

We use the Discrete Generalized Beta Distribution (DGBD) to quantify $c_{i}(r)$ for the whole range of $r$. However,
typically a scientist is mostly evaluated by his/her  highest ranked papers, say for $r \leq r^{*}$. Hence, in this
regime, we show that $c_{i}(r)$ can be parameterized by only two variables, $\beta$ and $h_{1}$, in order to
comprehensively capture a publication career. The emergence of a such a compact (two-parameter) and general
parametrization  
highlights an amazing statistical regularity in the scientific productivity of single individuals. Without endorsing the
extreme viewpoint "you are what you publish" or ``publish or perish'', such statistical regularity,  nevertheless,
highlights  an outstanding question on the role of social 
factors in ironing out individual details of human productivity. 
We believe that such question bears a great relevance to most fields of economic, natural and social sciences, where
productivity data are available.

\clearpage
\newpage

\begin{figure}[t]
\centering{\includegraphics[width=0.55\textwidth]{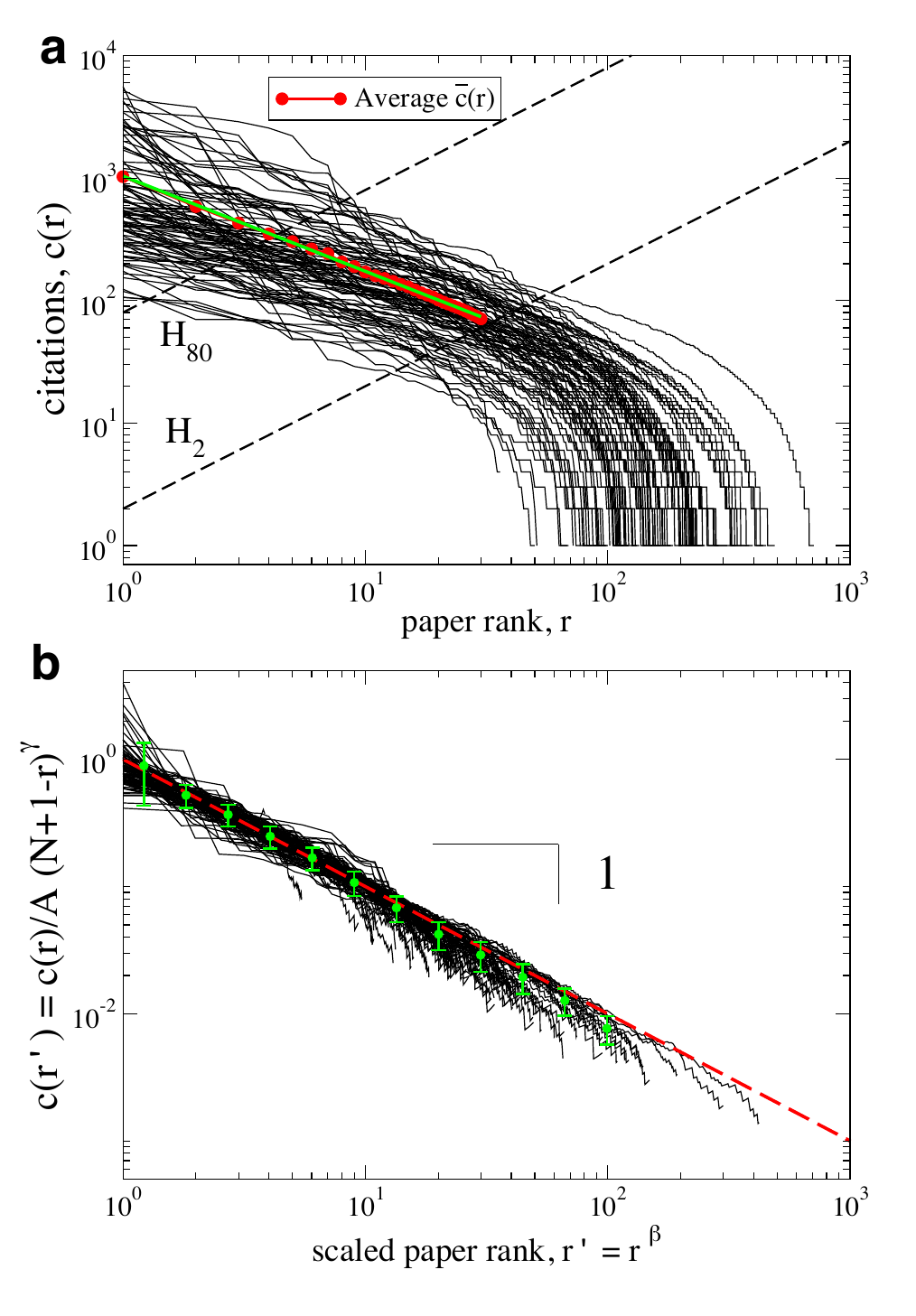}}
\caption{\label{ZAveR100} Zipf rank-citation curves plotted in panel (a) correspond to the the 100 dataset [B]
scientists. 
For reference, we plot the average $\overline{c}(r)$ of these 100 curves and find $\overline{c}(r) \sim r^{-\beta}$ with
 $\beta = 0.78 \pm 0.01$. The solid green line is a least-squares fit to $\overline{c}(r)$ over the range (a) $1 \leq r
\leq 30$. 
 We also plot the $H_{p}(r)$ lines  corresponding to $p=2$ and $p=80$ for reference. 
  (b) We re-scale the curves in panel (a), plotting $c(r') \equiv c(r)/ A(N+1-r)^{\gamma}$,  where we use the 
multilinear least-squares regression values for each individual $c_{i}(r)$ profile. Using the  rescaled rank value $r' \equiv
r^{\beta}$, we show excellent data collapse along the expected curve $c(r') = 1/r'$.  Green data points correspond to the average $c(r')$ value with 1$\sigma$ error bars  calculated using all 100 $c_{i}(r')$ curves separated into logarithmically spaced bins. } 
\end{figure}

\begin{figure}[t]
\centering{\includegraphics[width=0.55\textwidth]{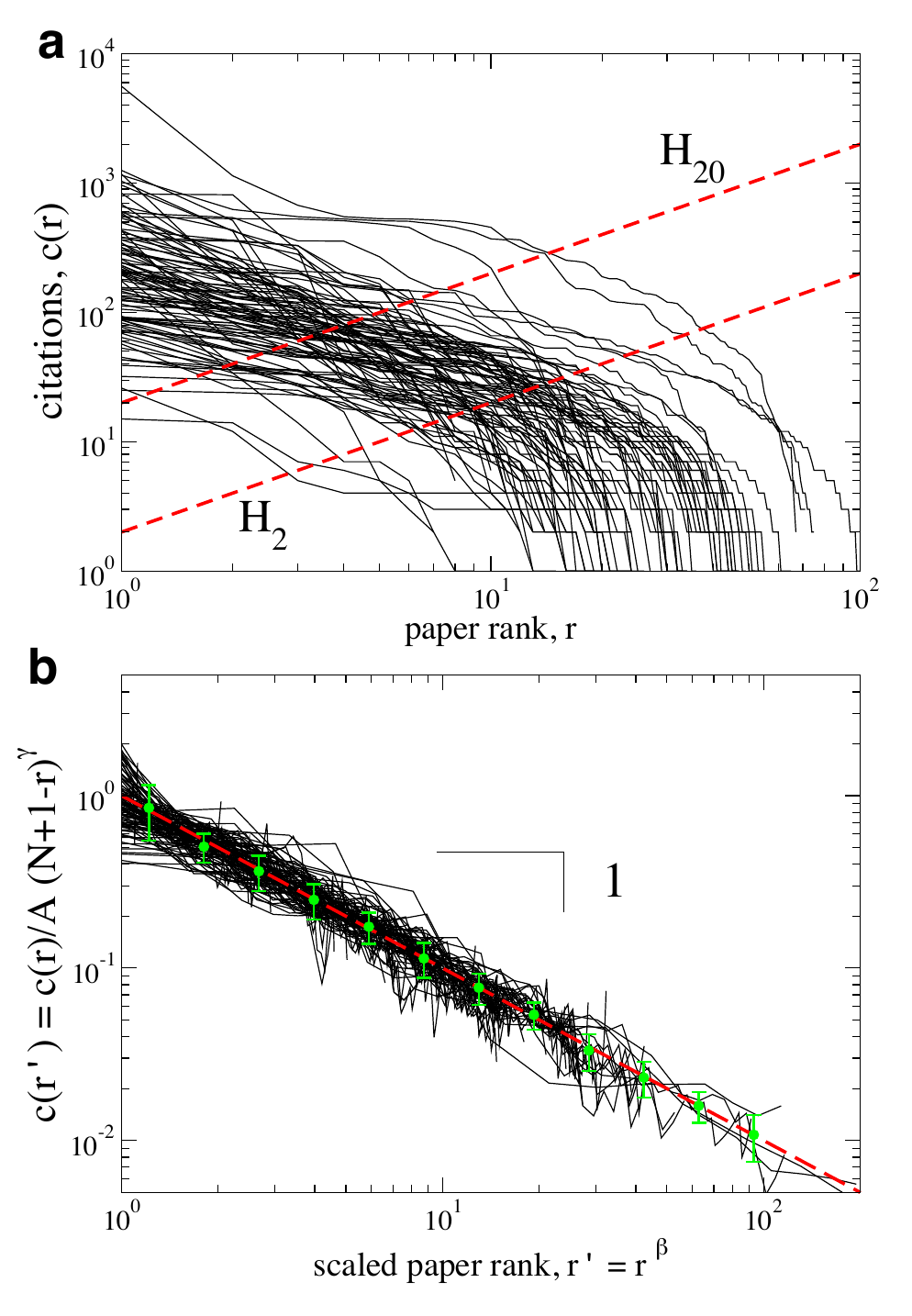}}
\caption{\label{ZAveAsst100} Zipf rank-citation curves plotted in panel (a) correspond to the the 100 dataset [C]
assistant professor scientists. 
For reference, we plot the $H_{p}(r)$ lines  corresponding to $p=1$.. 
  (b) We re-scale the curves in panel (a), plotting $c(r') \equiv c(r)/ A(N+1-r)^{\gamma}$,  where we use the 
multilinear least-squares regression values for each individual $c_{i}(r)$ profile. Using the  rescaled rank value $r' \equiv
r^{\beta}$, we show excellent data collapse along the expected curve $c(r') = 1/r'$ even for young careers.
Green data points correspond to the average $c(r')$ value with 1$\sigma$ error bars  calculated using all 100 $c_{i}(r')$ curves separated into logarithmically spaced bins. 
We note that young careers might possibly be analogous to advanced careers in other disciplines where overall publication rates
are lower. } 
\end{figure}

\begin{figure}[t]
\centering{\includegraphics[width=0.9\textwidth]{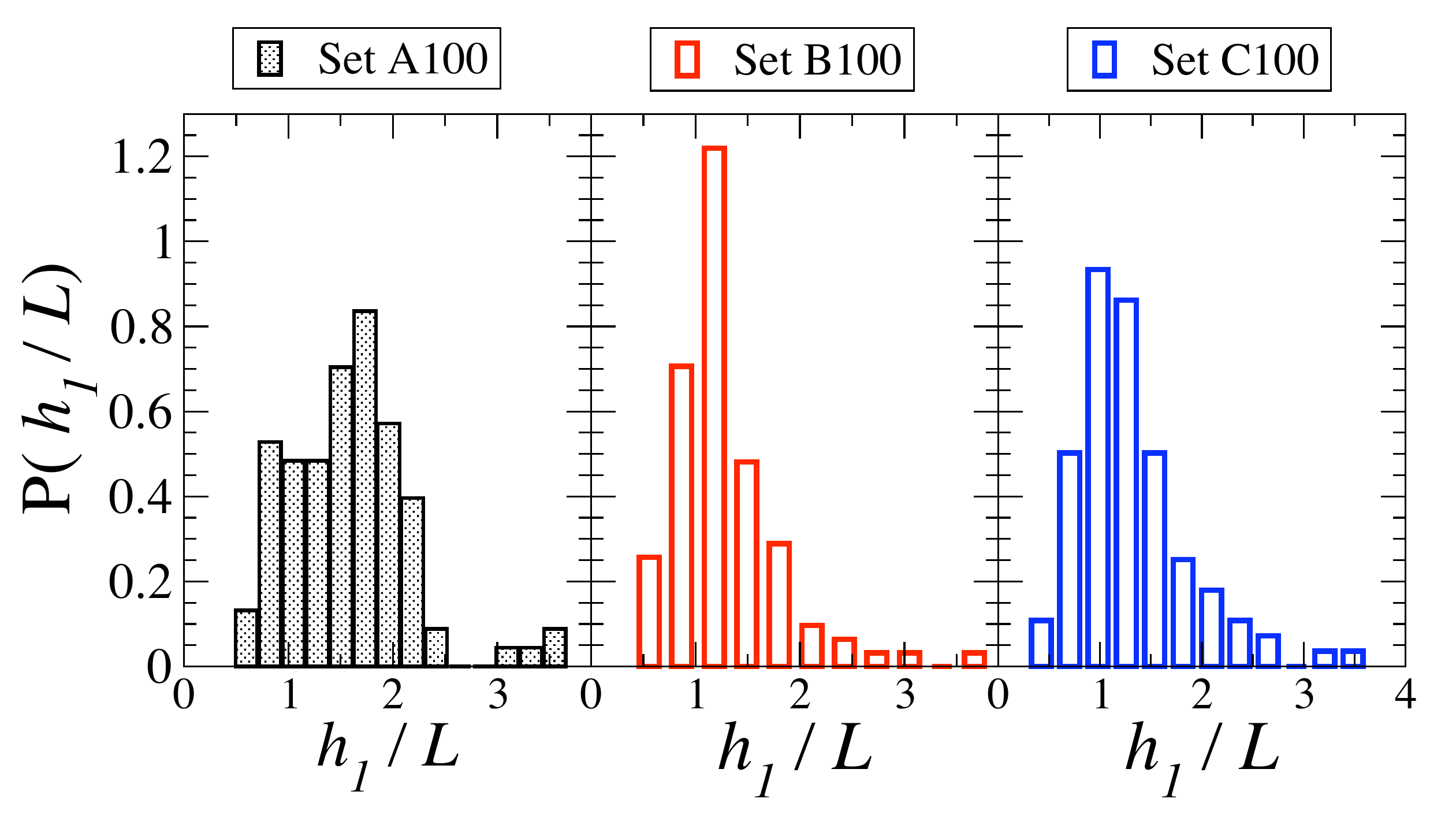}}
\caption{\label{H1byCLPDF}  Probability distribution of $h / L$ values calculated for scientists in datasets [A], [B] and [C].} 
\end{figure}

\begin{figure}[t]
\centering{\includegraphics[width=0.9\textwidth]{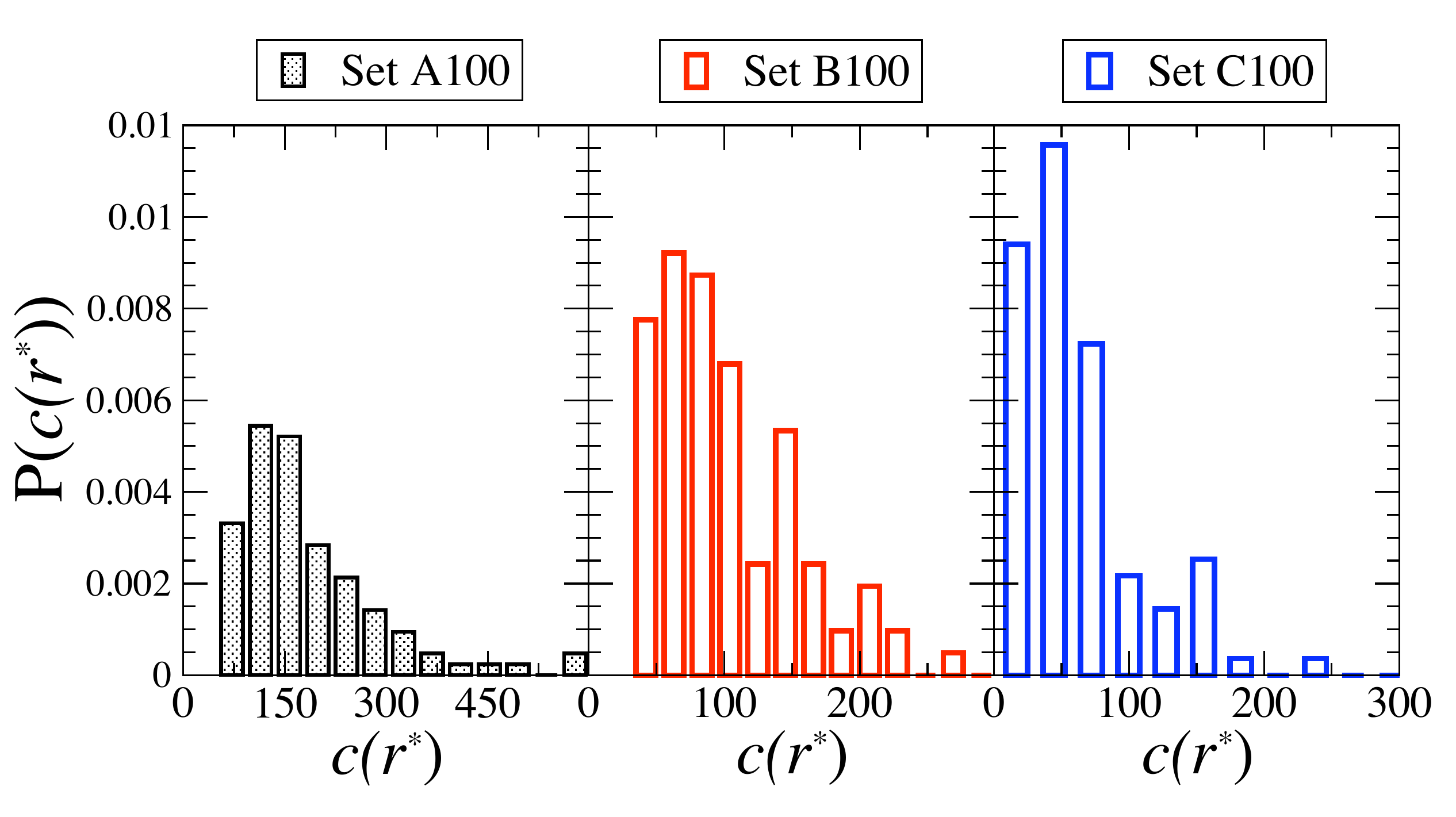}}
\caption{\label{CstarPDF}  Probability distribution of $c(r^{*})$ values calculated for scientists in datasets [A], [B] and [C].} 
\end{figure}

\begin{figure}[t]
\centering{\includegraphics[width=0.9\textwidth]{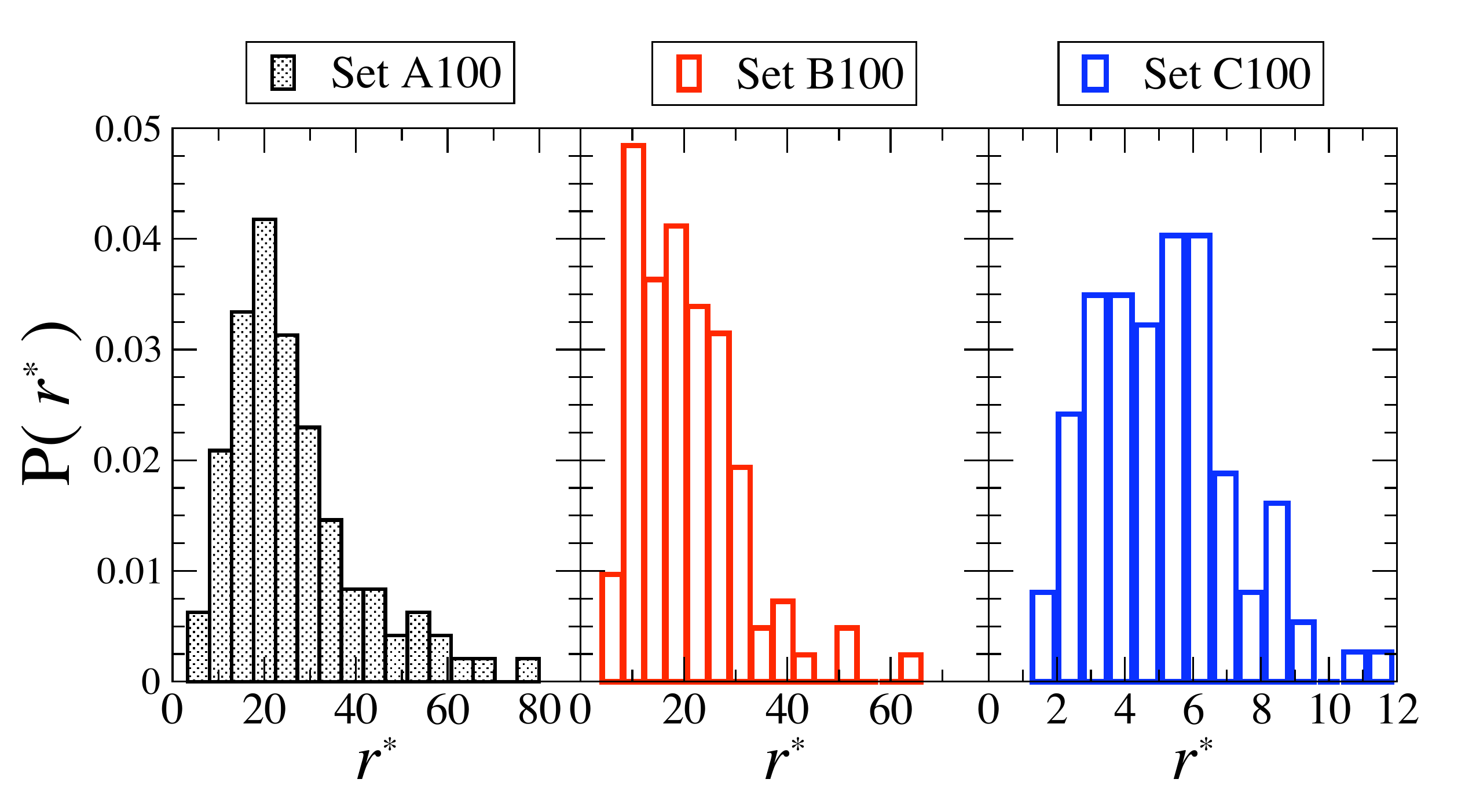}}
\caption{\label{rstarPDF}  Probability distribution of $r^{*}$ for scientists in
datasets [A], [B] and [C].} 
\end{figure}

\begin{figure}[t]
\centering{\includegraphics[width=0.9\textwidth]{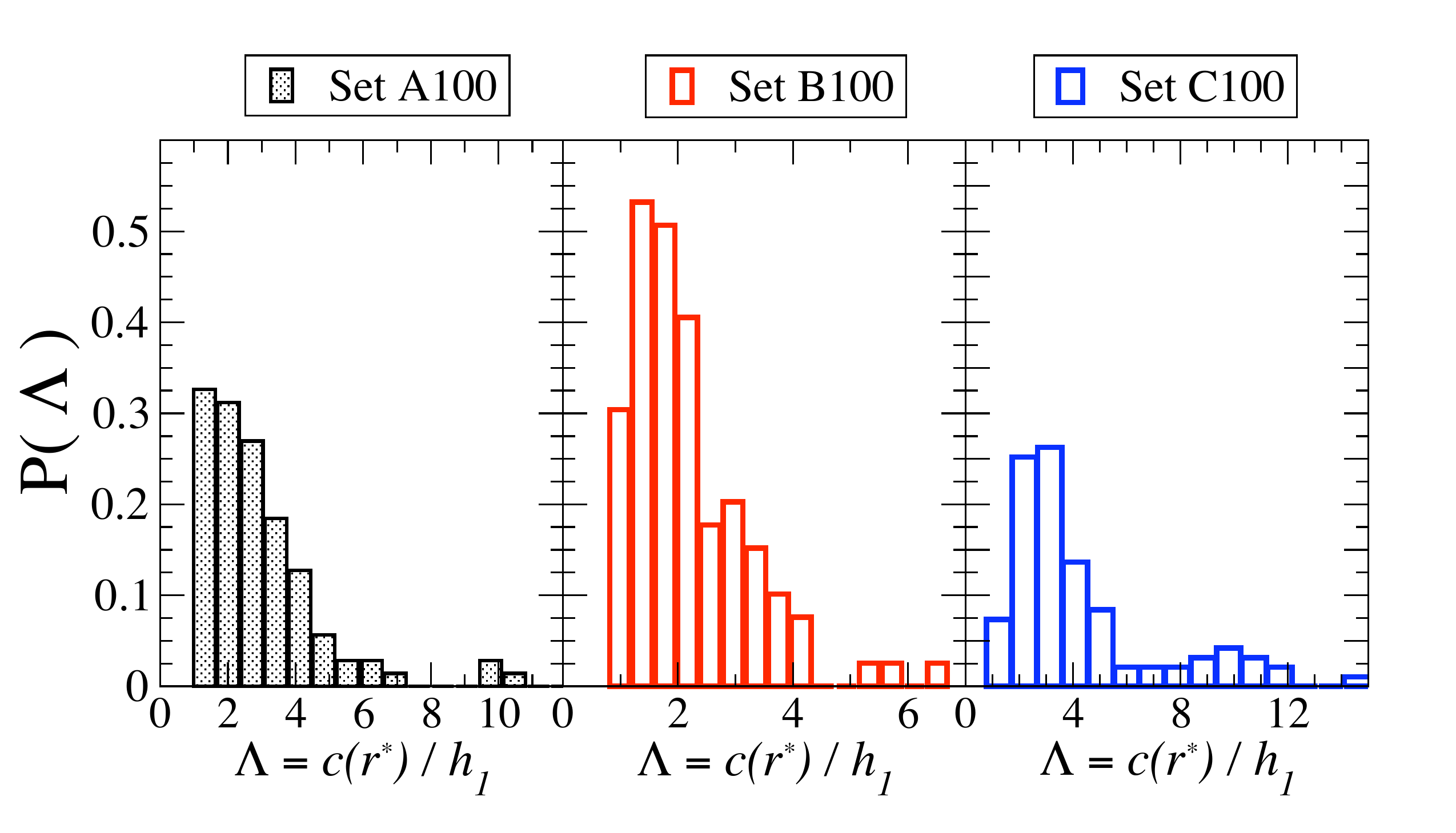}}
\caption{\label{PeakPDF}  Probability distribution of peak-values  $\Lambda \equiv c(r^{*}) / h$ for scientists in
datasets [A], [B] and [C].} 
\end{figure}

\begin{figure}[t]
\centering{\includegraphics[width=0.6\textwidth]{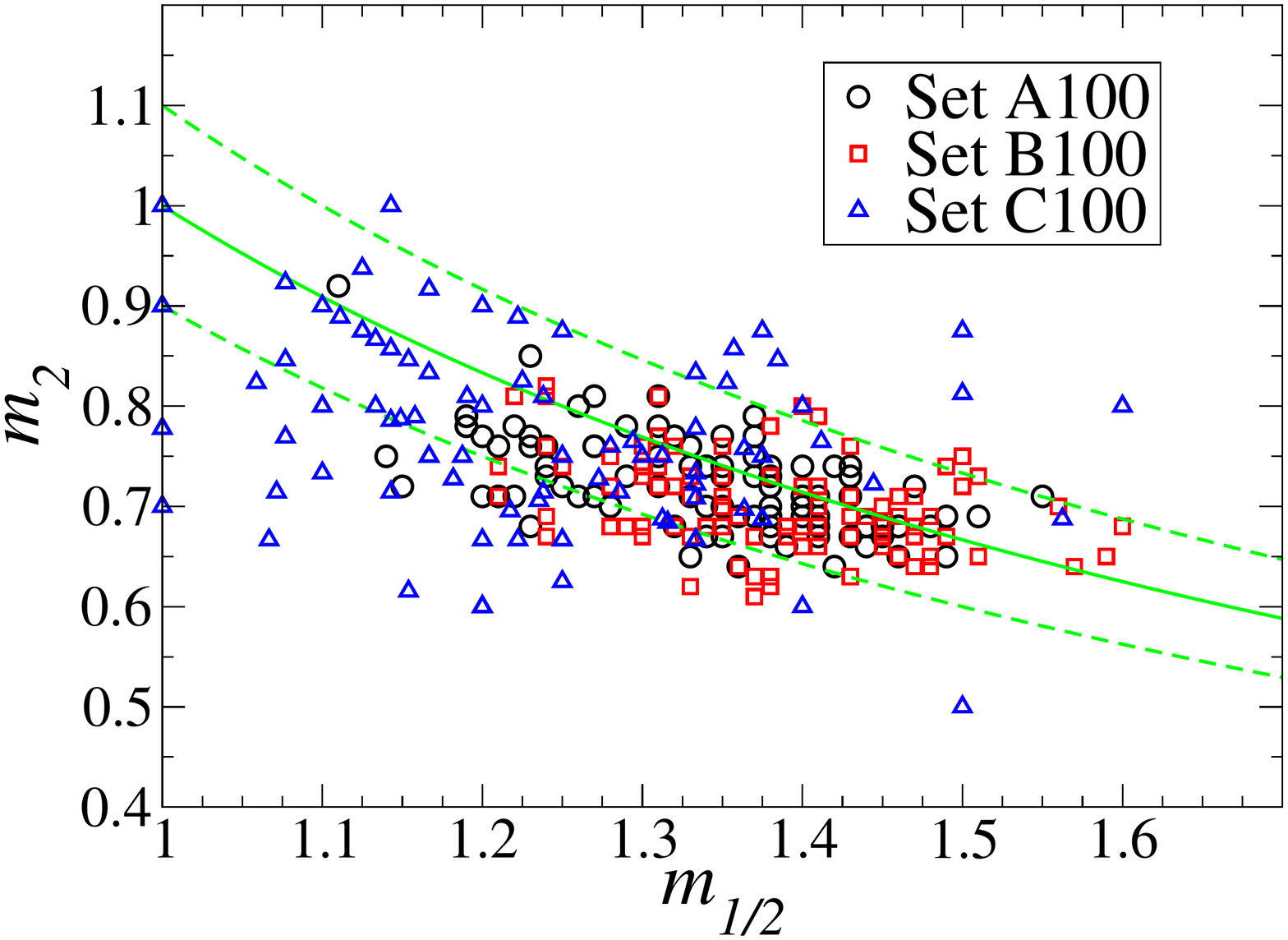}}
\caption{\label{m12m2} Scatter plot of $m_{1/2}= h_{1/2}/h$ and $m_{2}= h_{2}/h$ for for scientists datasets [A], [B] and [C]. The quantity $I_{2} \equiv m_{1/2} m_{2} \approx 0.97 \pm 0.07$ for all scientists analyzed. We plot 3
green curves corresponding to $I_{2}= 0.9$, $I_{2}=1.0$, and $I_{2}=1.1$ for comparison. } 
\end{figure}

\begin{figure}[t]
\centering{\includegraphics[width=0.6\textwidth]{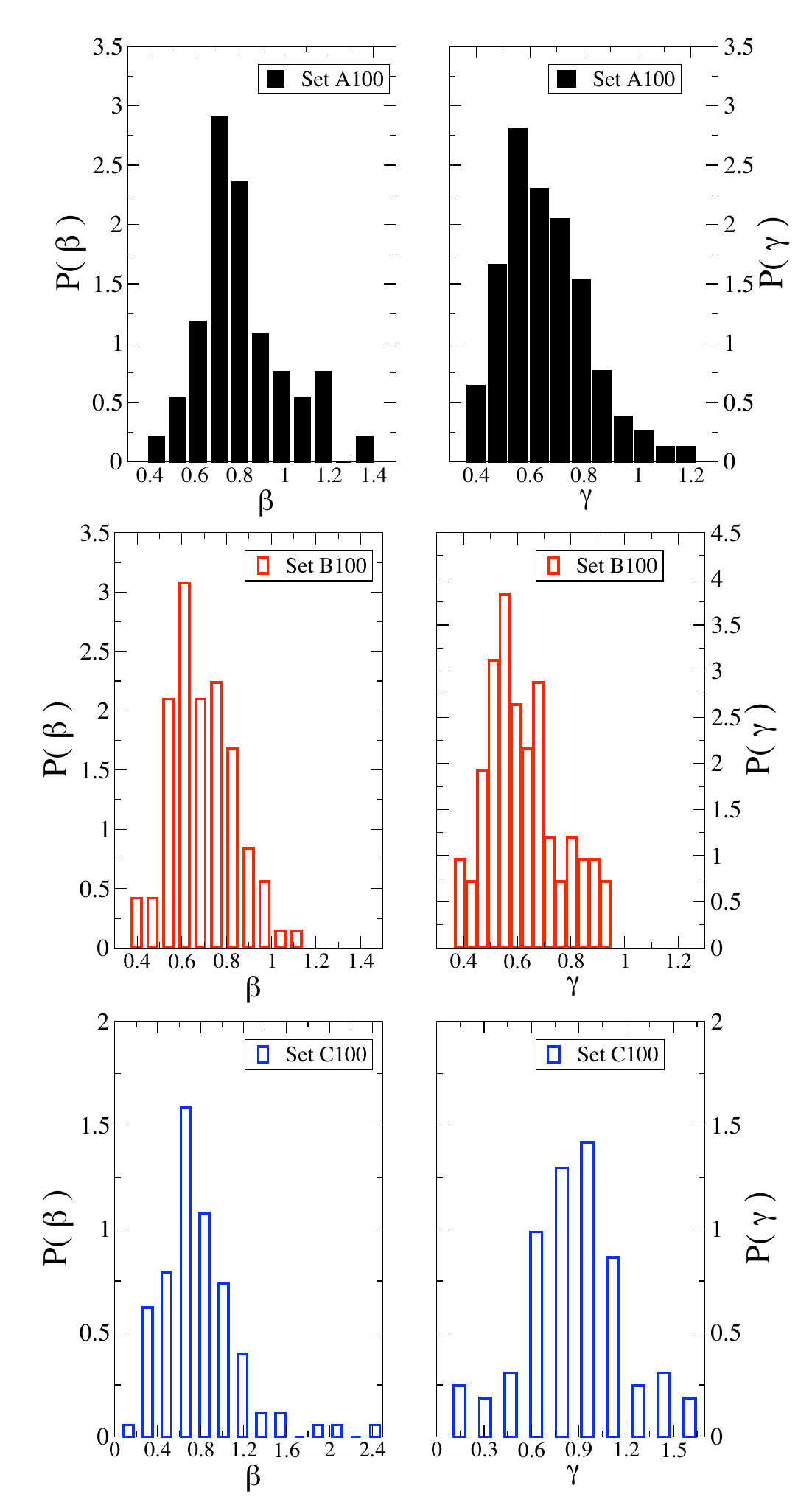}}
\caption{\label{BetaPDF} Probability distribution of $\beta$ and $\gamma$ values  calculated using multilinear least-squares regression of $\ln c_{i}(r) \equiv \ln A - \beta \ln r + \gamma \ln [r_{1}+1-r]$ for scientists in datasets [A], [B] and [C].  } 
\end{figure}

\begin{figure}[t]
\centering{\includegraphics[width=0.6\textwidth]{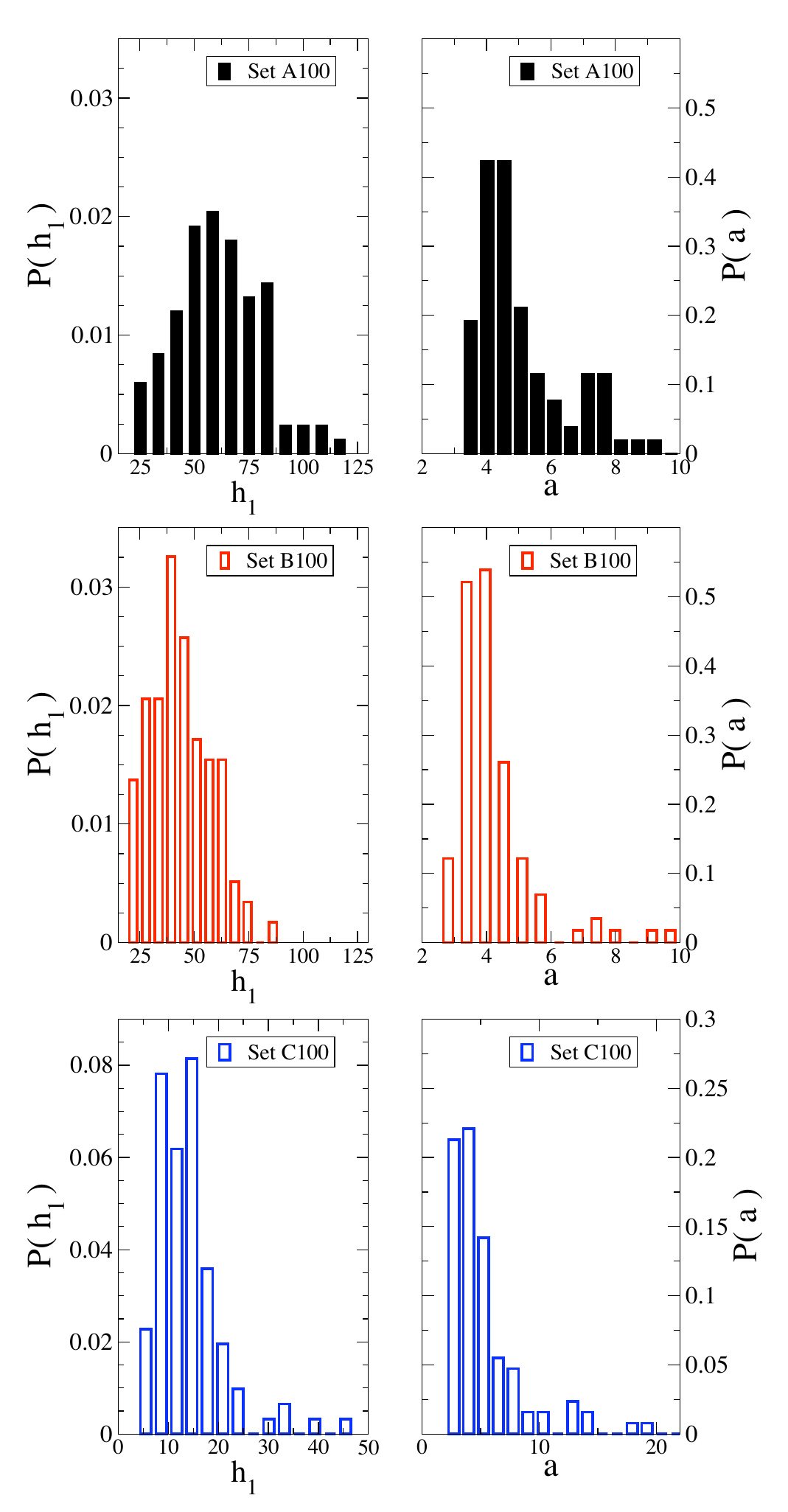}}
\caption{\label{H1aPDF}  Probability distribution of $h$ and $a$ values calculated for scientists in datasets [A], [B] and [C]. } 
\end{figure}

\begin{figure}
\centering{\includegraphics[width=0.49\textwidth]{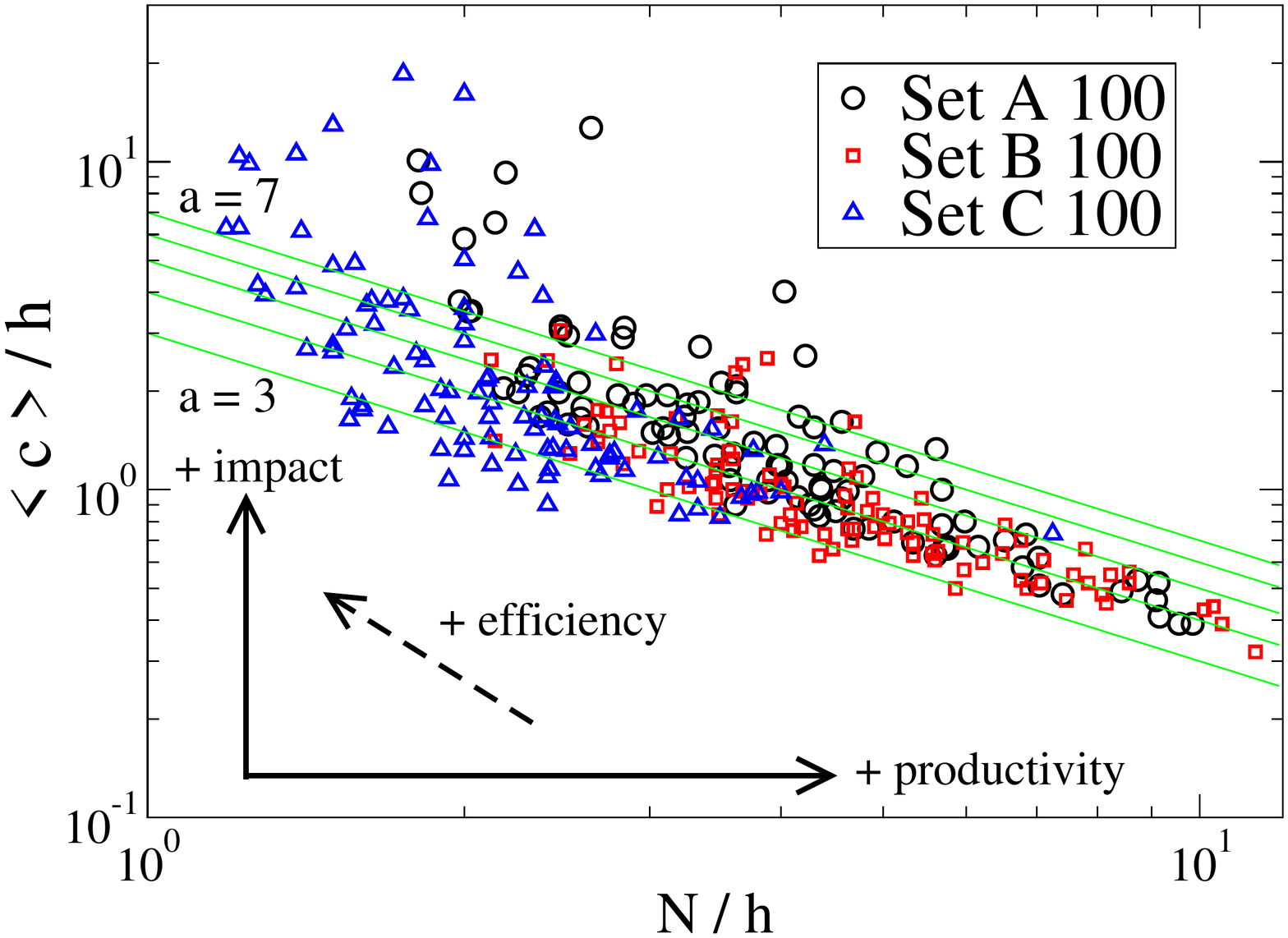}}
\caption{\label{NAAveC}   Statistical regularities in the impact-productivity space of scientists.
Scatter plot of $N$ and $\langle c \rangle$, in units of $h$, for the 300 scientists analyzed. 
The quantity $\langle c \rangle$ represents the impact per paper, while the quantity $N/h$ is inversely related to
efficiency, since small values correspond to scientists with many citations arising from a  relatively small number of
papers. Because each quantity is measured in units of $h$, only individuals with similar $h$ can be compared.
  Dotted green lines correspond to hyperbolic curves hyperbolic curves 
$\langle c \rangle / h = a /(N/h)$ with values $a = \{ 3,4,5,6,7\}$ (bottom to top).   } 
\end{figure}

\begin{figure}[t]
\centering{\includegraphics[width=0.6\textwidth]{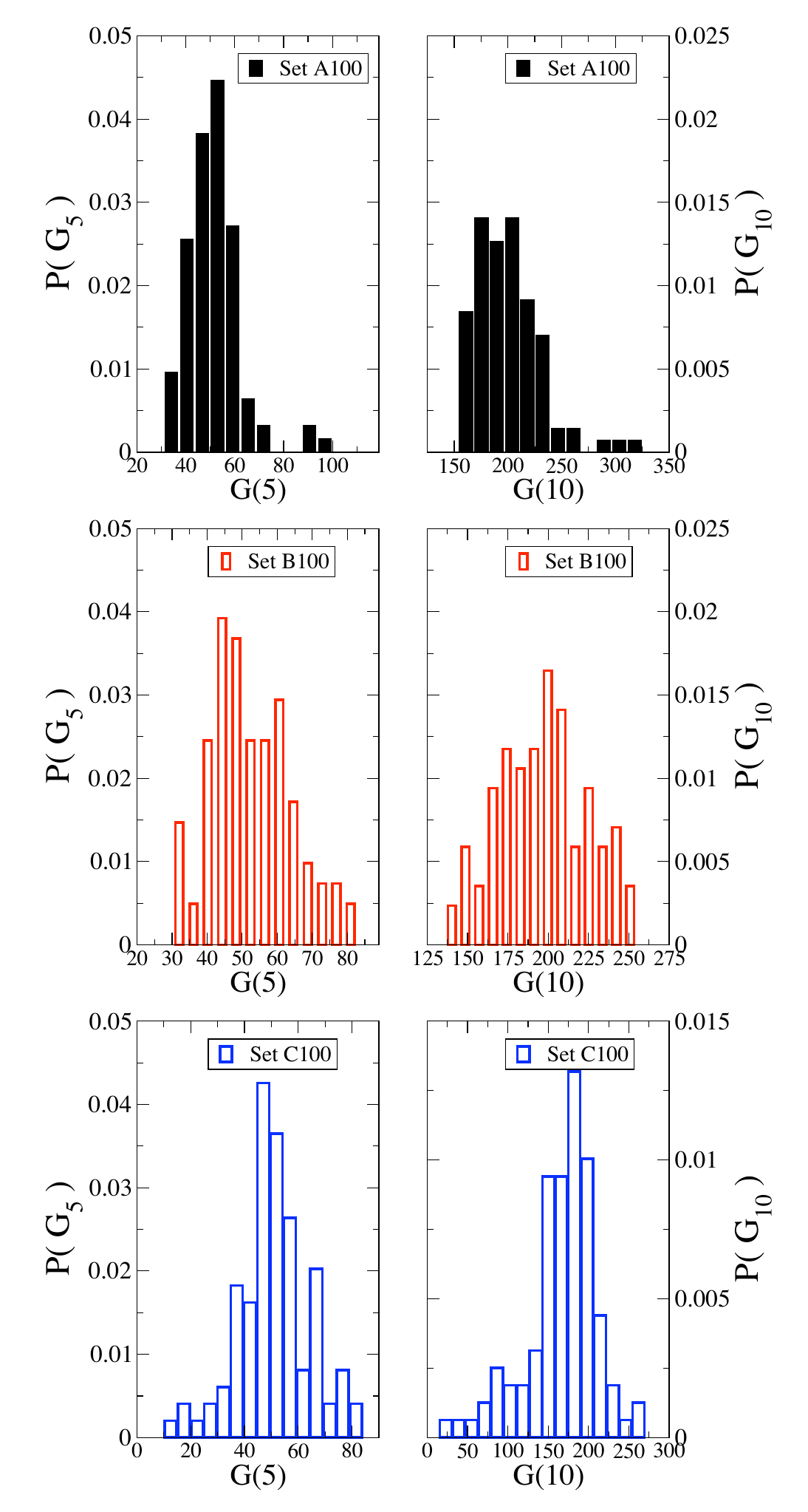}}
\caption{\label{GapPDF}  Probability distribution of $G(5)$ and $G(10)$ values calculated for scientists datasets [A], [B] and [C].
 }
\end{figure}


\begin{figure}[t]
\centering{\includegraphics[width=0.6\textwidth]{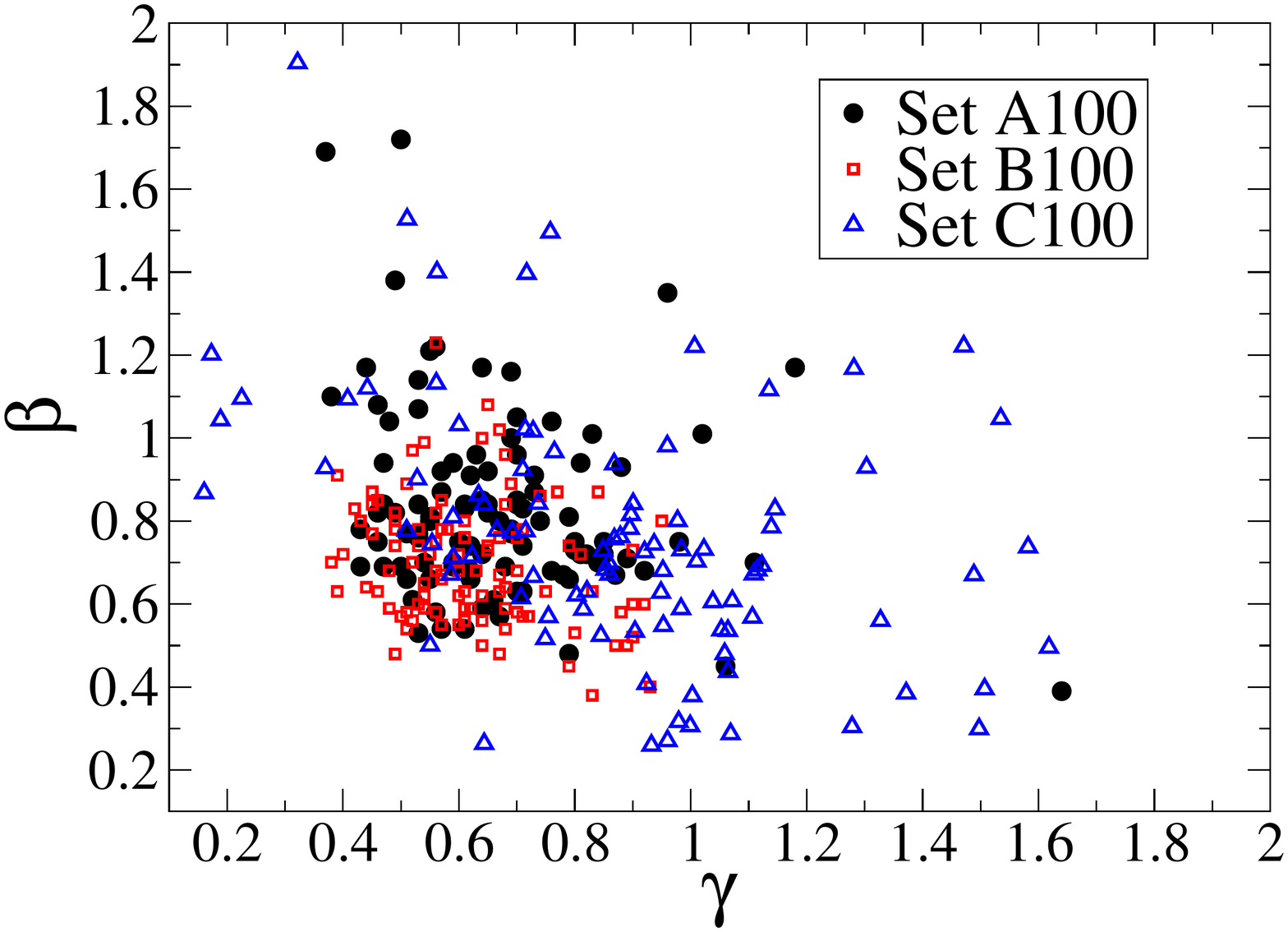}}
\caption{\label{GammaBetaM} Scatter plot of best-fit $\beta$ and $\gamma$ values.  } 
\end{figure}

\begin{table}[h]
\centering{ {\footnotesize
\begin{tabular}{@{\vrule height .5 pt depth4pt  width0pt}lccccccccccccccccccccc}
\noalign{
\vskip-1pt}
Name & $N$ & $C$ & $c(1)$ & $c(r^{*})$ & $r^{*}$ & $h_1$ & $h_2$ & $h_{80}$ & $\beta_{xy}$ & $\beta$ & $\gamma$ & $c(r^{*})/h_1$ & $N/h_1$ & $\langle c \rangle/h_1$ & $a$ & Gap(5) & Gap(10) & $m_{1/2}$ & $m_{2}$ & $P$ & $h_{1}/L$ \\
\hline
\hline
A, E & 116 & 13848 & 3339 & 145 & 16 & 41 & 30 & 6 & 1.29 & 1.17 & 0.64 & 3.55 & 2.83 & 2.91 & 8.24 & 37 & 165 & 1.37 & 0.73 & 7.67 & 0.71 \\
A, I & 205 & 15073 & 783 & 181 & 18.3 & 57 & 45 & 6 & 0.83 & 0.68 & 0.76 & 3.19 & 3.6 & 1.29 & 4.64 & 44 & 168 & 1.37 & 0.79 & 3.84 & 1.9 \\
A, A & 363 & 14996 & 708 & 92 & 33.9 & 63 & 44 & 4 & 0.54 & 0.66 & 0.55 & 1.46 & 5.76 & 0.66 & 3.78 & 51 & 206 & 1.4 & 0.7 & 2.49 & 1.58 \\
A, BJ & 185 & 18658 & 6267 & 133 & 22.1 & 51 & 38 & 7 & 1.18 & 0.94 & 0.59 & 2.63 & 3.63 & 1.98 & 7.17 & 45 & 176 & 1.37 & 0.75 & 6.25 & 0.86 \\
A, PW & 344 & 65575 & 4661 & 379 & 32.5 & 103 & 78 & 13 & 1.06 & 0.91 & 0.73 & 3.69 & 3.34 & 1.85 & 6.18 & 66 & 228 & 1.27 & 0.76 & 5.6 & 1.72 \\
A, A & 111 & 12514 & 1689 & 306 & 9.9 & 48 & 38 & 6 & 1 & 0.75 & 0.98 & 6.38 & 2.31 & 2.35 & 5.43 & 71 & 257 & 1.19 & 0.79 & 5.14 & 0.81 \\
B, P & 173 & 16709 & 3037 & 154 & 20.9 & 53 & 39 & 5 & 0.8 & 0.96 & 0.63 & 2.91 & 3.26 & 1.82 & 5.95 & 57 & 207 & 1.34 & 0.74 & 5.23 & 1.61 \\
B, J & 141 & 25350 & 5636 & 305 & 16.1 & 57 & 44 & 9 & 1.32 & 0.94 & 0.81 & 5.36 & 2.47 & 3.15 & 7.8 & 55 & 190 & 1.32 & 0.77 & 7.32 & 1.08 \\
B, CP & 60 & 10960 & 2486 & 264 & 8.3 & 28 & 21 & 6 & 1.94 & 1.35 & 0.96 & 9.45 & 2.14 & 6.52 & 14 & 73 & 262 & 1.14 & 0.75 & 13.8 & 0.58 \\
B, CWJ & 265 & 14474 & 1835 & 101 & 28.4 & 58 & 39 & 4 & 0.62 & 0.8 & 0.55 & 1.75 & 4.57 & 0.94 & 4.3 & 47 & 182 & 1.38 & 0.67 & 3.22 & 2.07 \\
B, CH & 78 & 17715 & 3849 & 400 & 9.8 & 39 & 33 & 8 & 1.6 & 1.01 & 1.02 & 10.3 & 2 & 5.82 & 11.6 & 66 & 229 & 1.23 & 0.85 & 11.3 & 0.95 \\
B, G & 91 & 15707 & 6280 & 240 & 11.5 & 46 & 33 & 5 & 0.95 & 1.01 & 0.83 & 5.24 & 1.98 & 3.75 & 7.42 & 89 & 292 & 1.15 & 0.72 & 7.17 & 1.28 \\
B, K & 190 & 21887 & 12906 & 82 & 23.8 & 45 & 31 & 5 & 1.02 & 1.17 & 0.44 & 1.84 & 4.22 & 2.56 & 10.8 & 46 & 186 & 1.38 & 0.69 & 9.94 & 1.1 \\
B, M & 218 & 16279 & 1445 & 148 & 22.7 & 55 & 39 & 6 & 0.97 & 0.84 & 0.65 & 2.7 & 3.96 & 1.36 & 5.38 & 48 & 203 & 1.4 & 0.71 & 4.49 & 1.67 \\
C, N & 140 & 9022 & 2261 & 117 & 15.4 & 43 & 32 & 3 & 0.56 & 0.77 & 0.69 & 2.74 & 3.26 & 1.5 & 4.88 & 38 & 168 & 1.4 & 0.74 & 4.11 & 0.86 \\
C, R & 267 & 18716 & 5147 & 128 & 25.1 & 66 & 49 & 4 & 0.47 & 0.74 & 0.62 & 1.95 & 4.05 & 1.06 & 4.3 & 52 & 214 & 1.33 & 0.74 & 3.51 & 1.83 \\
C, DM & 162 & 16307 & 6264 & 136 & 19.4 & 52 & 36 & 5 & 0.87 & 0.91 & 0.62 & 2.62 & 3.12 & 1.94 & 6.03 & 52 & 187 & 1.38 & 0.69 & 5.27 & 1.53 \\
C, DJ & 194 & 13801 & 1572 & 166 & 18.2 & 56 & 39 & 5 & 0.8 & 0.74 & 0.71 & 2.98 & 3.46 & 1.27 & 4.4 & 54 & 210 & 1.34 & 0.7 & 3.64 & 1.44 \\
C, SW & 469 & 25808 & 1444 & 115 & 44.4 & 82 & 56 & 6 & 0.65 & 0.77 & 0.53 & 1.4 & 5.72 & 0.67 & 3.84 & 50 & 197 & 1.32 & 0.68 & 2.78 & 3.57 \\
C, JI & 295 & 19894 & 1514 & 136 & 30.4 & 71 & 50 & 6 & 0.74 & 0.75 & 0.6 & 1.93 & 4.15 & 0.95 & 3.95 & 58 & 205 & 1.38 & 0.7 & 3 & 3.38 \\
C, ML & 752 & 50269 & 1588 & 153 & 63.8 & 107 & 69 & 9 & 0.81 & 0.7 & 0.54 & 1.44 & 7.03 & 0.62 & 4.39 & 40 & 193 & 1.42 & 0.64 & 2.79 & 1.73 \\
C, PB & 220 & 14257 & 1878 & 125 & 23.4 & 55 & 37 & 5 & 0.84 & 0.84 & 0.61 & 2.29 & 4 & 1.18 & 4.71 & 50 & 174 & 1.45 & 0.67 & 3.75 & 1.53 \\
D, S & 594 & 19992 & 2119 & 65 & 57.6 & 65 & 44 & 4 & 0.54 & 0.69 & 0.43 & 1 & 9.14 & 0.52 & 4.73 & 46 & 191 & 1.46 & 0.68 & 2.54 & 1.71 \\
D, SD & 108 & 8339 & 744 & 189 & 10.8 & 45 & 33 & 4 & 0.75 & 0.67 & 0.87 & 4.22 & 2.4 & 1.72 & 4.12 & 60 & 221 & 1.29 & 0.73 & 3.63 & 0.87 \\
E, DE & 235 & 13741 & 780 & 170 & 15.9 & 65 & 44 & 4 & 0.54 & 0.48 & 0.79 & 2.63 & 3.62 & 0.9 & 3.25 & 49 & 217 & 1.38 & 0.68 & 2.47 & 1.71 \\
E, JH & 347 & 15475 & 891 & 109 & 29.6 & 65 & 45 & 4 & 0.52 & 0.7 & 0.59 & 1.68 & 5.34 & 0.69 & 3.66 & 47 & 197 & 1.37 & 0.69 & 2.66 & 1.41 \\
E, VJ & 129 & 11496 & 1630 & 194 & 13.8 & 46 & 34 & 5 & 0.92 & 0.81 & 0.79 & 4.24 & 2.8 & 1.94 & 5.43 & 37 & 160 & 1.35 & 0.74 & 4.81 & 0.88 \\
E, M & 58 & 16166 & 12906 & 55 & 11.9 & 22 & 17 & 3 & 1.13 & 1.69 & 0.37 & 2.51 & 2.64 & 12.7 & 33.4 & 51 & 199 & 1.23 & 0.77 & 32.9 & 1.22 \\
F, RP & 69 & 21058 & 1715 & 1288 & 3.8 & 38 & 35 & 9 & 1.72 & 0.39 & 1.64 & 33.9 & 1.82 & 8.03 & 14.6 & 100 & 325 & 1.11 & 0.92 & 14.5 & 0.76 \\
F, ME & 362 & 33076 & 1490 & 231 & 29.5 & 93 & 66 & 8 & 0.75 & 0.63 & 0.71 & 2.49 & 3.89 & 0.98 & 3.82 & 56 & 216 & 1.33 & 0.71 & 2.94 & 1.79 \\
F, MPA & 145 & 16913 & 2260 & 190 & 18.1 & 59 & 42 & 7 & 1.06 & 0.96 & 0.7 & 3.23 & 2.46 & 1.98 & 4.86 & 74 & 238 & 1.2 & 0.71 & 4.44 & 2.11 \\
F, DS & 137 & 16532 & 1834 & 275 & 14 & 61 & 43 & 6 & 0.87 & 0.72 & 0.85 & 4.51 & 2.25 & 1.98 & 4.44 & 46 & 189 & 1.28 & 0.7 & 3.96 & 1.97 \\
G, H & 193 & 23540 & 2425 & 241 & 20.7 & 77 & 52 & 7 & 0.84 & 0.8 & 0.74 & 3.14 & 2.51 & 1.58 & 3.97 & 59 & 245 & 1.23 & 0.68 & 3.53 & 2.03 \\
G, C & 128 & 19273 & 6250 & 197 & 16.4 & 51 & 40 & 5 & 0.77 & 1.05 & 0.7 & 3.88 & 2.51 & 2.95 & 7.41 & 47 & 189 & 1.29 & 0.78 & 6.93 & 1.55 \\
G, SL & 157 & 20303 & 2548 & 203 & 19.4 & 61 & 44 & 6 & 0.85 & 1 & 0.69 & 3.33 & 2.57 & 2.12 & 5.46 & 45 & 202 & 1.25 & 0.72 & 5 & 1.2 \\
G, AC & 1064 & 44312 & 1602 & 103 & 80.1 & 108 & 71 & 6 & 0.49 & 0.69 & 0.47 & 0.96 & 9.85 & 0.39 & 3.8 & 58 & 243 & 1.39 & 0.66 & 2.31 & 2.12 \\
G, DJ & 217 & 24264 & 1722 & 300 & 17.3 & 67 & 51 & 8 & 0.99 & 0.75 & 0.85 & 4.49 & 3.24 & 1.67 & 5.41 & 36 & 171 & 1.33 & 0.76 & 4.86 & 1.52 \\
H, FDM & 89 & 13658 & 1823 & 274 & 11.8 & 44 & 34 & 6 & 1.13 & 0.93 & 0.88 & 6.25 & 2.02 & 3.49 & 7.05 & 40 & 177 & 1.32 & 0.77 & 6.58 & 1.29 \\
H, BI & 272 & 32647 & 2978 & 241 & 26.9 & 78 & 59 & 10 & 1.08 & 0.84 & 0.7 & 3.1 & 3.49 & 1.54 & 5.37 & 48 & 201 & 1.27 & 0.76 & 4.71 & 1.73 \\
H, DR & 200 & 18673 & 2482 & 210 & 19.1 & 64 & 46 & 5 & 0.66 & 0.83 & 0.71 & 3.3 & 3.13 & 1.46 & 4.56 & 46 & 173 & 1.31 & 0.72 & 3.92 & 1.31 \\
H, TW & 398 & 21854 & 1889 & 125 & 35.1 & 70 & 50 & 5 & 0.6 & 0.69 & 0.59 & 1.79 & 5.69 & 0.78 & 4.46 & 51 & 179 & 1.43 & 0.71 & 3.17 & 1.59 \\
H, H & 78 & 4287 & 535 & 121 & 9 & 33 & 25 & 3 & 0.74 & 0.7 & 0.84 & 3.69 & 2.36 & 1.67 & 3.94 & 54 & 221 & 1.21 & 0.76 & 3.49 & 1.18 \\
H, SE & 200 & 13256 & 1423 & 157 & 18.7 & 56 & 40 & 5 & 0.77 & 0.78 & 0.69 & 2.8 & 3.57 & 1.18 & 4.23 & 47 & 195 & 1.27 & 0.71 & 3.55 & 1.19 \\
H, JE & 186 & 10380 & 535 & 128 & 18.2 & 52 & 38 & 4 & 0.64 & 0.63 & 0.7 & 2.47 & 3.58 & 1.07 & 3.84 & 44 & 177 & 1.38 & 0.73 & 2.95 & 1.53 \\
I, F & 249 & 14235 & 1231 & 131 & 22.7 & 52 & 36 & 6 & 1.06 & 0.85 & 0.64 & 2.52 & 4.79 & 1.1 & 5.26 & 55 & 198 & 1.38 & 0.69 & 4.42 & 1.13 \\
I, Y & 241 & 12384 & 1598 & 102 & 24.6 & 52 & 38 & 4 & 0.64 & 0.87 & 0.57 & 1.96 & 4.63 & 0.99 & 4.58 & 36 & 163 & 1.38 & 0.73 & 3.66 & 1.16 \\
J, R & 229 & 26017 & 1742 & 286 & 19.9 & 74 & 58 & 8 & 0.86 & 0.75 & 0.8 & 3.87 & 3.09 & 1.54 & 4.75 & 46 & 188 & 1.22 & 0.78 & 4.21 & 1.72 \\
J, S & 185 & 12356 & 3836 & 75 & 25.1 & 43 & 33 & 5 & 0.95 & 1.04 & 0.48 & 1.76 & 4.3 & 1.55 & 6.68 & 39 & 172 & 1.37 & 0.77 & 5.74 & 0.98 \\
K, HJ & 241 & 16011 & 1228 & 215 & 16.7 & 62 & 48 & 6 & 0.77 & 0.66 & 0.79 & 3.48 & 3.89 & 1.07 & 4.17 & 45 & 166 & 1.35 & 0.77 & 3.49 & 1.68 \\
K, G & 211 & 11298 & 1949 & 89 & 25.1 & 47 & 34 & 4 & 0.72 & 0.81 & 0.55 & 1.9 & 4.49 & 1.14 & 5.11 & 34 & 154 & 1.47 & 0.72 & 3.84 & 1.57 \\
\hline
\hline
$\langle x \rangle $ & 275 & 20368 & 2686 & 183 & 26 & 61 & 44 & 6 & 0.85 & 0.83 & 0.67 & 3.37 & 4.23 & 1.88 & 6.04 & 51 & 201 & 1.34 & 0.72 & 5.18 & 1.58 \\
$\sigma$ & 190 & 11381 & 2436 & 158 & 14.4 & 20 & 14 & 2 & 0.31 & 0.23 & 0.19 & 3.87 & 1.9 & 2 & 4.09 & 11 & 30 & 0.09 & 0.05 & 4.27 & 0.59 \\\hline
\hline
\end{tabular}}}
\caption{  Career citation statistics for 100 dataset [A]  scientists: 1-50. 
}
\label{table:top100a}
\end{table}

\begin{table}[h]
\centering{ {\footnotesize
\begin{tabular}{@{\vrule height .5 pt depth4pt  width0pt}lccccccccccccccccccccc}
\noalign{
\vskip-1pt}
Name & $N$ & $C$ & $c(1)$ & $c(r^{*})$ & $r^{*}$ & $h_1$ & $h_2$ & $h_{80}$ & $\beta_{xy}$ & $\beta$ & $\gamma$ & $c(r^{*})/h_1$ & $N/h_1$ & $\langle c \rangle/h_1$ & $a$ & Gap(5) & Gap(10) & $m_{1/2}$ & $m_{2}$ & $P$ & $h_{1}/L$ \\
\hline
\hline
L, RB & 79 & 7751 & 2271 & 147 & 10.8 & 32 & 24 & 5 & 1.35 & 1.04 & 0.76 & 4.62 & 2.47 & 3.07 & 7.57 & 43 & 190 & 1.31 & 0.75 & 7.06 & 1.07 \\
L, PA & 344 & 32668 & 3228 & 208 & 30.8 & 80 & 59 & 9 & 0.96 & 0.8 & 0.67 & 2.61 & 4.3 & 1.19 & 5.1 & 40 & 166 & 1.38 & 0.74 & 4.26 & 1.82 \\
L, EH & 234 & 20139 & 1862 & 167 & 25.1 & 62 & 46 & 6 & 0.81 & 0.82 & 0.65 & 2.7 & 3.77 & 1.39 & 5.24 & 57 & 197 & 1.42 & 0.74 & 4.29 & 1.19 \\
L, SG & 379 & 27530 & 1355 & 160 & 34.9 & 84 & 62 & 6 & 0.58 & 0.66 & 0.62 & 1.91 & 4.51 & 0.86 & 3.9 & 53 & 177 & 1.43 & 0.74 & 2.74 & 2.33 \\
L, MD & 151 & 11231 & 876 & 178 & 14.5 & 50 & 37 & 5 & 0.84 & 0.73 & 0.8 & 3.58 & 3.02 & 1.49 & 4.49 & 50 & 211 & 1.24 & 0.74 & 3.97 & 3.13 \\
M, AH & 455 & 17708 & 641 & 93 & 37.9 & 67 & 46 & 4 & 0.51 & 0.58 & 0.56 & 1.4 & 6.79 & 0.58 & 3.94 & 43 & 175 & 1.49 & 0.69 & 2.34 & 1.91 \\
M, ND & 216 & 10409 & 2741 & 83 & 20.5 & 41 & 29 & 5 & 1.1 & 1.07 & 0.53 & 2.03 & 5.27 & 1.18 & 6.19 & 62 & 216 & 1.22 & 0.71 & 5.52 & 0.8 \\
M, RN & 371 & 18413 & 1919 & 87 & 40.6 & 62 & 42 & 5 & 0.73 & 0.84 & 0.47 & 1.41 & 5.98 & 0.8 & 4.79 & 36 & 173 & 1.48 & 0.68 & 3.29 & 1.77 \\
N, DR & 191 & 21742 & 1371 & 265 & 18.3 & 73 & 52 & 8 & 0.97 & 0.72 & 0.81 & 3.63 & 2.62 & 1.56 & 4.08 & 57 & 215 & 1.26 & 0.71 & 3.62 & 2.09 \\
O, E & 438 & 22310 & 2973 & 101 & 41.6 & 76 & 49 & 5 & 0.62 & 0.77 & 0.51 & 1.34 & 5.76 & 0.67 & 3.86 & 43 & 199 & 1.36 & 0.64 & 2.75 & 1.85 \\
O, SR & 146 & 5051 & 1236 & 59 & 14.8 & 26 & 21 & 3 & 0.9 & 1.14 & 0.53 & 2.31 & 5.62 & 1.33 & 7.47 & 55 & 183 & 1.31 & 0.81 & 6.7 & 0.65 \\
P, G & 529 & 29994 & 2768 & 108 & 51.1 & 81 & 55 & 7 & 0.79 & 0.82 & 0.49 & 1.34 & 6.53 & 0.7 & 4.57 & 53 & 194 & 1.41 & 0.68 & 3.23 & 1.84 \\
P, SSP & 330 & 19184 & 1760 & 108 & 34.7 & 58 & 41 & 7 & 1.09 & 0.84 & 0.53 & 1.87 & 5.69 & 1 & 5.7 & 40 & 157 & 1.55 & 0.71 & 4.18 & 2 \\
P, M & 435 & 29719 & 5147 & 123 & 43.6 & 85 & 57 & 5 & 0.52 & 0.76 & 0.53 & 1.45 & 5.12 & 0.8 & 4.11 & 48 & 196 & 1.41 & 0.67 & 2.94 & 2.36 \\
P, JB & 298 & 26621 & 2719 & 170 & 31.5 & 75 & 48 & 7 & 0.92 & 0.83 & 0.61 & 2.27 & 3.97 & 1.19 & 4.73 & 54 & 218 & 1.36 & 0.64 & 3.77 & 1.79 \\
P, JP & 250 & 62338 & 12906 & 173 & 31.4 & 62 & 46 & 9 & 1.26 & 1.38 & 0.49 & 2.8 & 4.03 & 4.02 & 16.2 & 63 & 224 & 1.34 & 0.74 & 15.4 & 1.63 \\
P, A & 169 & 10053 & 3849 & 50 & 25.9 & 37 & 25 & 3 & 0.74 & 1.1 & 0.38 & 1.37 & 4.57 & 1.61 & 7.34 & 57 & 203 & 1.46 & 0.68 & 6.23 & 0.79 \\
P, LN & 784 & 24901 & 460 & 89 & 52.4 & 82 & 56 & 3 & 0.26 & 0.53 & 0.53 & 1.09 & 9.56 & 0.39 & 3.7 & 45 & 174 & 1.44 & 0.68 & 2.04 & 1.86 \\
P, JC & 620 & 23513 & 1330 & 75 & 59.4 & 71 & 46 & 6 & 0.81 & 0.78 & 0.43 & 1.07 & 8.73 & 0.53 & 4.66 & 57 & 243 & 1.49 & 0.65 & 2.82 & 1.31 \\
P, HD & 71 & 8721 & 1807 & 333 & 6.7 & 35 & 27 & 4 & 0.93 & 0.7 & 1.11 & 9.53 & 2.03 & 3.51 & 7.12 & 51 & 202 & 1.2 & 0.77 & 6.81 & 1.06 \\
R, L & 105 & 12124 & 3491 & 100 & 16.6 & 37 & 26 & 4 & 0.97 & 1.21 & 0.55 & 2.7 & 2.84 & 3.12 & 8.86 & 64 & 212 & 1.35 & 0.7 & 8.23 & 1.54 \\
R, TM & 345 & 25117 & 2112 & 193 & 27.2 & 81 & 58 & 6 & 0.63 & 0.69 & 0.68 & 2.39 & 4.26 & 0.9 & 3.83 & 45 & 190 & 1.38 & 0.72 & 3.03 & 1.84 \\
S, JJ & 265 & 7662 & 868 & 52 & 29.2 & 43 & 29 & 3 & 0.63 & 0.82 & 0.46 & 1.22 & 6.16 & 0.67 & 4.14 & 55 & 202 & 1.35 & 0.67 & 2.95 & 0.84 \\
S, LM & 154 & 9510 & 3062 & 71 & 21.1 & 37 & 27 & 5 & 1.19 & 1.08 & 0.46 & 1.93 & 4.16 & 1.67 & 6.95 & 38 & 165 & 1.35 & 0.73 & 6.02 & 0.95 \\
S, GA & 335 & 21292 & 1328 & 155 & 28.3 & 77 & 53 & 5 & 0.56 & 0.61 & 0.66 & 2.02 & 4.35 & 0.83 & 3.59 & 49 & 195 & 1.36 & 0.69 & 2.64 & 1.75 \\
S, DJ & 333 & 17958 & 589 & 129 & 28.8 & 71 & 49 & 5 & 0.62 & 0.59 & 0.64 & 1.82 & 4.69 & 0.76 & 3.56 & 46 & 183 & 1.41 & 0.69 & 2.43 & 2.03 \\
S, M & 415 & 19276 & 580 & 115 & 33.2 & 74 & 48 & 4 & 0.48 & 0.54 & 0.61 & 1.56 & 5.61 & 0.63 & 3.52 & 36 & 161 & 1.46 & 0.65 & 2.1 & 2.24 \\
S, JR & 174 & 24689 & 5636 & 208 & 18.9 & 52 & 41 & 8 & 1.26 & 1.16 & 0.69 & 4.01 & 3.35 & 2.73 & 9.13 & 57 & 230 & 1.19 & 0.79 & 8.7 & 0.98 \\
S, MO & 573 & 19269 & 1456 & 74 & 49.2 & 68 & 48 & 5 & 0.63 & 0.75 & 0.46 & 1.1 & 8.43 & 0.49 & 4.17 & 53 & 205 & 1.4 & 0.71 & 2.72 & 1.55 \\
S, YR & 637 & 26458 & 1038 & 114 & 44.1 & 86 & 59 & 4 & 0.37 & 0.54 & 0.57 & 1.33 & 7.41 & 0.48 & 3.58 & 54 & 197 & 1.4 & 0.69 & 2.1 & 1.87 \\
S, DJ & 242 & 15414 & 7118 & 71 & 27.8 & 49 & 32 & 4 & 0.77 & 0.94 & 0.47 & 1.45 & 4.94 & 1.3 & 6.42 & 58 & 239 & 1.33 & 0.65 & 5.48 & 2.23 \\
S, HE & 909 & 41505 & 892 & 115 & 68.9 & 100 & 68 & 7 & 0.62 & 0.61 & 0.52 & 1.15 & 9.09 & 0.46 & 4.15 & 53 & 216 & 1.44 & 0.68 & 2.36 & 2.22 \\
S, PJ & 173 & 19462 & 1700 & 201 & 19.9 & 58 & 44 & 7 & 1.01 & 0.87 & 0.73 & 3.48 & 2.98 & 1.94 & 5.79 & 58 & 232 & 1.24 & 0.76 & 5.15 & 1.66 \\
S, R & 46 & 8952 & 3491 & 82 & 9.8 & 21 & 16 & 3 & 1.2 & 1.72 & 0.5 & 3.95 & 2.19 & 9.27 & 20.3 & 45 & 186 & 1.33 & 0.76 & 19.9 & 0.88 \\
S, RH & 127 & 9186 & 1526 & 96 & 16.3 & 35 & 28 & 4 & 0.9 & 1.22 & 0.56 & 2.76 & 3.63 & 2.07 & 7.5 & 49 & 195 & 1.26 & 0.8 & 6.92 & 0.92 \\
T, J & 181 & 22501 & 1782 & 275 & 18.1 & 70 & 51 & 8 & 0.99 & 0.72 & 0.82 & 3.94 & 2.59 & 1.78 & 4.59 & 64 & 232 & 1.24 & 0.73 & 4.06 & 2.26 \\
T, M & 262 & 15755 & 1687 & 133 & 24.3 & 60 & 43 & 4 & 0.55 & 0.72 & 0.64 & 2.23 & 4.37 & 1 & 4.38 & 40 & 180 & 1.38 & 0.72 & 3.37 & 1.05 \\
T, DC & 493 & 17649 & 1602 & 82 & 41.6 & 70 & 46 & 4 & 0.51 & 0.66 & 0.51 & 1.18 & 7.04 & 0.51 & 3.6 & 38 & 168 & 1.44 & 0.66 & 2.23 & 1.59 \\
V, CM & 253 & 14935 & 2466 & 112 & 25.3 & 58 & 39 & 5 & 0.8 & 0.92 & 0.57 & 1.94 & 4.36 & 1.02 & 4.44 & 60 & 207 & 1.34 & 0.67 & 3.73 & 1.38 \\
W, S & 208 & 42287 & 5094 & 488 & 17.9 & 91 & 71 & 10 & 0.88 & 0.68 & 0.92 & 5.37 & 2.29 & 2.23 & 5.11 & 57 & 231 & 1.19 & 0.78 & 4.76 & 1.47 \\
W, DA & 330 & 16955 & 610 & 140 & 25.5 & 68 & 48 & 5 & 0.63 & 0.57 & 0.67 & 2.07 & 4.85 & 0.76 & 3.67 & 41 & 181 & 1.41 & 0.71 & 2.53 & 2.06 \\
W, KW & 742 & 24655 & 458 & 90 & 51.6 & 81 & 54 & 3 & 0.28 & 0.53 & 0.53 & 1.12 & 9.16 & 0.41 & 3.76 & 54 & 205 & 1.43 & 0.67 & 2.05 & 1.93 \\
W, SR & 124 & 9821 & 1511 & 144 & 15.2 & 48 & 34 & 4 & 0.72 & 0.85 & 0.7 & 3 & 2.58 & 1.65 & 4.26 & 80 & 282 & 1.21 & 0.71 & 3.76 & 1.66 \\
W, F & 263 & 26549 & 1722 & 254 & 22.2 & 81 & 63 & 7 & 0.68 & 0.67 & 0.78 & 3.14 & 3.25 & 1.25 & 4.05 & 52 & 194 & 1.31 & 0.78 & 3.44 & 2.19 \\
W, E & 264 & 65014 & 2034 & 860 & 14.6 & 121 & 92 & 13 & 0.89 & 0.45 & 1.06 & 7.11 & 2.18 & 2.04 & 4.44 & 37 & 167 & 1.23 & 0.76 & 4.09 & 3.56 \\
W, WK & 49 & 13348 & 3815 & 495 & 6.5 & 27 & 21 & 6 & 1.94 & 1.17 & 1.18 & 18.4 & 1.81 & 10.1 & 18.3 & 53 & 233 & 1.19 & 0.78 & 18.1 & 0.84 \\
Y, E & 172 & 17852 & 6022 & 153 & 19.6 & 49 & 36 & 6 & 1.06 & 0.92 & 0.65 & 3.14 & 3.51 & 2.12 & 7.44 & 40 & 171 & 1.43 & 0.73 & 6.6 & 1.44 \\
Y, CN & 194 & 23798 & 1537 & 318 & 16.2 & 67 & 54 & 8 & 0.93 & 0.71 & 0.89 & 4.76 & 2.9 & 1.83 & 5.3 & 53 & 218 & 1.27 & 0.81 & 4.84 & 1.03 \\
Z, P & 331 & 22263 & 1514 & 148 & 30.8 & 77 & 52 & 6 & 0.71 & 0.68 & 0.62 & 1.93 & 4.3 & 0.87 & 3.75 & 65 & 237 & 1.45 & 0.68 & 2.7 & 2.33 \\
Z, A & 581 & 36151 & 7861 & 109 & 54.3 & 85 & 59 & 4 & 0.37 & 0.69 & 0.5 & 1.29 & 6.84 & 0.73 & 5 & 47 & 183 & 1.51 & 0.69 & 3.19 & 2.24 \\
\hline
$\langle x \rangle $ & 275 & 20368 & 2686 & 183 & 26 & 61 & 44 & 6 & 0.85 & 0.83 & 0.67 & 3.37 & 4.23 & 1.88 & 6.04 & 51 & 201 & 1.34 & 0.72 & 5.18 & 1.58 \\
$\sigma$ & 190 & 11381 & 2436 & 158 & 14.4 & 20 & 14 & 2 & 0.31 & 0.23 & 0.19 & 3.87 & 1.9 & 2 & 4.09 & 11 & 30 & 0.09 & 0.05 & 4.27 & 0.59 \\
\hline
\hline
\end{tabular}}}
\caption{  Career citation statistics for 100 dataset [A] scientists: 51-100. 
}
\label{table:top100b}
\end{table}

\begin{table}[h]
\centering{ {\footnotesize
\begin{tabular}{@{\vrule height .5 pt depth4pt  width0pt}lccccccccccccccccccccc}
\noalign{
\vskip-1pt}
Name & $N$ & $C$ & $c(1)$ & $c(r^{*})$ & $r^{*}$ & $h_1$ & $h_2$ & $h_{80}$ & $\beta_{xy}$ & $\beta$ & $\gamma$ & $c(r^{*})/h_1$ & $N/h_1$ & $\langle c \rangle/h_1$ & $a$ & Gap(5) & Gap(10) & $m_{1/2}$ & $m_{2}$ & $P$ & $h_{1}/L$ \\
\hline
\hline
A, P & 125 & 5167 & 668 & 94 & 13 & 36 & 26 & 3 & 0.71 & 0.84 & 0.68 & 2.64 & 3.47 & 1.15 & 3.99 & 63 & 236 & 1.31 & 0.72 & 3.41 & 0.86 \\
A, DE & 469 & 18982 & 1819 & 89 & 40.9 & 66 & 46 & 6 & 0.81 & 0.82 & 0.49 & 1.36 & 7.11 & 0.61 & 4.36 & 44 & 184 & 1.41 & 0.7 & 3.12 & 1.5 \\
B, RZ & 143 & 4946 & 200 & 102 & 9.3 & 41 & 28 & 2 & 0.4 & 0.38 & 0.83 & 2.51 & 3.49 & 0.84 & 2.94 & 41 & 165 & 1.34 & 0.68 & 2.22 & 1.24 \\
B, BB & 252 & 6928 & 520 & 71 & 20.6 & 45 & 33 & 2 & 0.32 & 0.62 & 0.6 & 1.59 & 5.6 & 0.61 & 3.42 & 64 & 215 & 1.38 & 0.73 & 2.41 & 1.13 \\
B, WF & 73 & 2723 & 227 & 96 & 6.9 & 29 & 20 & 2 & 0.6 & 0.5 & 0.89 & 3.34 & 2.52 & 1.29 & 3.24 & 52 & 212 & 1.24 & 0.69 & 2.75 & 0.57 \\
B, AL & 170 & 25048 & 4461 & 203 & 21.5 & 61 & 45 & 8 & 1.14 & 1.08 & 0.65 & 3.34 & 2.79 & 2.42 & 6.73 & 47 & 226 & 1.25 & 0.74 & 6.21 & 2.9 \\
B, RH & 87 & 2589 & 298 & 63 & 10.4 & 25 & 18 & 2 & 0.68 & 0.76 & 0.67 & 2.54 & 3.48 & 1.19 & 4.14 & 61 & 207 & 1.32 & 0.72 & 3.38 & 0.83 \\
B, L & 112 & 1841 & 107 & 41 & 10.5 & 25 & 16 & 1 & 0.33 & 0.5 & 0.64 & 1.65 & 4.48 & 0.66 & 2.95 & 51 & 200 & 1.36 & 0.64 & 1.98 & 1.19 \\
B, K & 763 & 35274 & 2726 & 100 & 66.1 & 89 & 58 & 5 & 0.51 & 0.68 & 0.48 & 1.13 & 8.57 & 0.52 & 4.45 & 78 & 251 & 1.48 & 0.65 & 2.49 & 2.07 \\
B, KI & 64 & 1199 & 124 & 49 & 5.9 & 21 & 13 & 1 & 0.44 & 0.45 & 0.79 & 2.35 & 3.05 & 0.89 & 2.72 & 53 & 209 & 1.33 & 0.62 & 2.08 & 0.68 \\
B, RW & 311 & 7063 & 282 & 61 & 25 & 44 & 31 & 3 & 0.58 & 0.62 & 0.54 & 1.39 & 7.07 & 0.52 & 3.65 & 31 & 169 & 1.41 & 0.7 & 2.33 & 1.19 \\
B, AJ & 240 & 9685 & 1384 & 81 & 24.4 & 48 & 33 & 3 & 0.54 & 0.66 & 0.57 & 1.69 & 5 & 0.84 & 4.2 & 42 & 172 & 1.44 & 0.69 & 2.85 & 1.3 \\
B, JH & 334 & 8108 & 733 & 51 & 32.2 & 44 & 31 & 3 & 0.58 & 0.77 & 0.45 & 1.18 & 7.59 & 0.55 & 4.19 & 44 & 166 & 1.45 & 0.7 & 2.74 & 1.13 \\
B, SJ & 275 & 19230 & 1696 & 161 & 25 & 74 & 50 & 5 & 0.6 & 0.67 & 0.67 & 2.18 & 3.72 & 0.94 & 3.51 & 45 & 179 & 1.35 & 0.68 & 2.72 & 1.68 \\
B, RA & 384 & 9774 & 442 & 62 & 32 & 49 & 32 & 3 & 0.56 & 0.56 & 0.52 & 1.27 & 7.84 & 0.52 & 4.07 & 38 & 171 & 1.59 & 0.65 & 2.07 & 1.11 \\
C, EM & 108 & 6069 & 1306 & 103 & 12.2 & 34 & 25 & 4 & 1.01 & 1.02 & 0.67 & 3.04 & 3.18 & 1.65 & 5.25 & 57 & 215 & 1.21 & 0.74 & 4.8 & 1.1 \\
C, NJ & 107 & 2898 & 255 & 57 & 12 & 28 & 18 & 2 & 0.68 & 0.62 & 0.64 & 2.05 & 3.82 & 0.97 & 3.7 & 49 & 184 & 1.57 & 0.64 & 2.6 & 1.47 \\
C, NS & 140 & 2953 & 166 & 59 & 11.2 & 30 & 20 & 1 & 0.23 & 0.48 & 0.67 & 1.99 & 4.67 & 0.7 & 3.28 & 48 & 186 & 1.47 & 0.67 & 2.11 & 0.71 \\
C, G & 125 & 10245 & 5600 & 74 & 16.9 & 34 & 27 & 3 & 0.68 & 0.99 & 0.54 & 2.2 & 3.68 & 2.41 & 8.86 & 48 & 174 & 1.41 & 0.79 & 7.98 & 0.83 \\
D, C & 208 & 19421 & 2693 & 165 & 22.1 & 58 & 43 & 7 & 1.03 & 1 & 0.64 & 2.86 & 3.59 & 1.61 & 5.77 & 51 & 199 & 1.31 & 0.74 & 5.13 & 2.32 \\
D, TJ & 361 & 15040 & 872 & 94 & 32.4 & 64 & 41 & 4 & 0.59 & 0.58 & 0.56 & 1.47 & 5.64 & 0.65 & 3.67 & 47 & 207 & 1.48 & 0.64 & 2.11 & 1.31 \\
D, G & 507 & 26718 & 1352 & 123 & 43.5 & 75 & 52 & 7 & 0.84 & 0.78 & 0.53 & 1.64 & 6.76 & 0.7 & 4.75 & 36 & 156 & 1.44 & 0.69 & 3.34 & 1.32 \\
E, JP & 101 & 5833 & 383 & 151 & 9.1 & 36 & 29 & 3 & 0.63 & 0.6 & 0.92 & 4.2 & 2.81 & 1.6 & 4.5 & 57 & 205 & 1.31 & 0.81 & 4.01 & 1.16 \\
E, RW & 188 & 12092 & 2535 & 139 & 16.6 & 40 & 32 & 6 & 1.2 & 0.96 & 0.68 & 3.5 & 4.7 & 1.61 & 7.56 & 49 & 176 & 1.4 & 0.8 & 6.77 & 1.08 \\
F, JC & 113 & 1854 & 174 & 37 & 11.4 & 26 & 16 & 1 & 0.33 & 0.55 & 0.6 & 1.46 & 4.35 & 0.63 & 2.74 & 59 & 212 & 1.38 & 0.62 & 1.86 & 0.84 \\
F, AR & 385 & 17615 & 4267 & 74 & 39.6 & 59 & 41 & 4 & 0.59 & 0.85 & 0.46 & 1.27 & 6.53 & 0.78 & 5.06 & 61 & 225 & 1.46 & 0.69 & 3.72 & 1.28 \\
F, PA & 99 & 5286 & 413 & 158 & 8.4 & 37 & 28 & 4 & 0.9 & 0.6 & 0.9 & 4.29 & 2.68 & 1.44 & 3.86 & 35 & 153 & 1.3 & 0.76 & 3.38 & 1.19 \\
F, KJ & 135 & 8154 & 406 & 176 & 10.4 & 46 & 35 & 4 & 0.7 & 0.52 & 0.9 & 3.84 & 2.93 & 1.31 & 3.85 & 46 & 185 & 1.35 & 0.76 & 3.3 & 1.02 \\
F, ED & 240 & 5711 & 532 & 45 & 26.5 & 37 & 24 & 2 & 0.48 & 0.63 & 0.46 & 1.22 & 6.49 & 0.64 & 4.17 & 66 & 223 & 1.51 & 0.65 & 2.27 & 1.48 \\
F, D & 347 & 15664 & 518 & 108 & 29.8 & 65 & 45 & 4 & 0.52 & 0.56 & 0.61 & 1.67 & 5.34 & 0.69 & 3.71 & 61 & 210 & 1.48 & 0.69 & 2.34 & 1.71 \\
G, GW & 210 & 13358 & 1199 & 131 & 22.1 & 61 & 41 & 3 & 0.41 & 0.73 & 0.65 & 2.16 & 3.44 & 1.04 & 3.59 & 43 & 196 & 1.3 & 0.67 & 2.84 & 1.56 \\
G, DC & 75 & 2057 & 342 & 69 & 7.9 & 21 & 17 & 2 & 0.72 & 0.86 & 0.74 & 3.32 & 3.57 & 1.31 & 4.66 & 54 & 190 & 1.24 & 0.81 & 4.16 & 0.84 \\
G, W & 322 & 7594 & 375 & 54 & 28.8 & 47 & 30 & 2 & 0.36 & 0.58 & 0.51 & 1.17 & 6.85 & 0.5 & 3.44 & 42 & 192 & 1.47 & 0.64 & 1.92 & 1.07 \\
G, SC & 132 & 4725 & 407 & 89 & 12.8 & 38 & 26 & 2 & 0.44 & 0.68 & 0.7 & 2.36 & 3.47 & 0.94 & 3.27 & 56 & 213 & 1.32 & 0.68 & 2.65 & 1.73 \\
G, AM & 284 & 6316 & 376 & 55 & 24.7 & 42 & 28 & 2 & 0.4 & 0.6 & 0.53 & 1.32 & 6.76 & 0.53 & 3.58 & 56 & 212 & 1.43 & 0.67 & 2.17 & 0.89 \\
G, P & 255 & 16210 & 2645 & 105 & 28 & 55 & 39 & 5 & 0.8 & 0.97 & 0.52 & 1.92 & 4.64 & 1.16 & 5.36 & 51 & 201 & 1.35 & 0.71 & 4.45 & 1.28 \\
H, P & 491 & 20518 & 2568 & 71 & 50.4 & 63 & 41 & 4 & 0.59 & 0.83 & 0.42 & 1.14 & 7.79 & 0.66 & 5.17 & 42 & 176 & 1.46 & 0.65 & 3.45 & 1.75 \\
H, S & 527 & 18490 & 1145 & 67 & 52.4 & 64 & 42 & 5 & 0.73 & 0.8 & 0.43 & 1.06 & 8.23 & 0.55 & 4.51 & 72 & 240 & 1.41 & 0.66 & 2.92 & 1.68 \\
H, SW & 146 & 25088 & 3444 & 387 & 14.6 & 69 & 50 & 8 & 1.01 & 0.73 & 0.9 & 5.62 & 2.12 & 2.49 & 5.27 & 62 & 217 & 1.28 & 0.72 & 4.84 & 1.53 \\
H, HJ & 383 & 8042 & 232 & 52 & 31.7 & 47 & 32 & 2 & 0.33 & 0.54 & 0.51 & 1.11 & 8.15 & 0.45 & 3.64 & 53 & 198 & 1.45 & 0.68 & 1.97 & 1.52 \\
H, F & 236 & 12176 & 821 & 124 & 20.8 & 57 & 41 & 4 & 0.59 & 0.59 & 0.68 & 2.19 & 4.14 & 0.91 & 3.75 & 44 & 170 & 1.4 & 0.72 & 2.7 & 1.16 \\
H, JJ & 163 & 27512 & 5232 & 370 & 15.8 & 68 & 52 & 8 & 0.97 & 0.87 & 0.84 & 5.45 & 2.4 & 2.48 & 5.95 & 48 & 184 & 1.24 & 0.76 & 5.53 & 1.33 \\
H, MS & 279 & 3331 & 292 & 25 & 27.8 & 27 & 19 & 1 & 0.25 & 0.7 & 0.38 & 0.94 & 10.3 & 0.44 & 4.57 & 30 & 136 & 1.56 & 0.7 & 2.44 & 0.77 \\
H, CE & 151 & 6326 & 403 & 92 & 15.7 & 42 & 30 & 3 & 0.6 & 0.63 & 0.67 & 2.2 & 3.6 & 1 & 3.59 & 33 & 152 & 1.43 & 0.71 & 2.7 & 1.56 \\
I, J & 147 & 5831 & 1028 & 71 & 17.4 & 40 & 27 & 3 & 0.68 & 0.85 & 0.57 & 1.78 & 3.68 & 0.99 & 3.64 & 79 & 244 & 1.3 & 0.68 & 2.95 & 1.25 \\
J, PK & 423 & 7661 & 212 & 46 & 32.3 & 42 & 27 & 2 & 0.42 & 0.48 & 0.49 & 1.11 & 10.1 & 0.43 & 4.34 & 52 & 198 & 1.57 & 0.64 & 1.81 & 1 \\
K, LP & 188 & 17057 & 2007 & 218 & 16.9 & 54 & 44 & 7 & 1.01 & 0.87 & 0.77 & 4.04 & 3.48 & 1.68 & 5.85 & 61 & 239 & 1.22 & 0.81 & 5.33 & 1.06 \\
K, E & 231 & 8413 & 839 & 76 & 23.6 & 47 & 32 & 3 & 0.56 & 0.72 & 0.55 & 1.63 & 4.91 & 0.77 & 3.81 & 56 & 201 & 1.47 & 0.68 & 2.62 & 1.88 \\
K, W & 172 & 18752 & 2763 & 228 & 17.3 & 63 & 46 & 6 & 0.81 & 0.74 & 0.79 & 3.63 & 2.73 & 1.73 & 4.72 & 61 & 226 & 1.3 & 0.73 & 4.16 & 2.33 \\
K, DV & 78 & 1112 & 106 & 29 & 9.8 & 19 & 12 & 1 & 0.48 & 0.68 & 0.56 & 1.54 & 4.11 & 0.75 & 3.08 & 49 & 189 & 1.37 & 0.63 & 2.15 & 0.76 \\
\hline
$\langle x \rangle $ &  217 & 9230 & 1024 & 96 & 20.7 & 44 & 31 & 3 & 0.62 & 0.7 & 0.62 & 2.19 & 4.92 & 1.01 & 4.23 & 52 & 196 & 1.39 & 0.7 & 3.19 & 1.28 \\
$\sigma$ & 121 & 6860 & 1158 & 64 & 10.6 & 14 & 10 & 2 & 0.25 & 0.16 & 0.14 & 1.08 & 2.01 & 0.54 & 1.23 & 11 & 27 & 0.09 & 0.05 & 1.36 & 0.5 \\
\hline
\hline
\end{tabular}}}
\caption{  Career citation statistics for 100 dataset [B] scientists: 1-50. 
}
\label{table:random100a}
\end{table}

\begin{table}[h]
\centering{ { \footnotesize
\begin{tabular}{@{\vrule height .5 pt depth4pt  width0pt}lccccccccccccccccccccc}
\noalign{
\vskip-1pt}
Name & $N$ & $C$ & $c(1)$ & $c(r^{*})$ & $h_1$ & $h_2$ & $h_{80}$ & $\beta_{pq}$ & $\beta$ & $\gamma$ & $c(r^{*})/h_1$ & $N/h_1$ & $\langle c \rangle/h_1$ & $a$ & Gap(5) & Gap(10) & $m_{1/2}$ & $m_{2}$ & $P$ & $h_{1}/L$ \\
\hline
\hline
K, TR & 161 & 5394 & 549 & 66 & 18.5 & 35 & 26 & 3 & 0.71 & 0.75 & 0.56 & 1.89 & 4.6 & 0.96 & 4.4 & 39 & 153 & 1.49 & 0.74 & 3.18 & 1.17 \\
K, L & 268 & 10661 & 3016 & 61 & 29.7 & 49 & 31 & 2 & 0.35 & 0.84 & 0.45 & 1.25 & 5.47 & 0.81 & 4.44 & 59 & 215 & 1.43 & 0.63 & 3.26 & 0.98 \\
K, W & 395 & 4433 & 193 & 27 & 33.5 & 35 & 23 & 1 & 0.18 & 0.63 & 0.39 & 0.8 & 11.3 & 0.32 & 3.62 & 64 & 213 & 1.4 & 0.66 & 1.9 & 0.63 \\
K, WR & 111 & 15302 & 2535 & 301 & 11.3 & 45 & 37 & 6 & 1.03 & 0.8 & 0.95 & 6.69 & 2.47 & 3.06 & 7.56 & 56 & 211 & 1.24 & 0.82 & 7.11 & 1.41 \\
L, RB & 157 & 6244 & 571 & 98 & 14.6 & 39 & 30 & 3 & 0.6 & 0.76 & 0.7 & 2.54 & 4.03 & 1.02 & 4.11 & 42 & 165 & 1.31 & 0.77 & 3.41 & 0.87 \\
L, P & 255 & 6264 & 300 & 57 & 23.4 & 41 & 30 & 2 & 0.36 & 0.59 & 0.54 & 1.4 & 6.22 & 0.6 & 3.73 & 42 & 165 & 1.51 & 0.73 & 2.2 & 0.93 \\
L, MJ & 180 & 2110 & 298 & 40 & 12.4 & 21 & 16 & 2 & 0.77 & 0.82 & 0.56 & 1.94 & 8.57 & 0.56 & 4.78 & 43 & 159 & 1.43 & 0.76 & 3.65 & 0.88 \\
L, M & 240 & 7535 & 535 & 59 & 27.1 & 43 & 28 & 3 & 0.65 & 0.74 & 0.49 & 1.39 & 5.58 & 0.73 & 4.08 & 63 & 215 & 1.51 & 0.65 & 2.57 & 1.02 \\
L, AJ & 152 & 14577 & 2261 & 113 & 19.8 & 42 & 29 & 7 & 1.6 & 1.23 & 0.56 & 2.71 & 3.62 & 2.28 & 8.26 & 56 & 216 & 1.24 & 0.69 & 7.66 & 0.91 \\
L, RA & 190 & 5481 & 489 & 69 & 17.9 & 36 & 27 & 3 & 0.68 & 0.78 & 0.57 & 1.94 & 5.28 & 0.8 & 4.23 & 47 & 202 & 1.28 & 0.75 & 3.24 & 0.86 \\
L, H & 234 & 6277 & 279 & 60 & 22 & 42 & 27 & 2 & 0.42 & 0.55 & 0.57 & 1.45 & 5.57 & 0.64 & 3.56 & 59 & 203 & 1.48 & 0.64 & 2.08 & 1.83 \\
L, MS & 143 & 2379 & 319 & 41 & 13.6 & 24 & 17 & 2 & 0.72 & 0.89 & 0.51 & 1.71 & 5.96 & 0.69 & 4.13 & 48 & 178 & 1.33 & 0.71 & 3.25 & 0.5 \\
M, L & 264 & 13179 & 863 & 125 & 22.9 & 57 & 40 & 5 & 0.77 & 0.74 & 0.65 & 2.2 & 4.63 & 0.88 & 4.06 & 62 & 225 & 1.35 & 0.7 & 3.22 & 1.3 \\
M, BT & 244 & 9633 & 686 & 91 & 22.3 & 52 & 37 & 3 & 0.47 & 0.59 & 0.62 & 1.76 & 4.69 & 0.76 & 3.56 & 55 & 195 & 1.46 & 0.71 & 2.37 & 1.41 \\
M, P & 398 & 5915 & 372 & 36 & 34.3 & 38 & 25 & 2 & 0.46 & 0.72 & 0.4 & 0.96 & 10.5 & 0.39 & 4.1 & 46 & 190 & 1.45 & 0.66 & 2.31 & 0.9 \\
M, DE & 107 & 6011 & 865 & 139 & 10.2 & 40 & 29 & 3 & 0.63 & 0.63 & 0.83 & 3.48 & 2.68 & 1.4 & 3.76 & 44 & 169 & 1.35 & 0.73 & 3.25 & 1.08 \\
M, JE & 176 & 8053 & 572 & 91 & 19.4 & 44 & 30 & 4 & 0.83 & 0.8 & 0.61 & 2.09 & 4 & 1.04 & 4.16 & 49 & 189 & 1.41 & 0.68 & 3.28 & 1.22 \\
M, GE & 420 & 10571 & 862 & 52 & 40.6 & 52 & 33 & 3 & 0.54 & 0.64 & 0.44 & 1.02 & 8.08 & 0.48 & 3.91 & 68 & 238 & 1.38 & 0.63 & 2.09 & 1.33 \\
N, AHC & 158 & 3509 & 431 & 44 & 17.9 & 30 & 20 & 2 & 0.6 & 0.78 & 0.49 & 1.47 & 5.27 & 0.74 & 3.9 & 62 & 217 & 1.37 & 0.67 & 2.74 & 1.5 \\
O, V & 104 & 6588 & 663 & 164 & 9.6 & 40 & 31 & 3 & 0.58 & 0.58 & 0.88 & 4.12 & 2.6 & 1.58 & 4.12 & 34 & 147 & 1.38 & 0.78 & 3.54 & 3.64 \\
O, SA & 150 & 9554 & 538 & 176 & 11.7 & 53 & 39 & 4 & 0.62 & 0.5 & 0.87 & 3.34 & 2.83 & 1.2 & 3.4 & 55 & 201 & 1.3 & 0.74 & 2.86 & 1.2 \\
P, VM & 83 & 2089 & 254 & 50 & 10.4 & 24 & 18 & 2 & 0.68 & 0.68 & 0.63 & 2.09 & 3.46 & 1.05 & 3.63 & 43 & 150 & 1.5 & 0.75 & 2.69 & 1.41 \\
P, CJ & 184 & 8877 & 522 & 115 & 17.7 & 49 & 33 & 4 & 0.75 & 0.64 & 0.68 & 2.36 & 3.76 & 0.98 & 3.7 & 46 & 187 & 1.39 & 0.67 & 2.8 & 1.09 \\
P, PM & 204 & 8569 & 432 & 109 & 17.2 & 50 & 34 & 3 & 0.52 & 0.58 & 0.7 & 2.19 & 4.08 & 0.84 & 3.43 & 60 & 212 & 1.34 & 0.68 & 2.6 & 0.98 \\
P, VL & 137 & 2932 & 433 & 41 & 15.5 & 27 & 20 & 1 & 0.23 & 0.74 & 0.53 & 1.54 & 5.07 & 0.79 & 4.02 & 46 & 183 & 1.3 & 0.74 & 3.02 & 0.6 \\
P, CY & 118 & 3214 & 548 & 42 & 16.2 & 25 & 17 & 3 & 1.13 & 0.87 & 0.45 & 1.7 & 4.72 & 1.09 & 5.14 & 41 & 147 & 1.6 & 0.68 & 3.67 & 0.56 \\
R, AR & 113 & 5257 & 295 & 101 & 12.3 & 36 & 25 & 3 & 0.74 & 0.63 & 0.75 & 2.83 & 3.14 & 1.29 & 4.06 & 51 & 196 & 1.36 & 0.69 & 3.27 & 1.24 \\
S, BEA & 284 & 4937 & 337 & 43 & 25 & 38 & 23 & 2 & 0.51 & 0.59 & 0.48 & 1.13 & 7.47 & 0.46 & 3.42 & 69 & 244 & 1.37 & 0.61 & 1.86 & 1.03 \\
S, RD & 121 & 4585 & 449 & 109 & 9.6 & 37 & 27 & 2 & 0.42 & 0.53 & 0.8 & 2.97 & 3.27 & 1.02 & 3.35 & 50 & 190 & 1.35 & 0.73 & 2.68 & 1.32 \\
S, F & 266 & 10047 & 636 & 82 & 25.6 & 53 & 36 & 3 & 0.48 & 0.67 & 0.57 & 1.56 & 5.02 & 0.71 & 3.58 & 41 & 173 & 1.4 & 0.68 & 2.48 & 1.89 \\
S, WD & 45 & 1330 & 154 & 78 & 4.2 & 21 & 15 & 1 & 0.36 & 0.4 & 0.93 & 3.75 & 2.14 & 1.41 & 3.02 & 33 & 148 & 1.43 & 0.71 & 2.51 & 0.68 \\
S, J & 77 & 3254 & 643 & 78 & 10 & 28 & 20 & 3 & 0.94 & 0.89 & 0.69 & 2.8 & 2.75 & 1.51 & 4.15 & 76 & 234 & 1.21 & 0.71 & 3.68 & 2.8 \\
S, L & 108 & 4026 & 440 & 86 & 11.3 & 30 & 22 & 2 & 0.54 & 0.78 & 0.71 & 2.89 & 3.6 & 1.24 & 4.47 & 49 & 180 & 1.33 & 0.73 & 3.82 & 0.97 \\
S, GF & 202 & 26489 & 3501 & 150 & 24.8 & 52 & 37 & 7 & 1.22 & 1.23 & 0.56 & 2.9 & 3.88 & 2.52 & 9.8 & 72 & 227 & 1.35 & 0.71 & 9.09 & 1.27 \\
S, D & 363 & 7894 & 238 & 53 & 30.5 & 45 & 30 & 2 & 0.36 & 0.57 & 0.5 & 1.2 & 8.07 & 0.48 & 3.9 & 55 & 190 & 1.49 & 0.67 & 2.08 & 1.55 \\
S, KR & 211 & 7371 & 482 & 81 & 20.2 & 48 & 32 & 3 & 0.56 & 0.68 & 0.6 & 1.7 & 4.4 & 0.73 & 3.2 & 58 & 195 & 1.33 & 0.67 & 2.38 & 1.3 \\
S, EA & 272 & 12743 & 882 & 103 & 26.1 & 50 & 36 & 5 & 0.87 & 0.78 & 0.58 & 2.07 & 5.44 & 0.94 & 5.1 & 34 & 138 & 1.5 & 0.72 & 3.8 & 0.86 \\
S, S & 220 & 9322 & 2280 & 67 & 25.5 & 45 & 32 & 3 & 0.56 & 0.8 & 0.49 & 1.51 & 4.89 & 0.94 & 4.6 & 38 & 168 & 1.47 & 0.71 & 3.37 & 1.32 \\
S, A & 158 & 16325 & 1760 & 230 & 16 & 59 & 45 & 5 & 0.68 & 0.72 & 0.81 & 3.91 & 2.68 & 1.75 & 4.69 & 51 & 197 & 1.32 & 0.76 & 4.09 & 1.84 \\
S, S & 220 & 4178 & 577 & 35 & 23.6 & 31 & 21 & 3 & 0.9 & 0.91 & 0.39 & 1.14 & 7.1 & 0.61 & 4.35 & 53 & 200 & 1.29 & 0.68 & 3.12 & 1.15 \\
T, MA & 123 & 1304 & 119 & 28 & 11.4 & 21 & 14 & 1 & 0.4 & 0.65 & 0.54 & 1.35 & 5.86 & 0.5 & 2.96 & 63 & 211 & 1.24 & 0.67 & 2.08 & 0.68 \\
T, LJ & 138 & 3494 & 228 & 61 & 13.3 & 33 & 22 & 2 & 0.54 & 0.56 & 0.64 & 1.85 & 4.18 & 0.77 & 3.21 & 47 & 182 & 1.45 & 0.67 & 2.14 & 1.5 \\
T, D & 315 & 11639 & 569 & 95 & 25.5 & 59 & 40 & 3 & 0.42 & 0.59 & 0.61 & 1.61 & 5.34 & 0.63 & 3.34 & 83 & 254 & 1.39 & 0.68 & 2.23 & 1.97 \\
T, MS & 246 & 17261 & 2888 & 134 & 25.5 & 63 & 42 & 4 & 0.57 & 0.76 & 0.61 & 2.14 & 3.9 & 1.11 & 4.35 & 51 & 206 & 1.49 & 0.67 & 3.3 & 2.03 \\
V, JJM & 131 & 5524 & 491 & 99 & 12.8 & 42 & 29 & 2 & 0.38 & 0.57 & 0.72 & 2.37 & 3.12 & 1 & 3.13 & 58 & 203 & 1.43 & 0.69 & 2.33 & 1.4 \\
W, IA & 209 & 4156 & 383 & 49 & 19.1 & 35 & 23 & 2 & 0.51 & 0.7 & 0.52 & 1.4 & 5.97 & 0.57 & 3.39 & 53 & 203 & 1.4 & 0.66 & 2.3 & 1.3 \\
W, RE & 261 & 12111 & 880 & 103 & 24.6 & 54 & 39 & 4 & 0.62 & 0.72 & 0.6 & 1.92 & 4.83 & 0.86 & 4.15 & 43 & 159 & 1.41 & 0.72 & 3.05 & 1.08 \\
W, RB & 185 & 5611 & 375 & 73 & 17.5 & 40 & 27 & 3 & 0.68 & 0.63 & 0.61 & 1.84 & 4.63 & 0.76 & 3.51 & 64 & 233 & 1.35 & 0.68 & 2.45 & 1.03 \\
W, H & 240 & 11408 & 657 & 127 & 18.9 & 60 & 41 & 4 & 0.59 & 0.57 & 0.71 & 2.13 & 4 & 0.79 & 3.17 & 64 & 255 & 1.28 & 0.68 & 2.43 & 1.4 \\
W, JA & 120 & 2722 & 196 & 60 & 10.8 & 31 & 21 & 1 & 0.21 & 0.54 & 0.68 & 1.94 & 3.87 & 0.73 & 2.83 & 69 & 226 & 1.32 & 0.68 & 2.07 & 1.29 \\
\hline
$\langle x \rangle $ &  217 & 9230 & 1024 & 96 & 20.7 & 44 & 31 & 3 & 0.62 & 0.7 & 0.62 & 2.19 & 4.92 & 1.01 & 4.23 & 52 & 196 & 1.39 & 0.7 & 3.19 & 1.28 \\
$\sigma$ & 121 & 6860 & 1158 & 64 & 10.6 & 14 & 10 & 2 & 0.25 & 0.16 & 0.14 & 1.08 & 2.01 & 0.54 & 1.23 & 11 & 27 & 0.09 & 0.05 & 1.36 & 0.5 \\
\hline
\hline
\end{tabular}}}
\caption{  Career citation statistics for 100 dataset [B] scientists: 51-100. 
}
\label{table:random100b}
\end{table}

\begin{table}[h]
\centering{ {\footnotesize
\begin{tabular}{@{\vrule height .5 pt depth4pt  width0pt}lccccccccccccccccccccc}
\noalign{
\vskip-1pt}
Name & $N$ & $C$ & $c(1)$ & $c(r^{*})$ & $r^{*}$ & $h_1$ & $h_2$ & $h_{80}$ & $\beta_{xy}$ & $\beta$ & $\gamma$ & $c(r^{*})/h_1$ & $N/h_1$ & $\langle c \rangle/h_1$ & $a$ & Gap(5) & Gap(10) & $m_{1/2}$ & $m_{2}$ & $P$ & $h_{1}/L$ \\
\hline
\hline
A, AG & 64 & 1009 & 135 & 28 & 9.4 & 16 & 12 & 1 & 0.66 & 0.81 & 0.59 & 1.78 & 4 & 0.99 & 3.94 & 50 & 174 & 1.31 & 0.75 & 746 & 0.84 \\
A, MW & 32 & 268 & 53 & 9 & 7.07 & 10 & 7 & 0 & 0.18 & 1.12 & 0.44 & 0.98 & 3.2 & 0.84 & 2.68 & 69 & 207 & 1 & 0.7 & 230 & 0.91 \\
A, A & 50 & 1169 & 112 & 61 & 6.06 & 18 & 13 & 1 & 0.95 & 0.7 & 0.86 & 3.42 & 2.78 & 1.3 & 3.61 & 34 & 161 & 1.33 & 0.72 & 982 & 1.38 \\
A, J & 17 & 250 & 83 & 26 & 3 & 8 & 7 & 1 & 0.18 & 0.54 & 1.06 & 3.27 & 2.13 & 1.84 & 3.91 & 40 & 124 & 1.38 & 0.88 & 221 & 1 \\
A, BP & 18 & 2472 & 825 & 157 & 4.07 & 13 & 12 & 3 & 1.63 & 1.17 & 1.28 & 12.14 & 1.38 & 10.56 & 14.63 & 65 & 127 & 1.08 & 0.92 & 2447 & 0.76 \\
A, NP & 39 & 1370 & 208 & 89 & 4.73 & 17 & 14 & 1 & 1.24 & 0.69 & 1.12 & 5.25 & 2.29 & 2.07 & 4.74 & 45 & 176 & 1.35 & 0.82 & 1246 & 1.55 \\
B, A & 51 & 2202 & 440 & 68 & 7.86 & 21 & 15 & 3 & 1.1 & 0.97 & 0.76 & 3.25 & 2.43 & 2.06 & 4.99 & 53 & 198 & 1.29 & 0.71 & 2003 & 2.1 \\
B, DR & 55 & 1245 & 208 & 39 & 8.28 & 18 & 13 & 1 & 0.57 & 0.71 & 0.59 & 2.2 & 3.06 & 1.26 & 3.84 & 47 & 170 & 1.44 & 0.72 & 962 & 1.38 \\
B, M & 71 & 10032 & 1153 & 323 & 8.27 & 40 & 33 & 6 & 1.69 & 0.73 & 0.98 & 8.08 & 1.78 & 3.53 & 6.27 & 38 & 189 & 1.23 & 0.83 & 9477 & 3.08 \\
B, BA & 55 & 1345 & 260 & 40 & 8.54 & 16 & 13 & 2 & 0.95 & 0.84 & 0.64 & 2.51 & 3.44 & 1.53 & 5.25 & 45 & 155 & 1.5 & 0.81 & 1073 & 1.6 \\
B, MD & 17 & 162 & 25 & 11 & 5.33 & 8 & 5 & 0 & 0.43 & 0.93 & 0.37 & 1.45 & 2.13 & 1.19 & 2.53 & 55 & 153 & 1.25 & 0.63 & 133 & 0.73 \\
B, BB & 35 & 646 & 216 & 39 & 4.83 & 14 & 10 & 1 & 0.43 & 0.76 & 0.87 & 2.82 & 2.5 & 1.32 & 3.3 & 70 & 227 & 1.14 & 0.71 & 589 & 0.82 \\
B, SK & 35 & 729 & 198 & 35 & 5.9 & 12 & 10 & 1 & 0.91 & 0.92 & 0.71 & 2.92 & 2.92 & 1.74 & 5.06 & 42 & 158 & 1.33 & 0.83 & 639 & 1.09 \\
B, D & 13 & 894 & 422 & 78 & 3.48 & 7 & 7 & 2 & 1.72 & 1.12 & 1.14 & 11.23 & 1.86 & 9.82 & 18.24 & 15 & 69 & 1.14 & 1 & 885 & 0.7 \\
B, M & 17 & 578 & 225 & 59 & 3.26 & 11 & 8 & 1 & 1.35 & 0.67 & 1.11 & 5.45 & 1.55 & 3.09 & 4.78 & 57 & 171 & 1.18 & 0.73 & 552 & 1.22 \\
B, J & 19 & 252 & 43 & 33 & 2.25 & 10 & 6 & 0 & 0.29 & 0.27 & 0.96 & 3.31 & 1.9 & 1.33 & 2.52 & 62 & 184 & 1.2 & 0.6 & 225 & 1.43 \\
B, R & 24 & 644 & 428 & 17 & 5.63 & 9 & 6 & 1 & 1.1 & 1.4 & 0.56 & 1.98 & 2.67 & 2.98 & 7.95 & 53 & 183 & 1.33 & 0.67 & 608 & 0.64 \\
C, I & 27 & 1492 & 600 & 43 & 6.32 & 12 & 9 & 2 & 2.92 & 1.5 & 0.76 & 3.62 & 2.25 & 4.6 & 10.36 & 55 & 192 & 1.25 & 0.75 & 1430 & 0.86 \\
C, AL & 81 & 4249 & 833 & 111 & 9.12 & 33 & 23 & 2 & 0.71 & 0.62 & 0.8 & 3.38 & 2.45 & 1.59 & 3.9 & 49 & 178 & 1.36 & 0.7 & 3547 & 2.36 \\
C, NJ & 64 & 1432 & 343 & 29 & 10.54 & 17 & 13 & 1 & 0.95 & 1.13 & 0.56 & 1.76 & 3.76 & 1.32 & 4.96 & 55 & 192 & 1.29 & 0.76 & 1222 & 1.13 \\
D, AJ & 27 & 620 & 200 & 26 & 6.07 & 11 & 8 & 1 & 1.35 & 1.03 & 0.6 & 2.38 & 2.45 & 2.09 & 5.12 & 47 & 171 & 1.27 & 0.73 & 547 & 1.22 \\
D, C & 37 & 712 & 124 & 39 & 5.38 & 14 & 12 & 1 & 0.66 & 0.67 & 0.86 & 2.79 & 2.64 & 1.37 & 3.63 & 50 & 166 & 1.36 & 0.86 & 589 & 2 \\
D, M & 32 & 1452 & 431 & 68 & 5.84 & 16 & 12 & 2 & 1.1 & 0.94 & 0.87 & 4.28 & 2 & 2.84 & 5.67 & 49 & 186 & 1.38 & 0.75 & 1343 & 1.78 \\
D, RD & 15 & 1940 & 1036 & 36 & 5.66 & 10 & 9 & 3 & 1.1 & 1.9 & 0.32 & 3.68 & 1.5 & 12.93 & 19.4 & 25 & 82 & 1.2 & 0.9 & 1918 & 0.63 \\
D, R & 31 & 987 & 142 & 74 & 3.84 & 16 & 14 & 1 & 0.49 & 0.48 & 1.06 & 4.64 & 1.94 & 1.99 & 3.86 & 49 & 175 & 1.25 & 0.88 & 877 & 1.45 \\
D, MVG & 11 & 623 & 152 & 160 & 1.23 & 9 & 7 & 1 & 3.11 & 0.05 & 2.1 & 17.79 & 1.22 & 6.29 & 7.69 & 21 & 56 & 1 & 0.78 & 622 & 1 \\
E, DA & 24 & 631 & 146 & 45 & 4.41 & 15 & 9 & 1 & 1.1 & 0.61 & 0.71 & 3.04 & 1.6 & 1.75 & 2.8 & 75 & 234 & 1.2 & 0.6 & 583 & 1.25 \\
E, H & 24 & 793 & 151 & 72 & 3.55 & 14 & 11 & 1 & 0.77 & 0.57 & 1.11 & 5.21 & 1.71 & 2.36 & 4.05 & 57 & 195 & 1.14 & 0.79 & 743 & 1.75 \\
F, A & 56 & 738 & 89 & 29 & 7.08 & 16 & 11 & 1 & 0.35 & 0.67 & 0.73 & 1.83 & 3.5 & 0.82 & 2.88 & 52 & 187 & 1.31 & 0.69 & 552 & 1.45 \\
F, F & 33 & 1042 & 216 & 49 & 6.09 & 16 & 12 & 1 & 1.1 & 0.84 & 0.74 & 3.1 & 2.06 & 1.97 & 4.07 & 56 & 192 & 1.19 & 0.75 & 944 & 1.14 \\
F, GA & 36 & 592 & 87 & 37 & 4.74 & 15 & 10 & 1 & 0.43 & 0.59 & 0.98 & 2.47 & 2.4 & 1.1 & 2.63 & 67 & 208 & 1.2 & 0.67 & 493 & 1.5 \\
F, DP & 74 & 17020 & 5611 & 473 & 8.3 & 47 & 37 & 6 & 1.37 & 0.79 & 1.14 & 10.07 & 1.57 & 4.89 & 7.7 & 52 & 249 & 1.15 & 0.79 & 16575 & 3.62 \\
G, VM & 23 & 458 & 175 & 26 & 5.09 & 12 & 8 & 1 & 0.11 & 0.75 & 0.55 & 2.17 & 1.92 & 1.66 & 3.18 & 59 & 189 & 1.25 & 0.67 & 405 & 1 \\
G, ML & 23 & 1029 & 244 & 130 & 2.64 & 14 & 12 & 1 & 1.1 & 0.5 & 1.62 & 9.3 & 1.64 & 3.2 & 5.25 & 67 & 144 & 1.14 & 0.86 & 999 & 1.08 \\
G, M & 14 & 1576 & 965 & 13 & 6.1 & 7 & 6 & 2 & 2.32 & 2.5 & 0.07 & 1.98 & 2 & 16.08 & 32.16 & 43 & 103 & 1.14 & 0.86 & 1556 & 0.7 \\
G, GH & 53 & 1720 & 318 & 70 & 6.72 & 21 & 17 & 2 & 0.59 & 0.73 & 0.92 & 3.37 & 2.52 & 1.55 & 3.9 & 52 & 183 & 1.24 & 0.81 & 1497 & 1.24 \\
H, H & 50 & 2499 & 364 & 89 & 7.43 & 21 & 17 & 3 & 0.88 & 0.98 & 0.96 & 4.28 & 2.38 & 2.38 & 5.67 & 48 & 185 & 1.19 & 0.81 & 2302 & 1.75 \\
H, F & 39 & 1013 & 208 & 44 & 6.49 & 16 & 11 & 1 & 0.77 & 0.77 & 0.69 & 2.79 & 2.44 & 1.62 & 3.96 & 48 & 180 & 1.38 & 0.69 & 864 & 1.33 \\
H, M & 12 & 54 & 15 & 3 & 5.24 & 5 & 3 & 0 & 0 & 1.2 & 0.17 & 0.77 & 2.4 & 0.9 & 2.16 & 35 & 66 & 1.2 & 0.6 & 47 & 0.29 \\
H, ER & 15 & 413 & 91 & 71 & 1.91 & 10 & 8 & 1 & 0.66 & 0.3 & 1.5 & 7.15 & 1.5 & 2.75 & 4.13 & 48 & 108 & 1.1 & 0.8 & 398 & 1.25 \\
I, A & 16 & 289 & 64 & 37 & 2.77 & 10 & 9 & 0 & 0.05 & 0.56 & 1.33 & 3.76 & 1.6 & 1.81 & 2.89 & 69 & 150 & 1 & 0.9 & 279 & 1.25 \\
I, MF & 30 & 1442 & 586 & 98 & 4.3 & 15 & 13 & 1 & 0.95 & 0.8 & 0.98 & 6.59 & 2 & 3.2 & 6.41 & 65 & 186 & 1.13 & 0.87 & 1402 & 1.07 \\
J, P & 22 & 1075 & 244 & 121 & 3.09 & 13 & 11 & 1 & 1.92 & 0.54 & 1.05 & 9.31 & 1.69 & 3.76 & 6.36 & 63 & 188 & 1.15 & 0.85 & 1046 & 1.08 \\
J, E & 22 & 1469 & 332 & 165 & 2.58 & 17 & 14 & 2 & 0.84 & 0.39 & 1.37 & 9.75 & 1.29 & 3.93 & 5.08 & 84 & 214 & 1.06 & 0.82 & 1439 & 1.55 \\
J, AN & 29 & 466 & 79 & 38 & 2.93 & 15 & 10 & 0 & 0.43 & 0.32 & 0.98 & 2.58 & 1.93 & 1.07 & 2.07 & 79 & 257 & 1.07 & 0.67 & 422 & 1.88 \\
K, E & 21 & 1934 & 377 & 149 & 4.22 & 15 & 12 & 3 & 1.63 & 0.84 & 0.9 & 9.97 & 1.4 & 6.14 & 8.6 & 67 & 210 & 1.13 & 0.8 & 1904 & 1.25 \\
K, HG & 50 & 655 & 73 & 29 & 6.17 & 15 & 9 & 0 & 0.53 & 0.59 & 0.81 & 1.95 & 3.33 & 0.87 & 2.91 & 51 & 193 & 1.4 & 0.6 & 464 & 1.25 \\
K, J & 28 & 1711 & 495 & 79 & 5.84 & 16 & 12 & 2 & 1.63 & 1.02 & 0.71 & 4.94 & 1.75 & 3.82 & 6.68 & 50 & 197 & 1.25 & 0.75 & 1621 & 1.33 \\
K, EA & 27 & 336 & 39 & 29 & 2.86 & 12 & 8 & 0 & 0.11 & 0.31 & 1 & 2.47 & 2.25 & 1.04 & 2.33 & 53 & 199 & 1.33 & 0.67 & 265 & 1.5 \\
K, I & 19 & 384 & 88 & 37 & 3.59 & 10 & 8 & 1 & 0.66 & 0.68 & 0.95 & 3.76 & 1.9 & 2.02 & 3.84 & 52 & 170 & 1.2 & 0.8 & 354 & 0.91 \\
\hline
$\langle x \rangle$ & 33 & 1334 & 326 & 71 & 5.1 & 14 & 11 & 1 & 0.99 & 0.79 & 0.89 & 4.91 & 2.26 & 3.35 & 6.15 & 51 & 167 & 1.22 & 0.77 & 1225 & 1.29 \\
$\sigma$ & 19 & 2022 & 596 & 70 & 2.04 & 7 & 5 & 1 & 0.67 & 0.38 & 0.36 & 4.24 & 0.86 & 4.57 & 6.21 & 14 & 45 & 0.18 & 0.09 & 1948 & 0.56 \\
\hline
\hline
\end{tabular}}}
\caption{  Career citation statistics for 100 dataset [C] scientists: 1-50. 
}
\label{table:asst100a}
\end{table}

\begin{table}[h]
\centering{ { \footnotesize
\begin{tabular}{@{\vrule height .5 pt depth4pt  width0pt}lccccccccccccccccccccc}
\noalign{
\vskip-1pt}
Name & $N$ & $C$ & $c(1)$ & $c(r^{*})$ & $r^{*}$ & $h_1$ & $h_2$ & $h_{80}$ & $\beta_{xy}$ & $\beta$ & $\gamma$ & $c(r^{*})/h_1$ & $N/h_1$ & $\langle c \rangle/h_1$ & $a$ & Gap(5) & Gap(10) & $m_{1/2}$ & $m_{2}$ & $P$ & $h_{1}/L$ \\
\hline
\hline
K, SM & 15 & 395 & 71 & 56 & 2.17 & 10 & 9 & 0 & 0.53 & 0.29 & 1.07 & 5.68 & 1.5 & 2.63 & 3.95 & 58 & 166 & 1.1 & 0.9 & 376 & 1.43 \\
K, AA & 11 & 1028 & 383 & 160 & 2.61 & 9 & 8 & 2 & 2.32 & 0.67 & 1.49 & 17.81 & 1.22 & 10.38 & 12.69 & 49 & 152 & 1.11 & 0.89 & 1020 & 0.82 \\
K, IN & 42 & 1815 & 567 & 71 & 6.52 & 20 & 15 & 2 & 0.74 & 0.78 & 0.72 & 3.6 & 2.1 & 2.16 & 4.54 & 53 & 190 & 1.3 & 0.75 & 1638 & 1.25 \\
L, A & 21 & 289 & 69 & 25 & 3.86 & 9 & 7 & 0 & 0.18 & 0.68 & 0.85 & 2.86 & 2.33 & 1.53 & 3.57 & 48 & 152 & 1.33 & 0.78 & 251 & 0.82 \\
L, LJ & 18 & 815 & 290 & 83 & 3.33 & 9 & 8 & 2 & 1.35 & 0.93 & 1.3 & 9.28 & 2 & 5.03 & 10.06 & 49 & 146 & 1.22 & 0.89 & 794 & 0.82 \\
L, RL & 44 & 879 & 91 & 43 & 5.49 & 16 & 11 & 1 & 0.77 & 0.52 & 0.84 & 2.71 & 2.75 & 1.25 & 3.43 & 40 & 166 & 1.56 & 0.69 & 658 & 0.89 \\
L, BJ & 19 & 591 & 156 & 23 & 5.43 & 8 & 7 & 1 & 1.72 & 1.4 & 0.72 & 2.88 & 2.38 & 3.89 & 9.23 & 33 & 145 & 1.5 & 0.88 & 547 & 1 \\
L, J & 36 & 881 & 93 & 41 & 5.96 & 17 & 13 & 1 & 0.57 & 0.5 & 0.55 & 2.44 & 2.12 & 1.44 & 3.05 & 45 & 167 & 1.41 & 0.76 & 688 & 0.81 \\
L, Y & 14 & 648 & 177 & 47 & 4.55 & 11 & 8 & 1 & 1.35 & 0.9 & 0.53 & 4.31 & 1.27 & 4.21 & 5.36 & 29 & 87 & 1.18 & 0.73 & 627 & 1 \\
M, O & 22 & 151 & 32 & 5 & 6.76 & 5 & 4 & 0 & 0.66 & 1.1 & 0.23 & 1.11 & 4.4 & 1.37 & 6.04 & 36 & 128 & 1.6 & 0.8 & 112 & 0.33 \\
M, V & 80 & 2038 & 206 & 62 & 8.71 & 24 & 17 & 2 & 0.59 & 0.76 & 0.88 & 2.58 & 3.33 & 1.06 & 3.54 & 41 & 183 & 1.33 & 0.71 & 1670 & 2.18 \\
M, BA & 19 & 378 & 202 & 17 & 5.26 & 9 & 6 & 1 & 0.29 & 1.09 & 0.41 & 1.93 & 2.11 & 2.21 & 4.67 & 53 & 118 & 1.33 & 0.67 & 348 & 0.82 \\
M, L & 18 & 966 & 345 & 140 & 2.22 & 13 & 10 & 1 & 0.91 & 0.39 & 1.51 & 10.81 & 1.38 & 4.13 & 5.72 & 49 & 79 & 1.08 & 0.77 & 951 & 1.63 \\
M, P & 15 & 723 & 180 & 100 & 2.77 & 10 & 8 & 2 & 2.32 & 0.74 & 1.58 & 10.06 & 1.5 & 4.82 & 7.23 & 68 & 124 & 1.1 & 0.8 & 712 & 1 \\
M, D & 24 & 1029 & 283 & 53 & 5.15 & 12 & 11 & 2 & 0.77 & 1.22 & 1.01 & 4.49 & 2 & 3.57 & 7.15 & 56 & 187 & 1.17 & 0.92 & 990 & 0.92 \\
M, B & 16 & 132 & 60 & 5 & 4.7 & 5 & 3 & 0 & 1.1 & 1.53 & 0.51 & 1.16 & 3.2 & 1.65 & 5.28 & 46 & 106 & 1.2 & 0.6 & 122 & 0.42 \\
M, E & 38 & 374 & 77 & 14 & 7.15 & 10 & 8 & 0 & 0.66 & 0.78 & 0.51 & 1.48 & 3.8 & 0.98 & 3.74 & 39 & 157 & 1.4 & 0.8 & 266 & 1.43 \\
M, OI & 36 & 716 & 90 & 53 & 3.71 & 15 & 11 & 1 & 0.77 & 0.44 & 1.06 & 3.55 & 2.4 & 1.33 & 3.18 & 34 & 165 & 1.33 & 0.73 & 597 & 1.07 \\
M, AE & 44 & 1695 & 303 & 56 & 7.67 & 18 & 12 & 2 & 2.32 & 1.02 & 0.73 & 3.15 & 2.44 & 2.14 & 5.23 & 57 & 200 & 1.33 & 0.67 & 1524 & 1.5 \\
N, D & 45 & 1427 & 231 & 77 & 5.46 & 19 & 15 & 1 & 0.74 & 0.7 & 1.01 & 4.09 & 2.37 & 1.67 & 3.95 & 61 & 204 & 1.16 & 0.79 & 1303 & 1.73 \\
N, A & 25 & 759 & 116 & 69 & 2.62 & 16 & 14 & 1 & 0.49 & 0.26 & 0.93 & 4.32 & 1.56 & 1.9 & 2.96 & 60 & 186 & 1.13 & 0.88 & 707 & 1.45 \\
N, V & 7 & 1536 & 1174 & 147 & 2.8 & 6 & 6 & 2 & 2.32 & 0.97 & 2.16 & 24.56 & 1.17 & 36.57 & 42.67 & 10 & 15 & 1 & 1 & 1535 & 1 \\
N, Z & 54 & 1175 & 258 & 44 & 6.98 & 19 & 13 & 1 & 0.57 & 0.73 & 0.85 & 2.35 & 2.84 & 1.15 & 3.25 & 49 & 194 & 1.32 & 0.68 & 977 & 1.19 \\
O, AL & 28 & 516 & 93 & 65 & 1.96 & 14 & 10 & 1 & 0.43 & 0.3 & 1.28 & 4.65 & 2 & 1.32 & 2.63 & 79 & 198 & 1.07 & 0.71 & 486 & 1.27 \\
O, SB & 33 & 505 & 92 & 27 & 5.59 & 12 & 8 & 1 & 0.66 & 0.86 & 0.63 & 2.3 & 2.75 & 1.28 & 3.51 & 45 & 186 & 1.25 & 0.67 & 440 & 0.86 \\
P, N & 49 & 5544 & 550 & 241 & 6.33 & 30 & 22 & 5 & 1.92 & 0.83 & 1.15 & 8.05 & 1.63 & 3.77 & 6.16 & 82 & 270 & 1.1 & 0.73 & 5374 & 2.73 \\
P, NB & 26 & 225 & 45 & 14 & 4.92 & 8 & 6 & 0 & 0.29 & 0.72 & 0.62 & 1.84 & 3.25 & 1.08 & 3.52 & 40 & 155 & 1.38 & 0.75 & 169 & 0.67 \\
P, AT & 24 & 2094 & 659 & 153 & 4.11 & 13 & 11 & 3 & 2.8 & 1.22 & 1.47 & 11.8 & 1.85 & 6.71 & 12.39 & 75 & 198 & 1.08 & 0.85 & 2072 & 0.87 \\
P, MG & 14 & 206 & 57 & 24 & 2.33 & 9 & 6 & 0 & 0.29 & 0.26 & 0.64 & 2.75 & 1.56 & 1.63 & 2.54 & 38 & 172 & 1.22 & 0.67 & 188 & 1 \\
P, A & 48 & 1000 & 106 & 48 & 5.83 & 18 & 13 & 1 & 0.57 & 0.57 & 0.75 & 2.67 & 2.67 & 1.16 & 3.09 & 41 & 170 & 1.33 & 0.72 & 808 & 1.38 \\
P, F & 21 & 998 & 317 & 63 & 4.83 & 13 & 11 & 1 & 0.77 & 0.78 & 0.67 & 4.91 & 1.62 & 3.66 & 5.91 & 38 & 141 & 1.38 & 0.85 & 928 & 1 \\
P, S & 62 & 1576 & 231 & 51 & 8.22 & 23 & 16 & 1 & 0.38 & 0.67 & 0.59 & 2.26 & 2.7 & 1.11 & 2.98 & 51 & 207 & 1.22 & 0.7 & 1306 & 2.56 \\
S, T & 121 & 3788 & 240 & 81 & 11.89 & 33 & 25 & 2 & 0.81 & 0.74 & 0.94 & 2.48 & 3.67 & 0.95 & 3.48 & 35 & 150 & 1.36 & 0.76 & 2950 & 2.36 \\
S, TR & 60 & 933 & 133 & 55 & 4.81 & 16 & 12 & 1 & 0.66 & 0.55 & 0.95 & 3.44 & 3.75 & 0.97 & 3.64 & 36 & 156 & 1.38 & 0.75 & 769 & 1.14 \\
S, D & 19 & 1906 & 686 & 171 & 3.39 & 16 & 15 & 2 & 0.74 & 0.53 & 0.9 & 10.7 & 1.19 & 6.27 & 7.45 & 52 & 159 & 1.13 & 0.94 & 1878 & 2.29 \\
S, MD & 17 & 548 & 194 & 26 & 5.66 & 12 & 9 & 1 & 1.1 & 1.04 & 0.19 & 2.24 & 1.42 & 2.69 & 3.81 & 49 & 160 & 1.17 & 0.75 & 520 & 1.2 \\
S, L & 22 & 476 & 72 & 43 & 3.62 & 12 & 10 & 0 & 0.43 & 0.61 & 1.04 & 3.6 & 1.83 & 1.8 & 3.31 & 65 & 201 & 1.17 & 0.83 & 436 & 2 \\
S, OG & 22 & 444 & 83 & 48 & 2.66 & 13 & 8 & 1 & 0.66 & 0.38 & 1 & 3.72 & 1.69 & 1.55 & 2.63 & 69 & 218 & 1.15 & 0.62 & 407 & 1.08 \\
S, GT & 19 & 284 & 43 & 27 & 3.5 & 9 & 7 & 0 & 0.18 & 0.52 & 0.75 & 3.05 & 2.11 & 1.66 & 3.51 & 44 & 158 & 1.33 & 0.78 & 244 & 0.9 \\
S, M & 47 & 1265 & 126 & 68 & 4.51 & 21 & 15 & 1 & 0.43 & 0.41 & 0.92 & 3.25 & 2.24 & 1.28 & 2.87 & 55 & 197 & 1.29 & 0.71 & 1047 & 1.75 \\
S, AM & 51 & 1431 & 288 & 70 & 5.4 & 21 & 15 & 1 & 0.43 & 0.61 & 1.07 & 3.35 & 2.43 & 1.34 & 3.24 & 49 & 213 & 1.24 & 0.71 & 1241 & 1.5 \\
T, N & 14 & 2073 & 1256 & 45 & 4.72 & 8 & 7 & 2 & 1.72 & 2.12 & 0.8 & 5.71 & 1.75 & 18.51 & 32.39 & 54 & 88 & 1.13 & 0.88 & 2058 & 0.89 \\
T, AP & 29 & 85 & 26 & 3 & 6.58 & 4 & 2 & 0 & 2.32 & 0.87 & 0.16 & 0.91 & 7.25 & 0.73 & 5.31 & 37 & 131 & 1.5 & 0.5 & 47 & 0.57 \\
T, H & 22 & 347 & 88 & 27 & 4.07 & 11 & 8 & 1 & 0.66 & 0.63 & 0.82 & 2.52 & 2 & 1.43 & 2.87 & 54 & 195 & 1.18 & 0.73 & 303 & 1.22 \\
V, O & 22 & 653 & 123 & 55 & 3.93 & 12 & 9 & 1 & 1.1 & 0.73 & 1.02 & 4.65 & 1.83 & 2.47 & 4.53 & 56 & 176 & 1.17 & 0.75 & 606 & 0.92 \\
V, MG & 41 & 804 & 129 & 43 & 5.75 & 17 & 12 & 1 & 0.66 & 0.78 & 0.89 & 2.58 & 2.41 & 1.15 & 2.78 & 70 & 237 & 1.24 & 0.71 & 707 & 1.21 \\
W, RH & 57 & 2377 & 364 & 107 & 6.27 & 25 & 19 & 2 & 1 & 0.68 & 1.12 & 4.31 & 2.28 & 1.67 & 3.8 & 49 & 189 & 1.28 & 0.76 & 2110 & 1.92 \\
W, M & 18 & 469 & 90 & 38 & 4.07 & 10 & 9 & 1 & 1.1 & 0.82 & 0.9 & 3.85 & 1.8 & 2.61 & 4.69 & 54 & 136 & 1 & 0.9 & 440 & 1.25 \\
Y, A & 21 & 1176 & 536 & 126 & 3.09 & 9 & 7 & 2 & 1.72 & 1.05 & 1.53 & 14.11 & 2.33 & 6.22 & 14.52 & 58 & 157 & 1 & 0.78 & 1168 & 1.29 \\
Z, MW & 20 & 3151 & 592 & 311 & 3.52 & 16 & 15 & 4 & 2.02 & 0.63 & 0.95 & 19.48 & 1.25 & 9.85 & 12.31 & 16 & 31 & 0 & 0.94 & 3125 & 2 \\
\hline
$\langle x \rangle$ & 33 & 1334 & 326 & 71 & 5.1 & 14 & 11 & 1 & 0.99 & 0.79 & 0.89 & 4.91 & 2.26 & 3.35 & 6.15 & 51 & 167 & 1.22 & 0.77 & 1225 & 1.29 \\
$\sigma$ & 19 & 2022 & 596 & 70 & 2.04 & 7 & 5 & 1 & 0.67 & 0.38 & 0.36 & 4.24 & 0.86 & 4.57 & 6.21 & 14 & 45 & 0.18 & 0.09 & 1948 & 0.56 \\
\hline
\hline
\end{tabular}}}
\caption{  Career citation statistics for 100 dataset [C] scientists: 51-100. 
}
\label{table:asst100b}
\end{table}


\end{widetext}

\end{document}